\documentclass[twocolumn,preprintnumbers,amsmath,amssymb,amsfonts,final,a4paper,aps,citeautoscript,footinbib,prb,superscriptaddress]{revtex4-1}
\usepackage{graphicx}
\usepackage{tabularx} %to redefine size in tables and arrays+for newcolumntype
\usepackage[caption=false]{subfig} %to put several file inside a single figure
\usepackage{amsmath,amssymb,amsfonts,pifont,fancybox,float,mathtools}
\usepackage{graphicx}
\usepackage{epsfig}

\usepackage{times}
\usepackage{color}
\usepackage[pdftex,dvipsnames]{xcolor}  % Coloured text etc.

\usepackage{epsfig}
\usepackage[colorlinks,bookmarks=true,citecolor=blue,linkcolor=blue,urlcolor=blue, breaklinks=true]{hyperref}
\usepackage{simplewick}
\usepackage[latin1]{inputenc}
\usepackage{mathrsfs}
\usepackage[vcentermath,noautoscale]{youngtab}
\newcommand{\bolN}{\boldsymbol{N}}

\usepackage{epstopdf}
\graphicspath{{./figures/}{./Figs/}}
\begin{document}

%%%%%%%%%%%%%%%%%%%%%%%%%%%%%%%%%%%%%%%%%%%%%%%%%%%%%%%
\title{Even-odd effects in the $J_1-J_2$ SU($N$) Heisenberg spin chain}
\author{L. Herviou} 
\affiliation{Institute of Physics, Ecole Polytechnique F\'ed\'erale de Lausanne (EPFL), CH-105 Lausanne, Switzerland.}
\author{S. Capponi} 
\affiliation{Laboratoire de Physique Th\'eorique,
  Universit\'e de Toulouse, CNRS, UPS, France.}
\author{P. Lecheminant} 
\affiliation{Laboratoire de Physique Th\'eorique et
 Mod\'elisation, CNRS, CY Cergy Paris Universit\'e,
95302 Cergy-Pontoise Cedex, France.}
\date{\today}

%%%%%%%%%%%%%%%%%%%%%%%%%%%%%%%%%%%%%%%%%%%%%%%%%%%%%%
\begin{abstract}
The zero-temperature phase diagram of the $J_1-J_2$ SU($N$) antiferromagnetic Heisenberg spin chain  is investigated by means of 
complementary field theory and numerical approaches  for general $N$.
A fully gapped SU($N$)  valence bond solid made of $N$ sites is formed above a critical value of $J_2/J_1$ for all $N$. 
We find that the extension of this $N$-merized phase for larger values of $J_2$ strongly depends on the parity of $N$. For even $N$, the phase smoothly interpolates to the large $J_2$ regime where the model can be viewed as a zigzag SU($N$) two-leg spin ladder.  The phase exhibits both a $N$-merized ground state and incommensurate spin-spin correlations.
In stark contrast to the even case, we show that the $N$-merized phase with odd $N$ has only a finite extent with no incommensuration. A gapless phase in the SU($N$)$_1$ universality class is stabilized for larger $J_2$ that stems from the existence of a massless renormalization group flow from SU($N$)$_2$ to SU($N$)$_1$ conformal field theories when $N$ is odd.

\end{abstract}
%%%%%%%%%%%%%%%%%%%%%%%%%%%%%%%%%%%%%%%%%%%%%%%%%%%%%%
\maketitle
%%%%%%%%%%%%%%%%%%%%%%%%%%%%%%%%%%%%%%%%%%%%%%%%%%%%%%
\section{Introduction}
\label{sec:intro}
%%%%%%%%%%%%%%%%%%%%%%%%%%%%%%%%%%%%%%%%%%%%%%%%%%%%%%
Since the proposal made by Anderson \cite{Anderson-73},  a central focus of quantum magnetism over the years  has been the study of the interplay between frustration and quantum fluctuations. Competing interactions or geometric frustration strongly enhance quantum fluctuations and the magnetic order can be destroyed by large low-energy excitations. As a result, novel quantum phases of matter, such as quantum spin liquids, can emerge in such systems.~\cite{Savary_2016} Quantum fluctuations are also amplified by considering lattice spin systems with a larger SU($N$)  internal symmetry than the SU(2) spin rotation.~\cite{Affleck-M-88,Read-S-PRL89,Wu-MPL-06}
Non-magnetic ground states are then more likely for large values of $N$ since the extensive degeneracy in the semiclassical limit is larger. 
In this respect, large-$N$ approaches on different two-dimensional SU($N$)  spin models found a variety of non-N\'eel ground states as valence-bond solid (VBS) states and Abelian and non-Abelian chiral spin liquid ground states with topological order.\cite{Affleck-M-88,Read-S-PRL89,Hermele-G-11}
The latter states are magnetic counterparts of the fractional quantum Hall phases and have excitations with fractional quantum numbers and fractional statistics or non-Abelian ones.  For smaller values of $N$ ($3 \le N \le 10$), numerical and variational approaches have identified chiral and algebraic spin liquids ground states in various two-dimensional SU($N$) quantum magnets.\cite{Corboz2012,Nataf-L-W-P-M-L-16,ChenPRB2021,Yamada-2021,Xu-Ping2021,Tu-2022}
In one dimension, the SU($N$) symmetry also plays a key role by stabilising exotic symmetry protected topological phases which are beyond the SU(2) case. \cite{Nonne-M-C-L-T-13,Morimoto-U-M-F-14,Bois-C-L-M-T-15,Capponi-L-T-16,Roy-Q-18,Ueda-M-M-18,Fromholz-C-L-P-T-19}

The simplest model to analyze the interplay between frustration and an SU($N$) symmetry in one dimension is 
the $J_1-J_2$ antiferromagnetic SU($N$) Heisenberg spin chain with Hamiltonian:
 \begin{equation}
H_{J_1-J_2} =  J_{1} \sum_{i,A}  S^{A}_{i}  S^{A}_{i+1} + J_{2} \sum_{i,A}  S^{A}_{i}  S^{A}_{i+2},
\label{hamJ1J2}
\end{equation}
where $S^{A}_{i}$ ($A=1, \ldots, N^2 -1$) denotes the SU($N$) spin operators 
on the $i$ th site of the chain which transform in the fundamental representation $\bolN$ of  the SU($N$) group.  
These spin operators are normalized as $\text{Tr}(S^{A} S^{B})=\delta^{AB}/2$ and the spin exchanges 
are antiferromagnetic ($J_{1,2} \ge 0$).  Model (\ref{hamJ1J2}) can be realized by loading ultracold fermionic alkaline-earth or ytterbium atoms in a zigzag optical lattice as in Ref. \onlinecite{Zhang-G-15} and driving them into the Mott-insulating regime with one atom per site. 
With the considerable progress made in ultracold SU($N$) fermionic experiments\cite{Taie-Y-S-T-12,Zhang-et-al-Sr-14,Pagano14,Hofrichter-SUN-Mott-16,OzawaPRL2018,Taie2022nat}, the  $J_1-J_2$ SU($N$) Heisenberg spin chain could be achieved in the near future. 

The phase diagram of model (\ref{hamJ1J2}) is well-known for $N=2$ and the main effect of the competing $J_2$ interaction is to produce a Kosterlitz-Thouless transition between the critical phase of the spin-1/2 Heisenberg chain and a spontaneously dimerized phase. \cite{Haldane-NNN-82}
In the presence of frustration, quantum fluctuations destroy the quasi-long-range order of a spin-1/2 Heisenberg chain by producing a spectral gap
and the system dimerizes as exemplified by the exactly soluble Majumdar-Ghosh point when $J_2 = J_1/2$. \cite{MajumdarGhosh69}
The ground state is two-fold degenerate and can be viewed as a valence bond solid made of two spins, breaking spontaneously the one-step translation symmetry T$_{a_0}$. The dimerized phase is stabilized when $J_2/J_1 \ge 0.2411$. \cite{OkamotoPLA92}
The gapless spinons of the spin-1/2 Heisenberg chain  with fractional $S=1/2$ quantum numbers  becomes fully gapped in the dimerized phase. They are 
still deconfined excitations and can be viewed as the domain walls between the two-degenerate ground states. \cite{Shastry-81}
This dimerized phase extends to the strong $J_2$ regime where the $J_1-J_2$ Heisenberg spin chain (\ref{hamJ1J2}) can be viewed as a two-leg  spin ladder with a triangular or zigzag geometry.\cite{White-A-96,Allen-S-97,Nersesyan1998,ItoiQin}

 In stark contrast to $N=2$, little is known for the phase diagram of model (\ref{hamJ1J2}) for general $N >2$ even in the large $N$ limit. When $J_2 =0$, model (\ref{hamJ1J2}), e.g. the SU($N$) Heisenberg spin chain, is the so-called Sutherland model which is integrable by means of Bethe ansatz.  \cite{Sutherland-75} 
It displays a quantum critical behavior in the SU($N$)$_1$ Wess-Zumino-Novikov-Witten (WZNW)  universality class with central charge $c=N-1$. \cite{Affleck-NP86,Affleck-88}
The predictions of the exact solution and of the conformal field theory (CFT) approach have been carefully checked numerically by means of various methods. \cite{Frischmuth-M-T-99,Assaraf-A-C-L-99,Fuhringer-R-T-G-S-08,AguadoPRB2009,Manmana-H-C-F-R-11,MessioPRL2012,Dufour-N-M-15,Nataf-M-18}
The low-lying gapless excitations occur in pairs with individual dispersion relations covering
a fraction of the Brillouin zone.\cite{Johannesson-86}
The elementary excitations of the model are then a generalization of the spinons of the spin-1/2 Heisenberg chain
and carry fractional quantum numbers. They transform in the conjugate ${\bf {\bar N}}$ representation of the SU($N$) group
and, in this respect, they may be viewed as an analog of antiquarks in quantum chromodynamics.\cite{Bouwknegt-S-96, SchurichtGreiterPRB06}
It has been shown that they display fractional statistics with angle $\theta = \pi/N$. \cite{Bouwknegt-S-96, SchurichtGreiterPRB06,Greiter-R-07}

What happens for these spinons excitations upon switching on a nonzero $J_2$ when $N>2$ ? The lesson gained from the $N=2$ case leads us to expect that 
the natural instability of the gapless phase with $c=N-1$  is the formation of a singlet cluster phase of $N$ sites, a $N$-merized phase. The latter is 
the natural generalization of the valence bond solid of $N=2$, since $N$ is the minimum number of spins needed to form an SU($N$) singlet for spins in the fundamental representation of the SU($N$) group.
The $J_1-J_2$ SU(3) spin chain model has been investigated numerically by means of the density-matrix renormalization group (DMRG) and exact diagonalizations (EDs). A spontaneous trimerized phase, a singlet cluster phase of three sites, has been revealed when $0.45 \le J_2/J_1\le 3.5$. \cite{corbozPRB2007}
The phase is threefold degenerate and breaks spontaneously T$_{a_0}$. The generalized spinons become massive deconfined excitations which correspond to the domain walls of the trimerized phase similarly to the $N=2$ case. \cite{Greiter-R-S-07,Greiter-R-07,Rachel-T-F-S-G-09}
Interestingly enough, it was found numerically that the trimerized phase has a finite extent and does not extend to the large $J_2$ regime  in stark contrast to the $N=2$ case.\cite{corbozPRB2007} A critical SU(3)$_1$ phase with $c=2$ is expected to show up with deconfined critical spinons at sufficiently large $J_2$.

\begin{figure}
\includegraphics[width=\linewidth,clip]{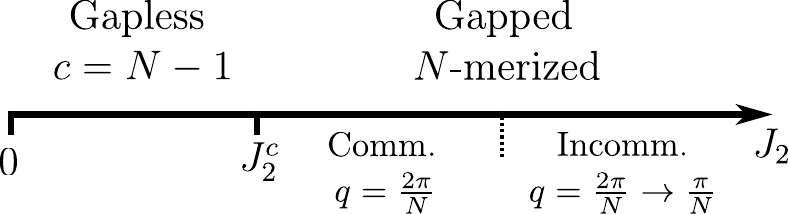}\\
\includegraphics[width=\linewidth,clip]{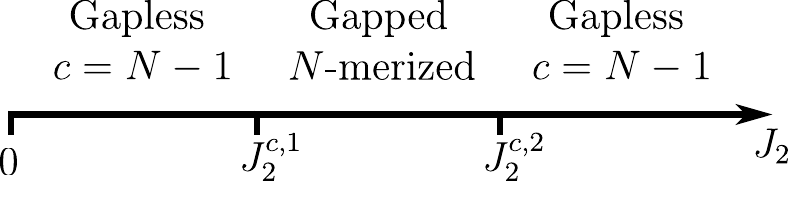}
\caption{Phase diagrams}
\label{fig:PDs}
\end{figure}

In this paper, we map out the phase diagram of model (\ref{hamJ1J2}) at zero temperature by means of complementary CFT techniques for $N>2$ and numerical approaches ED and infinite size DMRG (iDMRG) for $N=3,4$. At intermediate $J_2$ for all $N$, we find the existence of a  $N$-merized phase which is $N$-fold degenerate and breaks spontaneously T$_{a_0}$. The domain-wall excitations between consecutive degenerate ground states, i.e. $N$-merization kinks, have fractional quantum numbers and correspond to the deconfined gapped SU($N$) spinons which transform in the ${\bar \bolN}$-representation of  SU($N$). The extension of the $N$-merized  phase in the large $J_2$ regime can be investigated numerically for $N=3$ and $N=4$ as well as by a field theory approach which exploits the existence of a decoupling critical point when $J_2 \rightarrow \infty$. In this large $J_2$  regime, the model is best visualized as a two-leg zigzag spin ladder where the $J_1$ bonds couple two SU($N$) Heisenberg spin chains with spin-exchange $J_2$:
\begin{equation}
\begin{split}
H_{\text{zigzag}} =&  J_{2} \sum_{i,A}  \left( S^{A}_{1,i}  S^{A}_{1,i+1} 
+ S^{A}_{2,i}  S^{A}_{2,i+1} \right)  \\
&+ J_{1} \sum_{i,A}  S^{A}_{2,i}  \left( S^{A}_{1,i}  + S^{A}_{1,i+1} \right) ,
\end{split}
\label{eqn:def-2leg-ladder}
\end{equation}
where $S^{A}_{1,i}$ and $S^{A}_{2,i}$ ($A=1, \ldots, N^2 -1$) denote the SU($N$) spin operators 
on the $i$ th site on the chains 1 and 2 which transform in the fundamental representation $\bolN$ of  the SU($N$) group. The field theory analysis of the  two-leg SU($N$)  zigzag spin ladder (\ref{eqn:def-2leg-ladder})  in the  regime $J_1 \ll J_2$ reveals that the extension of the $N$-merized phase strongly depends on the parity of $N$. In the odd $N$ case, the existence of a massless renormalization group (RG) flow from SU($N$)$_2$ to SU($N$)$_1$ CFTs leads to the emergence of a gapless phase with SU($N$)$_1$ quantum criticality when $J_1 \ll J_2$. The $N$-merized phase has thus a finite extent $ J_2^{c, 1} \le J_2 \le J_2^{c, 2}$, surrounded by two SU($N$)$_1$  gapless phases when $N$ is odd. The original spinons of the Sutherland model experience a sequence of two transitions. A first one at $J_2=  J_2^{c, 1}$, where they become fully gapped deconfined excitations in the $N$-merized phase and then at $J_2=  J_2^{c, 2}$, where the gap closes and the spinons again become gapless. In contrast, the situation is very different in the even $N$ case. We show analytically and numerically for $N=4$ that the tetramerized phase smoothly interpolates to the strong-coupling large $J_2$ regime. In addition, we find that this phase for $J_2 > 2 J_1$ is characterized by an incommensurate behavior in the spin-spin correlation function as in the $N=2$ case whereas no such incommensuration is obtained for $N=3$. Our conjectured phase diagrams are summarized in Fig.~\ref{fig:PDs}.

The paper is organized as follows.  The low-energy approach appropriate in the weak-coupling  
region $J_2 \ll J_1$ and in the strong-coupling one $J_1 \ll J_2$ is presented in Sec.~\ref{sec:weakcoupling}.
The results of extensive ED and DMRG calculations in the $N=3$ and $N=4$ cases are described in Sec.~\ref{sec:dmrg}.
Finally, a summary of the main results is given in Sec.~\ref{sec:conclusion} together with five technical appendixes.

 %%%%%%%%%%%%%%%%%%%%%%%%%%%%%%%%%%%%%%%%%%%%%%%%%%%%%%%
\section{Low-energy description}
%%%%%%%%%%%%%%%%%%%%%%%%%%%%%%%%%%%%%%%%%%%%%%%%%%%%%%%
\label{sec:weakcoupling}

The phase diagram of model (\ref{hamJ1J2}) is investigated by a low-energy field theory approach using two different limits: the weak-coupling regime when 
$J_2 \ll J_1$ and the strong-coupling case $J_1 \ll J_2$ using the two-leg zigzag ladder geometry (\ref{eqn:def-2leg-ladder}).

\subsection{Weak-coupling regime: $J_2 \ll J_1$}

Let us first consider the weak-coupling  limit $J_2 \ll J_1$ by considering model (\ref{hamJ1J2}).
When $J_2 =0$, the Sutherland model (\ref{hamJ1J2}) displays a gapless behavior which is described by the SU($N$)$_1$ CFT perturbed by a marginally irrelevant current-current interaction with Hamiltonian density:\cite{Affleck-NP86,Affleck-88}
\begin{equation}
  \mathcal{H}_{\rm sutherland} = \frac{2\pi v}{N + 1} \left( : J^A_{R} J^A_{R}: + : J^A_{L} J^A_{L}: 
\right)  - \gamma J_R^{A} J_L^{A} ,
\label{Hcontsutherland}
\end{equation}
where $v$ is the spin velocity and  $J^{A}_{L,R}$ are the left and right SU($N$)$_1$ currents. In Eq. (\ref{Hcontsutherland}), a summation over repeated indices $A=1, \ldots, N^2 - 1$ is assumed as in the following. In the low-energy limit, a SU($N$)$_1$ quantum critical behavior with a central charge $c=N-1$ is stabilized and the marginally irrelevant current-current interaction ($\gamma >0$) leads to logarithmic corrections in correlation functions~\cite{Affleck-G-S-Z-89,Itoi-K-97,Majumdar_2002}.

The effect of the $J_2$ interaction of model  (\ref{hamJ1J2})  can be investigated in the weak-coupling limit since the lattice SU($N$) operators in the low-energy limit can be expressed in terms of the fields of the SU($N$)$_1$ CFT as: \cite{Affleck-88,Assaraf-A-C-L-99,James-K-L-R-T-18}
\begin{equation}
S^{A}_{n}/a_0 \simeq J^{A}_{L} +  J^{A}_{R} + \left(\mbox{e}^{ i 2k_F x} N^{A} + \mathrm{H.c.} \right)
+ \sum_{m=2}^{N-2} \mbox{e}^{ i 2m k_F x} n_m^{A},
\label{spinopappen}
\end{equation}
where $k_F = \pi/Na_0$, $x=n a_0$, $a_0$ being the lattice spacing.  The $2k_F$ part of the SU($N$) spin density involves the SU($N$)$_1$ WZNW field $g$ with scaling dimension $(N-1)/N$ which transforms in the $\bolN$-representation of the SU($N$) group:
\begin{equation}
N^{A} = i C  \; {\rm Tr} ( g T^A),
\label{WZNWcont}
\end{equation}
$C$ being a nonuniversal real constant and $T^A$ are the generators in the $\bolN$-representation normalized as: $\text{Tr}(T^{A} T^{B})=\delta^{AB}/2$.
The remaining $2m k_F$ parts of the decomposition (\ref{spinopappen}) are related to the $m=2, \ldots, N-2$ 
SU($N$)$_1$ primary fields $\Phi_{m}$ with scaling dimension $m(N-m)/N$ which transform
in the fully antisymmetric representation  of SU($N$)  made of a Young tableau with a single column and $m$ lines:
\begin{equation}
 n_m^{A} = i \alpha_m  \; {\rm Tr} ( \Phi_{m} T_m^A),
\label{spinopprimaryappen}
\end{equation}
where $T_m^A$ are SU($N$) generators in the $m$th fully antisymmetric representation of the SU($N$) group and $\alpha_m$ are nonuniversal real constants. The $2mk_F$ component of the spin-density (\ref{spinopappen}) satisfies the constraint: $n_m^{A \dagger} = n_{N-m}^{A}$.

Using the low-energy description (\ref{spinopappen}), one can derive the continuum limit of model (\ref{hamJ1J2}) in the weak-coupling regime $J_2 \ll J_1$. Its Hamiltonian density reads:
\begin{equation}
  \mathcal{H}_{J_1-J_2} = \frac{2\pi v}{N + 1} \left( : J^A_{R} J^A_{R}: + : J^A_{L} J^A_{L}: 
\right)  + \lambda  J_R^{A} J_L^{A} ,
\label{HcontJ1J2}
\end{equation}
where $\lambda \simeq 2 a_0 J_2 - \gamma$. Model (\ref{HcontJ1J2}) is an integrable field theory and its low-energy properties are well known.\cite{TsvelikJETP87,James-K-L-R-T-18}
When $\lambda <0$,  the interaction is marginally irrelevant and it scales to zero in the far infrared (IR) limit so that the SU($N$)$_1$ gapless phase of the Sutherland model extends to nonzero $J_2$. In contrast, when $\lambda >0$, a spectral gap is formed and the low-energy excitations are fully gapped. The nature of the phase can be determined by means of the continuum description of the $N$-merization operator (see for instance Appendix B of Ref. \onlinecite{Lecheminant-T-15}): 
\begin{eqnarray}
e^{- i \frac{2 \pi n}{N}} S^{A}_{n}  S^{A}_{n+1}  \sim {\rm Tr} g (x) .
\label{Nmerisation}
\end{eqnarray}
In the fully gapped phase, one has $\langle {\rm Tr} g \rangle \ne 0$ and a $N$-merized phase is stabilized which is $N$-fold degenerate, breaking spontaneously 
T$_{a_0}$ since according to the identifications \ref{spinopappen} and \ref{WZNWcont}:
\begin{equation}
g \rightarrow e^{i \frac{2 \pi}{N} } g, 
\label{Tao}
\end{equation}
under T$_{a_0}$. 

The existence of the  $N$-merized phase can also be obtained by a direct Abelian-bosonization approach of model (\ref{HcontJ1J2}) as shown in Appendix \ref{sec:AppendixA}. Physically, the lattice spins form SU($N$) singlets and since $N$ sites are needed to get an SU($N$) singlet, a 
$N$-merized phase or valence cluster phase of $N$ sites emerges when $J_2 > J_2^{c, 1}$ (e.g. $\lambda >0$). The nonuniversal value $J_2^{c, 1}$ of the transition cannot be determined within this low-energy approach and will be obtained by means of numerical approaches for $N=3$ and $N=4$ in Sec.~\ref{sec:dmrg}.
The original gapless spinons of the Sutherland model when $J_2=0$ become fully gapped. As discussed in Appendix \ref{sec:AppendixA}, 
they correspond to the domain-wall configurations between the $N$-degenerate ground states of the $N$-merized phase. It is shown that they transform in the ${\bf {\bar N}}$ representation of the SU($N$) group and thus share the same quantum numbers as the SU($N$) gapless spinons of the Sutherland model.

\subsection{Zigzag limit:$J_1 \ll J_2$}

We now consider the opposite limit $J_1 \ll J_2$.  A low-energy approach can still be derived by considering the  $J_1-J_2$ spin chain (\ref{hamJ1J2})  as 
a zigzag two-leg spin ladder with Hamiltonian (\ref{eqn:def-2leg-ladder}).

When $J_1=0$, we have two decoupled Sutherland models with central charge $c=2(N-1)$. A continuum limit can be performed by introducing two independent copies of the identification (\ref{spinopappen}):
\begin{eqnarray}
S^{A}_{l,n} &\simeq& J^{A}_{l L} +  J^{A}_{l R} +  \left(\mbox{e}^{ i 2k_F x} N_l^{A} + \mathrm{H.c.} \right) \nonumber \\
&+& \sum_{m=2}^{N-2} \mbox{e}^{ i 2m k_F x} n_{l m}^{A},
\label{spinopappenladder}
\end{eqnarray}
with $l=1,2$ and $x=n a_0$.
We introduce a small $J_1$ term which couples the two SU($N$) Sutherland models.  Using Eq. (\ref{spinopappenladder}), the leading part of the continuum limit of the $J_1$ term of the zigzag ladder Hamiltonian (\ref{eqn:def-2leg-ladder}) reads then as follows by keeping only nonoscillatory terms:
\begin{eqnarray}
 {\cal H}^{\rm zigzag}_{\rm cont} &\simeq&   J_{1} a_0   \left[ e^{2 i k_F x} N^{A}_{2} (x) 
+  e^{-2 i k_F x} N^{A \dagger}_{2} (x) \right]
\nonumber \\
&&
\left[ e^{2 i k_F x} N^{A}_{1} (x)  \left( 1 + e^{2 i k_F a_0} \right) + \mathrm{H.c.} \right]
\nonumber \\
&=&  2 a_0 J_1 \cos(\frac{\pi}{N})  \left( e^{i \pi/N} N^{A}_{1} N^{A \dagger}_{2} + \mathrm{H.c.} \right).
\label{2legzigzagcont}
\end{eqnarray}
In strong contrast to the $N=2$ case, the $2k_F$ backscattering for $N>2$ terms do not cancel exactly. \cite{White-A-96,Allen-S-97}
In this respect, the situation is very different from the $N=2$ case where the leading perturbation is only marginal with a  twist perturbation with nonzero conformal spin and current-current interactions. \cite{Nersesyan1998}
The interaction is now strongly relevant.
By exploiting the fact that $N_{l}^{A}$ can be expressed in terms of the two SU($N$)$_1$ WZNW fields $g_{1,2}$ with scaling dimension $(N-1)/N$: 
\begin{equation}
N_{l}^{A} = i C  \; {\rm Tr} ( g_{l} T^A),
\end{equation}
we get the leading part of the low-energy effective theory which governs the IR behavior of model (\ref{eqn:def-2leg-ladder})  in the regime $J_1 \ll J_2$:
\begin{eqnarray}
{\cal H}_{\text zigzag} &=& \frac{2\pi v}{N + 1} \sum_{l=1}^{2} \left( : J^A_{lR} J^A_{lR}: + : J^A_{lL} J^A_{lL}: \right)
\nonumber \\
&+& \lambda_1
\Big[ e^{i \pi/N} \mbox{Tr}(g_1g_2^{\dagger}) + \mathrm{H.c.}\Big]  \nonumber \\
&+& \lambda_2
\Big[e^{i \pi/N} \mbox{Tr}g_1\mbox{Tr}g_2^{\dagger} + \mathrm{H.c.}\Big] ,
 \label{cft}
\end{eqnarray}
where $\lambda_2 = - \lambda_1/N < 0$ and $\lambda_1 = a_0 J_{1} C^2 \cos(\pi/N) > 0$.  

Model (\ref{cft}) describes two SU($N$)$_1$ WZNW models perturbed by two strongly relevant perturbations
with the same scaling dimension $2(N-1)/N < 2$ which are expected to open a  spectral gap $\Delta \sim J_1^{N/2}$ (up to logarithmic 
corrections) as soon as the inter-chain coupling $J_1$ is switched on (the gap opens more slowly for larger $N$).  
The subdominant contribution to model (\ref{cft}) corresponds to the marginal current-current interaction:
\begin{equation}
{\cal H}_{cc} = \frac{1}{2} \left(J_{1} - \gamma \right) I^A_{ R} I^A_{ L}
 - \frac{1}{2} \left(J_{1} + \gamma \right) K^A_{ R} K^A_{ L},
\label{Vcc}
\end{equation}
where  $I^A_{ R,L} = J^A_{1 R,L} + J^A_{2 R,L}$ is a SU($N$)$_2$ current,  being the sum of
two SU($N$)$_1$  currents, and $K^A_{ R,L} = J^A_{1 R,L} - J^A_{2 R,L}$  is the so-called wrong current.\cite{Gogolin-N-T-book}
In the $N=4$ case, there is an additional marginal contribution which stems from the product $ n_{12}^{A} n_{22}^{A}$.

The effective Hamiltonian density \eqref{cft} and the marginal contribution \eqref{Vcc} thus describe the low-energy physics of 
the two-leg zigzag SU($N$) spin  ladder \eqref{eqn:def-2leg-ladder}  in the regime $J_1 \ll J_{2}$.   
It takes a rather similar form from the one obtained in Refs.~\onlinecite{Lecheminant-T-15,Weichselbaum-C-L-T-L-18} for the standard two-leg SU($N$) spin ladder, e.g. with a rectangular symmetry. However, there is a crucial difference here with the presence of the phase factor $e^{i \pi/N}$ in Eq. (\ref{cft}), stemming from the zigzag geometry, which makes the physics totally different. 

The next step of the approach to elucidate the IR properties of model  \eqref{cft}
is to exploit the existence of the following conformal embedding as in Refs.~\onlinecite{Lecheminant-T-15,Weichselbaum-C-L-T-L-18,CapponiSPTchiral2020}:
\begin{equation}
\text{SU($N$)}_1 \times \text{SU($N$)}_1 \sim \text{SU($N$)}_2 \times  \mathbb{Z}_N ,
\label{embedding}
\end{equation}
where ${\mathbb{Z}}_N$ is the parafermionic CFT with central charge $c=2(N-1)/(N+2)$ which
describes the universal properties of the phase transition of the two-dimensional $\mathbb{Z}_N$
clock model. \cite{Zamolodchikov-F-JETP-85} The low-temperature phase of the latter model is described 
by local order parameters $\sigma_k$ ($k=1,..,N-1$) which are primary fields of the $\mathbb{Z}_N$ CFT with scaling\/
dimension $d_k = k(N-k)/N(N+2)$. 
The SU($N$)$_2$ CFT has the central charge $c=2(N^2-1)/(N+2)$ and is generated by the currents  $I^A_{ R,L}$.
The two  SU($N$)$_1$  WZNW fields $g_{1,2 }$ can be expressed in the new 
SU($N$)$_2$ $\times$ ${\mathbb Z}_N$ basis as:\cite{Griffin-N-89,Lecheminant-T-15}
\begin{equation}
\begin{split}
& (g_1)_{\alpha \beta} \sim  G_{\alpha \beta} \; \sigma_1   \\
& (g_2)_{\alpha \beta} \sim  G_{\alpha \beta}  \;  \sigma_1^{\dagger} , 
\end{split}
\label{ident}
\end{equation}
where $\alpha, \beta = 1, \ldots, N$ and $G$ is the SU($N$)$_2$  WZNW field (in $\boldsymbol{N}$-representation)  
with scaling dimension $\Delta_G = (N^2-1)/N(N+2)$. \cite{DiFrancesco-M-S-book} 
Under T$_{a_0}$, $g_{1,2} \rightarrow e^{i \frac{2 \pi}{N} } g_{1,2}$ as in Eq. (\ref{Tao}) so that one has from (\ref{ident})
\begin{equation}
G \rightarrow e^{i \frac{2 \pi}{N} } G, 
\label{Taobis}
\end{equation}
under T$_{a_0}$. 

Using the Appendix \ref{AppCFT} (see also Ref. \onlinecite{Lecheminant-T-15}), model (\ref{cft}) can be expressed in the new basis (\ref{embedding}) as:
 \begin{eqnarray}
 {\cal H}_{\text zigzag} &=& \frac{2\pi v}{N + 2}  \left( : I^A_{R} I^A_{R}: + : I^A_{L} I^A_{L}: \right) + {\cal H}_{{\mathbb Z}_N}
\nonumber \\
&+& {\tilde \lambda}_1 \left(e^{i \pi/N} \Psi_{1L}  \Psi_{1R}  + \mathrm{H.c.} \right) \nonumber \\
&+& {\tilde \lambda}_2  \; {\rm Tr} (\Phi_{\rm adj}) \left( e^{i \pi/N} \sigma_2 + \mathrm{H.c.} \right) ,
\label{cftnewbasis}
\end{eqnarray}
where  ${\cal H}_{{\mathbb Z}_N}$ is the Hamiltonian density of the 
${\mathbb{Z}}_N$ parafermionic CFT. The latter is generated by the first ${\mathbb{Z}}_N$ parafermion
currents $\Psi_{1L,R}$  with conformal weights $h, {\bar h} = (N-1)/N$.
In Eq. (\ref{cftnewbasis}), $\Phi_{\rm adj}$ is the SU($N$)$_2$ primary field with scaling
dimension $2N/(N+2)$, which transforms in the adjoint representation of SU($N$) and the coupling constants are given by:
 \begin{eqnarray}
 {\tilde \lambda}_1  &=& \frac{(N^2 -1) \lambda_1}{N} = a_0 \frac{(N^2 -1)}{N}J_{1} C^2 \cos(\pi/N) > 0
\nonumber \\
 {\tilde \lambda}_2  &=& \lambda_2 = - a_0 J_{1} C^2 \frac{\cos(\pi/N)}{N} < 0 .
\label{couplings}
\end{eqnarray}

The effective field theory (\ref{cftnewbasis}) contains two different sectors, an SU($N$) singlet one, described
by the ${\mathbb{Z}}_N$ parafermions, and the second one depending on both SU($N$) and ${\mathbb{Z}}_N$ degrees of freedom. 
The two perturbations are strongly relevant with the same scaling dimension. Interestingly, the pure ${\mathbb{Z}}_N$ perturbation with coupling constant 
${\tilde \lambda}_1$ in Eq. (\ref{cftnewbasis}) is an integrable field theory. \cite{FateevZamoloPhysLettB91} Its Euclidean action is given by
\begin{equation}
{\cal S}_{\rm eff} = {\cal S}_{{\mathbb Z}_N} - g  \int d^2 x  \left(e^{i \pi/N} \Psi_{1L}  \Psi_{1R}  + \mathrm{H.c.} \right),
\label{effactionparaJperpF}
\end{equation}
where ${\cal S}_{{\mathbb Z}_N}$ is the action of the ${\mathbb{Z}}_N$  CFT and $g = -  {\tilde \lambda}_1  <0$.
The IR properties of the action (\ref{effactionparaJperpF}) strongly depend on the sign of the coupling constant $g$ and of the parity of $N$. \cite{FateevZamoloPhysLettB91}
When $g>0$, ${\cal S}_{\rm eff} $ displays an integrable massless RG flow to the minimal model series ${\cal M}_{N+1}$
with central charge $c= 1 - 6/(N+1)(N+2)$. \cite{FateevZamoloPhysLettB91} When $g<0$, as is the case here since $J_1 >0$, the IR behavior depends on the parity of $N$. 
When $N$ is even, the minus sign of the perturbation in Eq. (\ref{effactionparaJperpF}) can be absorbed into a redefinition 
of the first parafermionic current: $\Psi_{1L} \rightarrow - \Psi_{1L}$ which still satisfies the same
parafermionic algebra. This means that ${\cal S}_{\rm eff} $  exhibits a massless RG
flow to the minimal model series ${\cal M}_{N+1}$ in the far-IR when $g <0$ and $N$ even. 
However, when $N$ is odd, one cannot use the same transformation but the following redefinition: 
\begin{eqnarray}
\Psi_{1L} &\rightarrow&  {\tilde \Psi}_{1L} = - \mbox{e}^{i\pi/N} \Psi_{1L} \nonumber \\
\Psi_{1R} &\rightarrow& {\tilde \Psi}_{1R} =  \Psi_{1R} ,
\label{paratransNodd}
\end{eqnarray}
and ${\tilde \Psi}_{1L}$ is still a first parafermionic current (${\tilde \Psi}_{1L}^{N} \sim I$) when $N$ is odd. The action (\ref{effactionparaJperpF}) transforms then as follows:
\begin{equation}
{\cal S}_{\rm eff} = {\cal S}_{{\mathbb Z}_N} + g  \int d^2 x  \left( {\tilde \Psi}_{1L}  {\tilde \Psi}_{1R}  + \mathrm{H.c.} \right),
\label{effactionparaJperpFtrans}
\end{equation}
which is known to be a massive integrable field theory of ${\mathbb{Z}}_N$ parafermions when $g = - {\tilde \lambda}_1  <0$. \cite{Fateev-91}
We thus conclude that the ${\mathbb{Z}}_N$ perturbation (\ref{effactionparaJperpF}) when $g <0$, e.g. $J_1 >0$, displays different physical properties depending on the parity of $N$. When $N$ is even, model  (\ref{effactionparaJperpF}) is gapless in the long-distance limit with critical properties that are governed by the ${\cal M}_{N+1}$ universality class while it is fully massive in the odd $N$ case.
 
\subsubsection{Odd-$N$ case}
\label{low-energyoddN}

We first focus on the odd-$N$ case where model (\ref{effactionparaJperpF}) is a massive field theory with a mass gap $\Delta \sim J_1^{N/2}$. The exact spectrum consists of massive kink excitations that result from degenerate ground states labeled by an odd integer $s=1,3, \ldots, N$.\cite{Fateev-91}
After averaging over the ${\mathbb Z}_N$  degrees of freedom in Eq. (\ref{cftnewbasis}), we obtain an effective theory for the remaining SU($N$)$_2$ degrees of freedom that governs the physics of model (\ref{cftnewbasis}) at the energy scale $E \ll \Delta$:
 \begin{eqnarray}
 {\cal H}^{\text odd}_{\rm eff} &=& \frac{2\pi v}{N + 2}  \left( : I^A_{R} I^A_{R}: + : I^A_{L} I^A_{L}: \right) 
\nonumber \\
&+& {\tilde \lambda}_2  \; {\rm Tr} (\Phi_{\rm adj}) \left( e^{i \pi/N} \langle \sigma_2\rangle + \mathrm{H.c.} \right) .
\label{cftnewbasisZNaverage}
\end{eqnarray}
The vacuum expectation value $\langle \sigma_k \rangle$ of the ${\mathbb Z}_N$ spin operators in the field theory (\ref{effactionparaJperpFtrans}) with $g <0$  have been determined nonperturbatively in Ref. \onlinecite{BaseilhacNPB98}. To use their result, we need to find the transformation of $ \sigma_2$ under (\ref{paratransNodd}). By fusion, the transformation of the ${\mathbb Z}_N$ parafermion currents ($\Psi_{kL,R}, k = 1, \ldots, N -1$)  is found to be:
\begin{eqnarray}
\Psi_{kL} &\rightarrow&  {\tilde \Psi}_{kL} =\left( -1\right)^k \mbox{e}^{i k \pi/N} \Psi_{kL} \nonumber \\
\Psi_{kR} &\rightarrow& {\tilde \Psi}_{kR} =  \Psi_{kR} .
\label{paracurrtransNodd}
\end{eqnarray}
The  transformation of the ${\mathbb{Z}}_N$ spin fields $\sigma_k$ should be consistent with the fusion rules of the  ${\mathbb{Z}}_N$ parafermionic theory: \cite{Zamolodchikov-F-JETP-85}
$\sigma_k \mu_k \sim \Psi_{kL}$ and $\sigma_k \mu_k^{\dagger} \sim \Psi_{kR}$ ($\mu_k$ being
the ${\mathbb{Z}}_N$  disorder fields). 
We thus deduce:
\begin{eqnarray}
\sigma_2 &\rightarrow&  {\tilde \sigma}_2  = - e^{i  \pi/N} \sigma_2 \nonumber \\
\mu_2 &\rightarrow&   {\tilde \mu}_2  =  - e^{i \pi/N}   \mu_2 .
\label{sigmatransfo}
\end{eqnarray}
After the transformation (\ref{paratransNodd}), the low-energy Hamiltonian (\ref{cftnewbasisZNaverage}) becomes then
 \begin{eqnarray}
 {\cal H}^{\text odd}_{\rm eff} &=& \frac{2\pi v}{N + 2}  \left( : I^A_{R} I^A_{R}: + : I^A_{L} I^A_{L}: \right) 
\nonumber \\
&-& {\tilde \lambda}_2  \; {\rm Tr} (\Phi_{\rm adj}) \left(  \langle  {\tilde \sigma}_2 \rangle  + \mathrm{H.c.} \right) .
\label{Noddeffective}
\end{eqnarray}
The value $\langle  {\tilde \sigma}_2 \rangle$ in the field theory (\ref{effactionparaJperpFtrans}) can be found in Ref. \onlinecite{BaseilhacNPB98}
and we find $ \langle  {\tilde \sigma}_2 \rangle =  \langle  {\tilde \sigma}^{\dagger}_2 \rangle = \sigma > 0$.

Finally, the low-energy effective theory which governs the IR behavior of the two-leg SU($N$) zigzag spin ladder in the large $J_2$ limit 
reads as follows  when $N$ is odd:
\begin{eqnarray}
 {\cal H}^{\text odd}_{\rm eff} = \frac{2\pi v}{N + 2}  \left( : I^A_{R} I^A_{R}: + : I^A_{L} I^A_{L}: \right) 
+ \eta   {\rm Tr} (\Phi_{\rm adj})  ,
\label{Noddeffectivefin}
\end{eqnarray}
with $\eta = - 2 {\tilde \lambda}_2 \sigma >0$.  The effective field theory (\ref{Noddeffectivefin}) has been investigated in Ref. \onlinecite{Lecheminant-15}. While the adjoint perturbation is a strongly relevant perturbation with scaling dimension $2N/(N+2)$, a massless RG  flow from SU($N$)$_2$  to SU($N$)$_1$ CFT is predicted when $N$ is odd and $\eta >0$. An explicit proof in the $N=3$ case has been given in Ref. \onlinecite{Lecheminant-15}  by mapping model  (\ref{Noddeffectivefin}) with $N=3$ onto Gepner's parafermions \cite{Gepner-87}, and recently using RG interfaces. \cite{kikuchi-22}

We thus find that  the two-leg SU($N$) zigzag spin ladder in the regime $J_1 \ll J_2$ displays critical properties in the  SU($N$)$_1$  universality class when $N$ is odd as in the weak-coupling regime $J_2 \ll J_1$. An alternative approach in the $N=3$ case, based on a semiclassical approach, is presented in Appendix \ref{Semiclassical} which confirms the existence of this SU(3)$_1$ gapless phase in the zigzag regime. The fully gapped $N$-merized phase, found in the weak-coupling regime $J_2 \ll J_1$, has thus a finite extent as function of $J_2$ when $N$ is odd. The precise extension of this phase is clearly beyond the scope of the field-theory approach and will be investigated numerically in Sec. \ref{sec:dmrg}. The transition between the SU($N$)$_1$ phase, obtained in the regime $J_2 \gg J_1$, and the $N$-merized phase is similar to the one described in the weak-coupling regime $J_2 \ll J_1$. Along the massless RG flow, the SU($N$)$_2$  currents ${I}_{L,R}^A$  are transmuted in the far-IR regime to SU($N$)$_1$  currents ${\cal J}_{L,R}^A$ so that, taking into account the marginal contribution (\ref{Vcc}),  the transition is expected to be governed by 
\begin{equation}
  \mathcal{H}^{\text odd}_{\rm IR}  = \frac{2\pi v}{N + 1} \left( : {\cal J}^A_{R} {\cal J}^A_{R}: + : {\cal J}^A_{L} {\cal J}^A_{L}: 
\right)  + \lambda_{\rm eff}  {\cal J}_R^{A} {\cal J}_L^{A} ,
\label{HcontJ1J2trans}
\end{equation}
where $\lambda_{\rm eff} <0$ in the  regime $J_2 \gg J_1$ and should change its sign at the transition to produce the  fully gapped $N$-merized phase with an $N$-fold ground-state degeneracy. 

\subsubsection{Even-$N$ case}

We now turn to the even-$N$ case. In stark contrast to the odd-$N$ case, the ${\mathbb{Z}}_N$  field theory (\ref{effactionparaJperpF}) is not fully massive but describe an integrable massless RG flow to the minimal model series ${\cal M}_{N+1}$ with central charge $c= 1 - 6/(N+1)(N+2)$. Along the RG flow, some ${\mathbb{Z}}_N$ degrees of freedom acquire a gap but others remain gapless in the long-distance limit. It is thus tempting  to expect that the physical properties of the two-leg SU($N$) zigzag ladder \eqref{eqn:def-2leg-ladder} for even $N$ will be rather different from the odd $N$ case.

The strategy for even $N$ is thus to rewrite model (\ref{cftnewbasis}) in the IR
limit by exploiting the existence of the integrable massless RG flow. To this end, we need the ultraviolet (UV)-IR transmutation of the
${\mathbb{Z}}_{N}$  fields and most importantly the expression of the order fields $ \sigma_{1,2}$ at the IR ${\cal M}_{N+1}$ fixed point. Unfortunately,
to the best of our knowledge, we are not aware of such UV-IR transmutation when $N>4$.  In the special $N=4$ case, progress can be made by exploiting the fact that the ${\mathbb{Z}}_4$ CFT has central charge $c=1$ and it can be described by a bosonic field $\Phi$ on the orbifold line.\cite{Lecheminant-2002} The field theory (\ref{effactionparaJperpF}) turns out to be equivalent to the self-dual sine-Gordon model at $\beta^2 = 6 \pi$ with Hamiltonian density:\cite{Lecheminant-2002}
\begin{eqnarray}
{\cal H}_{\rm SDSG} &=& \frac{v}{2} \left[
\left(\partial_x \Phi\right)^2 +
\left(\partial_x \Theta\right)^2 \right]  \nonumber \\
&+& g\left[\cos\left(\sqrt{6\pi} \; \Phi\right)
+ \cos\left(\sqrt{6\pi} \; \Theta\right)\right],
\label{SDSG}
\end{eqnarray}
where $ \Theta$ is the dual field associated with $\Phi$. Model (\ref{SDSG}) is self-dual, being invariant under the $\Phi \leftrightarrow \Theta$ symmetry, with two strongly relevant perturbations with scaling dimension $3/2$. This self-duality symmetry produces a quantum critical behavior. Its nature can be inferred from the equivalence between (\ref{SDSG}) and the ${\mathbb{Z}}_4$ field theory (\ref{effactionparaJperpF}) which displays a massless RG flow from the $c=1$ ${\mathbb{Z}}_4$ CFT to the minimal model ${\cal M}_{5}$ with central charge $c=4/5$. Model (\ref{SDSG}) enjoys thus a $c=4/5$ quantum critical behavior.
The $\sigma_2$ primary field of the  ${\mathbb{Z}}_4$ CFT with scaling dimension $1/6$ can be bosonized and one has:\cite{Lecheminant-2002}  $\sigma_2 = \sqrt{2} \cos (\sqrt{2\pi/3} \; \Phi)$. The low-energy effective field theory (\ref{cftnewbasis}) reads then as follows for $N=4$:
\begin{eqnarray}
 {\cal H}^{N=4}_{\text zigzag} &=& \frac{\pi v}{3}  \left( : I^A_{R} I^A_{R}: + : I^A_{L} I^A_{L}: \right) \nonumber\\
 &+& \frac{v}{2} \left[
\left(\partial_x \Phi\right)^2 +
\left(\partial_x \Theta\right)^2 \right] 
\nonumber \\
&+& g\left[\cos\left(\sqrt{6\pi} \; \Phi\right)
+ \cos\left(\sqrt{6\pi} \; \Theta\right)\right] 
\nonumber \\
&+& 2 {\tilde \lambda}_2  \; {\rm Tr} (\Phi_{\rm adj}) \cos (\sqrt{2\pi/3} \; \Phi),
\label{cftnewbasisN=4}
\end{eqnarray}
with $g <0$ and ${\tilde \lambda}_2  <0$. The general structure of the field theory (\ref{cftnewbasisN=4}) leads us to expect that it is fully gapped in the ${\mathbb{Z}}_4$ and SU(4)$_2$ sectors. Indeed, we first note that  the self-dual symmetry $\Phi \leftrightarrow \Theta$ is now explicitly broken due to the presence of the $\cos (\sqrt{2\pi/3}  \;\Phi)$ term in Eq. (\ref{cftnewbasisN=4}). A mass gap for the ${\mathbb{Z}}_4$ degrees of freedom should be generated. Moreover, the SU(4)$_2$ perturbed by the relevant adjoint field is a massive field theory for either sign of its coupling constant and does not exhibit a massless flow as in the SU($2n+1$)$_2$ case.\cite{Lecheminant-15,kikuchi-22}
We may  conclude from these two facts that the two-leg SU(4) zigzag spin ladder in the regime $J_1 \ll J_2$ is a fully gapped phase. 

One can show the emergence of this gapped behavior by exploiting the fact that the UV-IR transmutation of some fields of the ${\mathbb{Z}}_4$ CFT along the massless RG flow (\ref{SDSG}) down to the ${\cal M}_{5}$  CFT  has been obtained in Refs. \onlinecite{Lecheminant-2002,VaeziParafermions}.  In particular, the leading contribution of the UV-IR transmutation of  the operator $\cos (\sqrt{2\pi/3}  \; \Phi)$ is:\cite{Lecheminant-2002,VaeziParafermions}
\begin{equation}
 \cos (\sqrt{2\pi/3}  \; \Phi) \sim \sigma_{{\mathbb{Z}}_3} + \mathrm{H.c.} ,
\label{UVIRtrans}
\end{equation}
where $\sigma_{{\mathbb{Z}}_3}$ is the ${\mathbb{Z}}_3$ spin field with scaling dimension $2/15$ of the ${\mathbb{Z}}_3$ CFT with central charge $c=4/5$. In the far-IR limit, the low-energy physics which governs the properties of the two-leg SU(4) zigzag spin ladder in the regime $J_1 \ll J_2$ is:
\begin{eqnarray}
 {\cal H}^{N=4}_{\text zigzag} &=& \frac{\pi v}{3}  \left( : I^A_{R} I^A_{R}: + : I^A_{L} I^A_{L}: \right) + {\cal H}_{{\mathbb Z}_3}  
 \nonumber \\
&+& 2 {\tilde \lambda}_2  \; {\rm Tr} (\Phi_{\rm adj}) \left(\sigma_{{\mathbb{Z}}_3} + \mathrm{H.c.}\right) .
\label{cftnewbasisN=4fn}
\end{eqnarray}
We use a simple mean-field analysis to investigate the IR properties of model
(\ref{cftnewbasisN=4fn}). We decouple the SU(4)$_2$ and ${\mathbb{Z}}_3$  sectors to get the mean-field Hamiltonian density: ${\cal H}_{\rm mf} = {\cal H}_1 + {\cal
H}_2$ with:
\begin{eqnarray}
   {\cal H}_1 &=& {\cal H}_{{\mathbb Z}_3}   
   + \kappa_1 \;   \left(\sigma_{{\mathbb{Z}}_3} + \mathrm{H.c.}\right) \label{meanfield} 
\\
   {\cal H}_2 &=& \frac{\pi v}{3}  \left( : I^A_{R} I^A_{R}: + : I^A_{L} I^A_{L}: \right)  + \kappa_2
     \; {\rm Tr} (\Phi_{\rm adj}), \nonumber
\end{eqnarray}
with the mean-field coupling constants
\begin{eqnarray}
\kappa_1 &=& 2 {\tilde \lambda}_2 \;
    \langle {\rm Tr} (\Phi_{\rm adj})  \rangle \nonumber \\
\kappa_2 &=& 2 {\tilde \lambda}_2 \;
  \langle   \sigma_{{\mathbb{Z}}_3} + \mathrm{H.c.} \rangle   .
\label{couplingsMF}
\end{eqnarray}

The Hamiltonian ${\cal H}_1$ describes the scaling limit of the two-dimensional ${\mathbb{Z}}_3$ Potts model at $T=T_c$ in a magnetic field. \cite{Delfino-08}
When $\kappa_1 < 0$, the magnetic field selects one specific color of the  ${\mathbb{Z}}_3$ Potts model and the model is fully gapped.
The Hamiltonian  ${\cal H}_2$ of  Eq. (\ref{meanfield}) describes an SU(4)$_2$ CFT perturbed by its adjoint primary field with scaling dimension $4/3$ which is a massive field theory for either sign of its coupling constant. \cite{kikuchi-22}
 Since $ \langle   \sigma_{{\mathbb{Z}}_3} + \mathrm{H.c.} \rangle > 0$ in the ground-state of ${\cal H}_1$, we have $\kappa_2  <0$ and a massive behavior occurs for ${\cal H}_2$ whose physical properties can be deduced from the identity:\cite{Knizhnik-Z-84}
\begin{eqnarray}
    {\rm Tr} (\Phi_{\rm adj}) &=&  {\rm Tr}( G^{+} T^{A} G T^{A} ) \notag \\
   &\sim&   {\rm Tr} \; G \;   {\rm Tr} \;  G^{+}
  - \frac{1}{4}  {\rm Tr} (G G^{+}) 
\textbf{ ,}\label{TrPhiadj}
\end{eqnarray}
where $G$ is the WZNW SU(4) matrix field. We thus find 
that the minimization of model ${\cal H}_2$
with  $\kappa_2 < 0$ selects  the center
group of SU(4):
\begin{equation}
   G = \exp(  i\pi k/2) I ,
\label{min1adj}
\end{equation}
with $k = 0, 1, 2,3$. From Eq. (\ref{Taobis}), we deduce that this solution breaks spontaneously T$_{a_0}$. The gapped phase is thus four-fold degenerate and the tetramerization phase of the weak-coupling limit $J_2 \ll J_1$ extends to the large $J_2$ regime in stark contrast to the odd-$N$ case.

%%%%%%%%%%%%%%%%%%%%%%%%%%%%%%%%%%%%%%%%%%%%%%%%%%%%%%%%%%%%%
\section{Numerical calculations}
\label{sec:dmrg}
%%%%%%%%%%%%%%%%2%%%%%%%%%%%%%%%%%%%%%%%%%%%%%%%%%%%%%%%%%%%%%

We have performed ED simulations on periodic chains using lattice symmetries as well as color conservation, which allows to reach $L=27$ for $N=3$ and $L=24$ for $N=4$ and to get the quantum numbers of the ground-state and its lowest excitations, hence suggesting possible symmetry breaking in the thermodynamic limit. In order to go beyond, we have used infinite-size DMRG (iDMRG) using the \href{https://itensor.github.io/ITensors.jl/stable/index.html}{ITensors.jl}~\cite{itensor} library and its subpackage \href{https://github.com/ITensor/ITensorInfiniteMPS.jl}{ITensorInfiniteMPS.jl} with a unit-cell of size $2N$. We used color conservation as well to accelerate the convergence. 
The maximal bond dimension $\chi$ which has been increased up to $8192$. Energies are converged below $10^{-9}$ for all $\chi$, and the entropies are converged below to $10^{-6}$, except for $\chi = 8192$ in the gapless phases or at large $J_2$.

Note that for convenience and up to some irrelevant constants and redefinitions of the energy scales, we can rewrite the Hamiltonian (\ref{hamJ1J2}) using permutation operators $\hat{\mathrm{P}}_{ij}$ which swap the two states on sites $i$ and $j$, due to the generalized Pauli identity \cite{Weichselbaum-C-L-T-L-18}:
\begin{equation}
2 \sum_{A}  S^{A}_{i}  S^{A}_{j} = \hat{\mathrm{P}}_{ij}-\frac{1}{N}. \label{eq:defPermut}
\end{equation}
We used these permutation operators to implement the Hamiltonian and fix $J_1 =1$.

\subsection{SU(3) case}

\begin{figure}[!htb]
\begin{center}
 \includegraphics[width=\linewidth,clip]{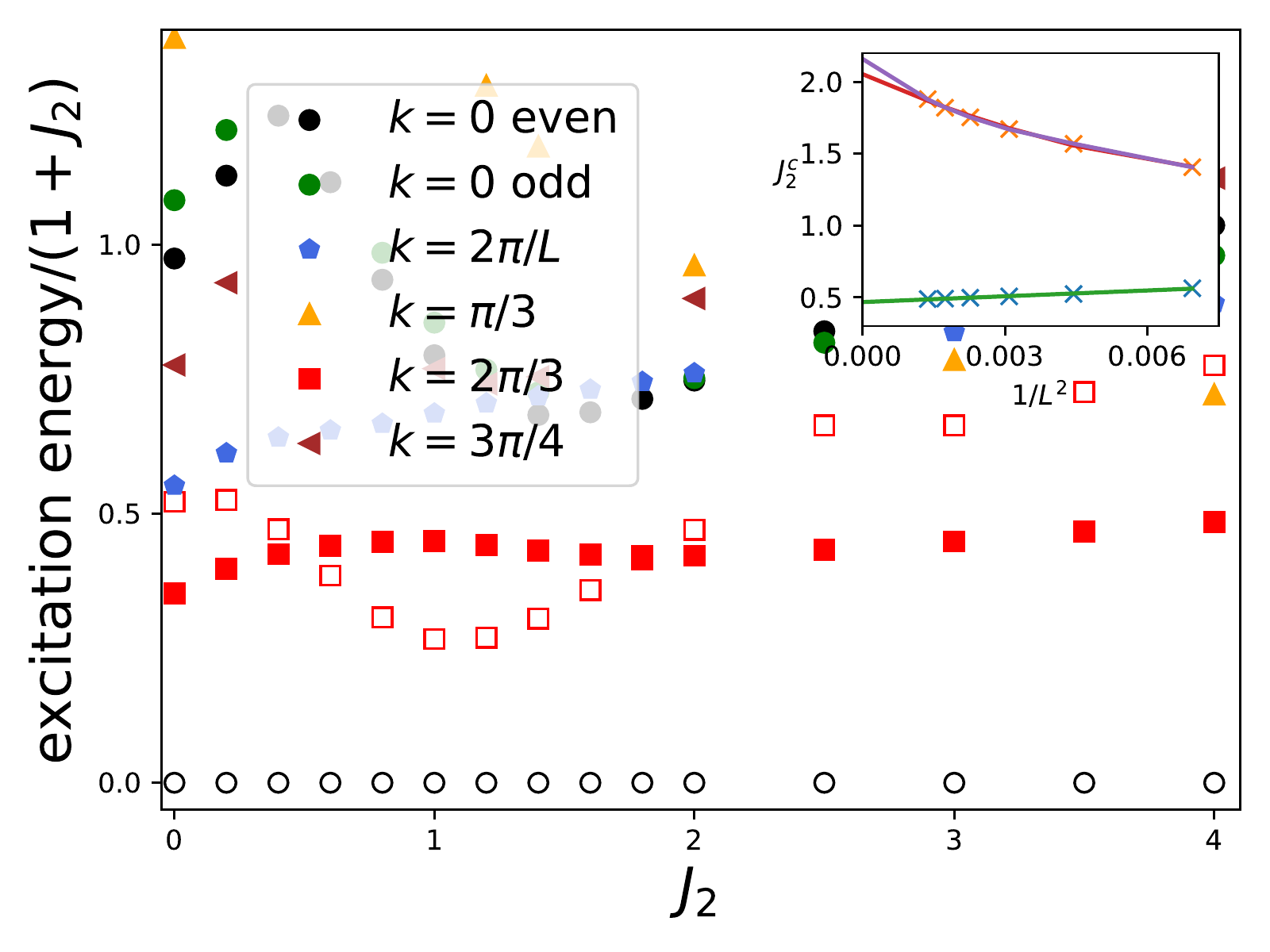}
\end{center}
\caption{Low-energy spectrum vs $J_2$ obtained by ED in the SU(3) case on the $L=24$ periodic chain, for some relevant momenta. Open (respectively filled) symbols denote SU(3) singlet (respectively nonsinglet) states. Inset: finite-size scaling of the level crossing between the lowest singlet or adjoint state at momentum $k=2\pi/3$.}
    \label{fig:ed_su3}
\end{figure}

We start the SU(3) case by revisiting former numerical results obtained by Corboz {\it et al.}~\cite{corbozPRB2007} Using ED (up to $L=21$) and level spectroscopy analysis on one hand, and finite DMRG on the other hand, they concluded to an intermediate trimerized phase for $0.45 \lessapprox J_2 \lessapprox 3.5$. In Fig.~\ref{fig:ed_su3}, we plot the relevant  low-energy excitations above the SU(3) singlet ground-state and their quantum numbers. In the expected critical phase (including the known integrable case for $J_2=0$), the lowest excitation is in the nontrivial adjoint representation and has a momentum $2\pi/3$ as expected (the finite-size gap is a finite-size effect).
However, for intermediate $J_2$, there are well-defined level crossings so that the lowest excitation is a two-fold degenerate (SU(3)) singlet at momentum $\pm 2\pi/3$, as expected for a singlet trimerized phase in the thermodynamic limit. In the inset of Fig.~\ref{fig:ed_su3}, we have attempted to extrapolate the critical values using finite-size scaling: the first one nicely converges to $J_{2}^{c,1}\approx 0.48$ while the other one has stronger finite-size effects so that we have less accuracy $J_2^{c,2}= 2.1(1)$. These values are in quantitative agreement with the previous results~\cite{corbozPRB2007}, as $J_2^{c,2}$ is plagued by very strong finite-size effects.
Note also that the first critical value has also been found in Ref.~\onlinecite{Rachel-T-F-S-G-09}.

%%%%%%%%%%%%%%%%%%%%
%%%%%%%%%%%%%%%%%%%%

\begin{figure}[!htb]
    \includegraphics[width=\linewidth,clip]{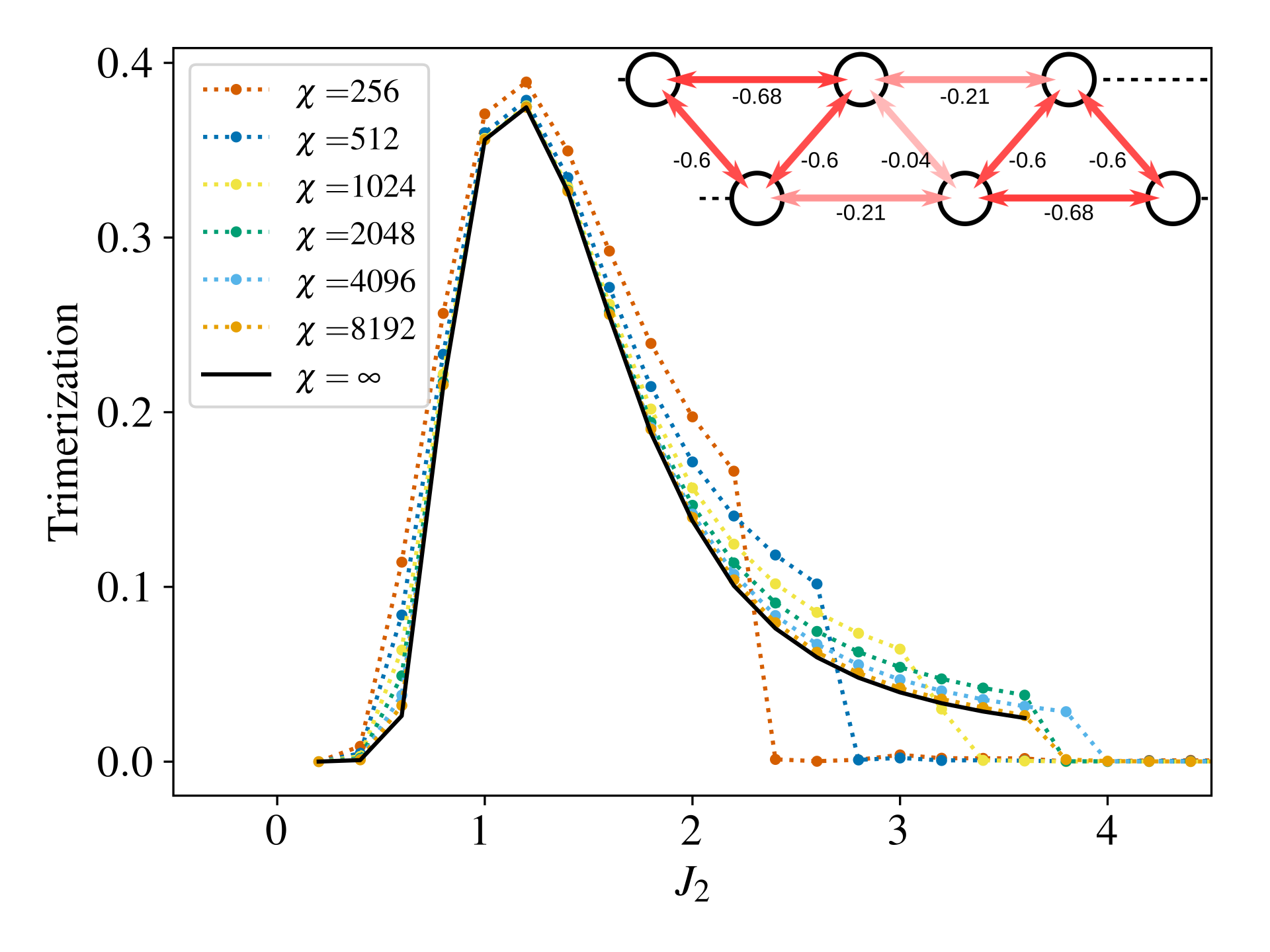}
    \caption{Trimerization $T_{i, 3}$ vs $J_2$ obtained by iDMRG in the SU(3) chain for various $\chi$ as well as its extrapolation to infinite bond dimension  using a naive second degree polynomial fit in the regime $J_2 \leq 3.6$ where we can reliably perform this extrapolation. The trimerized phase (see the inset) opens up at $J_2 \approx 0.5$, and extends, for the $\chi$ we have access to, to $J_2 \approx 3.8$. Then, crossovers with $\chi$, visible as kinks in the trimerization, prevent us to perform a reliable scaling. In inset, we represent the numerically obtained bond energies $\langle \hat{\mathrm{P}}_{ij} \rangle$ on each link where $ \hat{\mathrm{P}}_{ij}$ is defined in Eq.~\eqref{eq:defPermut}. The structure of the singlet is directy visible.}
    \label{fig:dmrg_su3}
\end{figure}

We have then performed iDMRG simulations with a compatible unit-cell of $6$ sites for various bond dimensions $\chi$. 
For intermediate $J_2$, there is a very clear trimerized pattern in the bond amplitudes (see e.g. inset of Fig.~\ref{fig:dmrg_su3}) so that we define an order parameter as
$$ T_{i, 3} = \langle \hat{\mathrm{P}}_{i,i+1} - \frac{1}{3} \sum\limits_{k=0}^2 \hat{\mathrm{P}}_{i+k,i+k+1}\rangle$$
which is measured in the ground-state obtained at fixed $\chi$.
Since the bond pattern can be shifted within our simulated unit cell, we choose to measure on bonds showing a weak/strong/strong pattern so that $T_{i, 3}$ is positive and largest. 
Note that iDMRG technically breaks translation invariance, even after convergence, due to the initial edges.
This effect, and the tendency to converge to less entangled superpositions allows us to directly use the trimerization as an order parameter, instead of a related correlation function.
Data are shown in Fig.~\ref{fig:dmrg_su3} and confirm that the trimerized singlet phase has a finite extent, in a range $0.5 \lessapprox J_2 \lessapprox 3.8$, larger than previously reported~\cite{corbozPRB2007}. 
To resolve this disagreement, we performed a naive extrapolation in $\chi^{-1}$ of the trimerization using a second-degree polynomial fit to evaluate the infinite $\chi$ behavior.
In the region $3.0 \lessapprox J_2 \lessapprox 4.0$, the trimerization strongly depends on $\chi$.
For too small bond dimensions, we do not resolve the gap and the matrix product states appear to belong to the large $J_2$ gapless phase, with a significant shift in the trimerization when varying $\chi$.
This crossover prevents us from reliably extrapolating any quantity for $J_2 \geq 3.4$.
Because we observe this behavior at lower $\chi$ on well-converged system, we do not expect that we underestimate the extent of the gapped phase.
The mismatch with the DMRG results of Ref.~\onlinecite{corbozPRB2007} is likely due to finite-size effects in the finite DMRG computation.
This effect also prevents us to make a precise interpolation of the transition toward a gapless phase. Nevertheless, our numerical data do support a finite $J_2$ range for the trimerized phase.

%%%%%%%%%%%%%%%%%%%%
%%%%%%%%%%%%%%%%%%%%

\begin{figure}[!htb]
\includegraphics[width=\linewidth,clip]{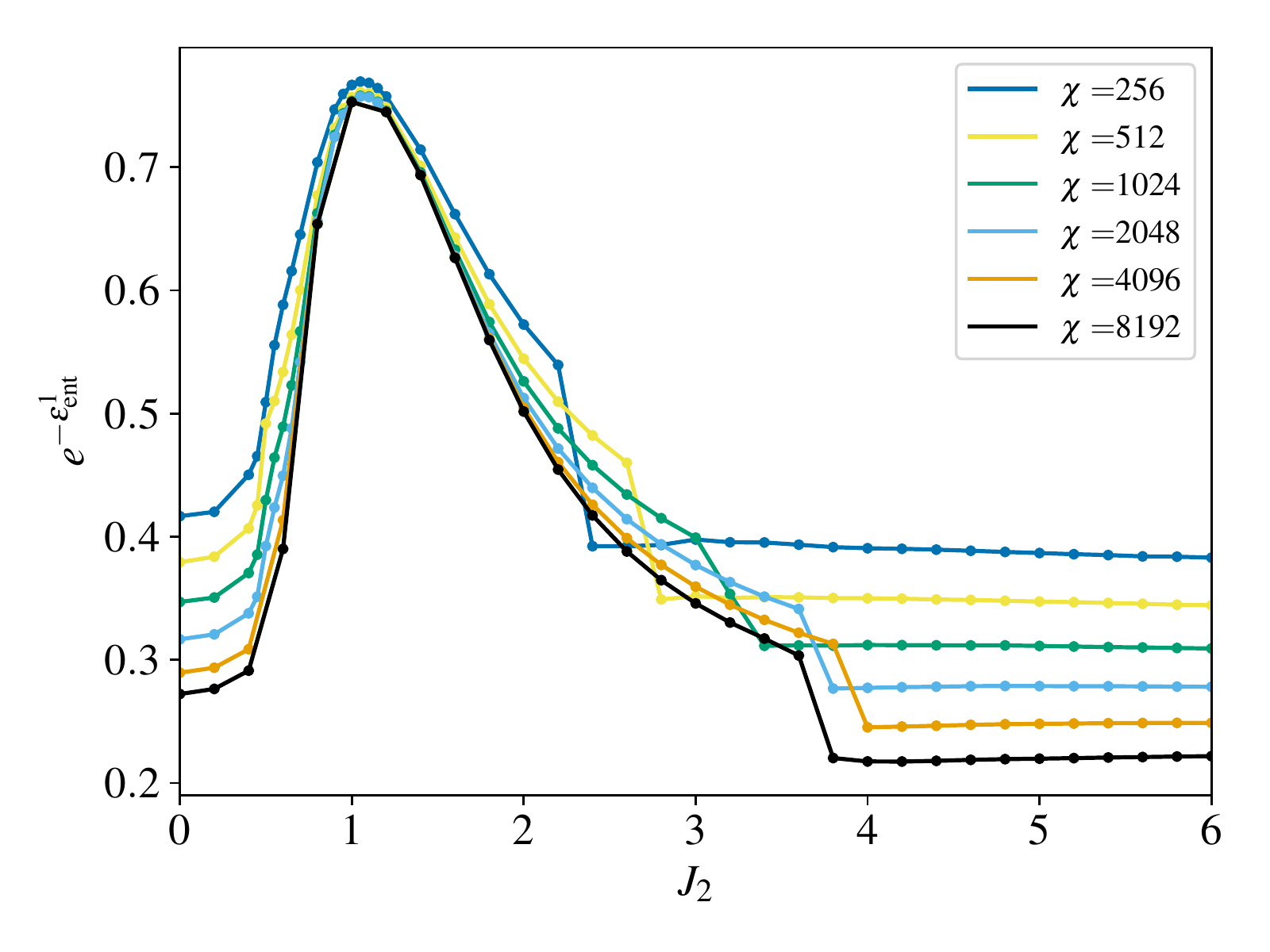}\\
\includegraphics[width=\linewidth,clip]{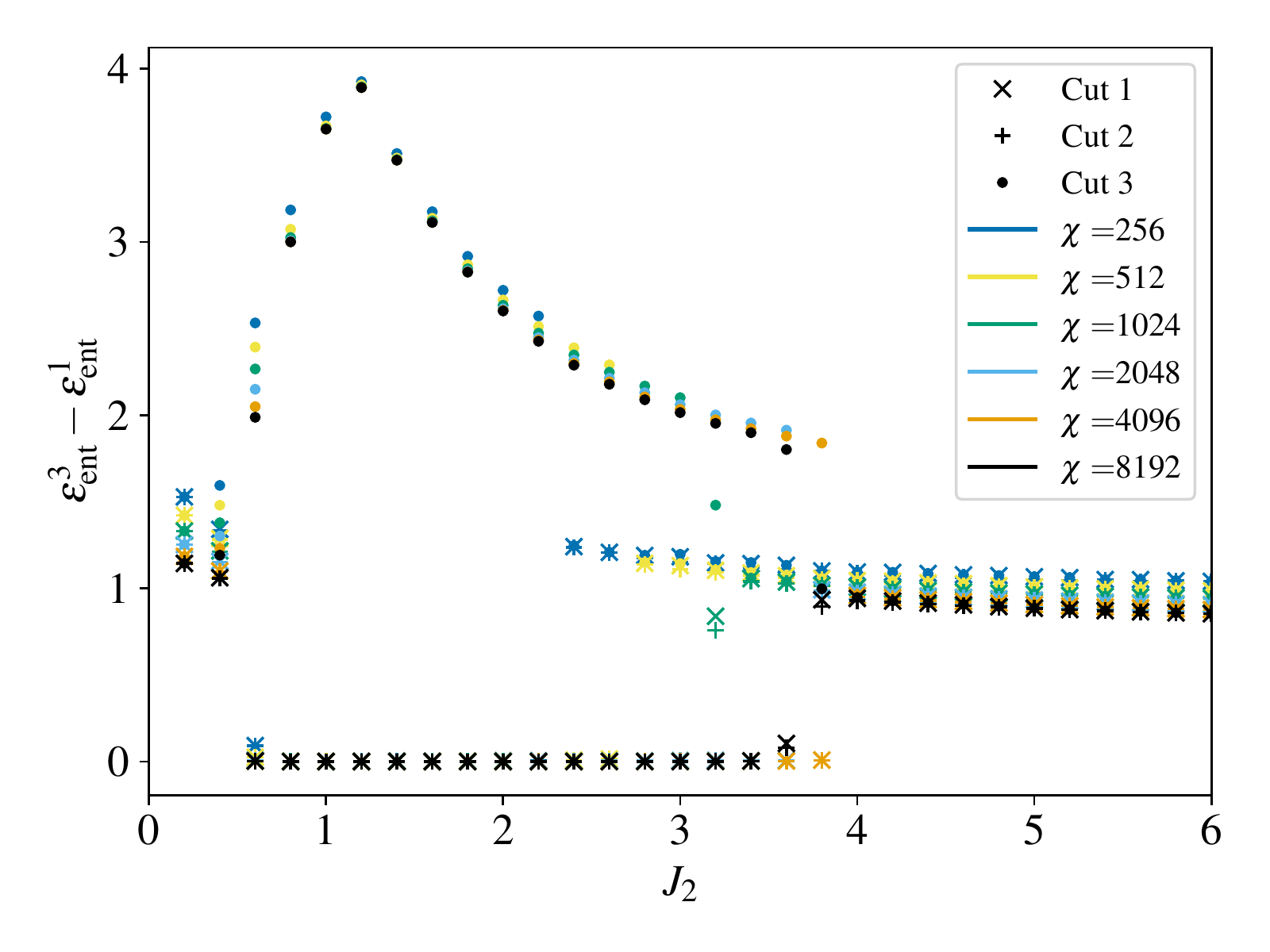}
\caption{Top: largest eigenvalue of the "mid-chain" reduced density matrix on the weakest bond of the unit-cell for the SU(3) chain, as defined in Eq.~\eqref{eq:defEntSpec}. The opening of the trimerized phase is characterized by a large peak where the matrix product state becomes close to a product state. At large $J_2$, the discontinuity indicates when our simulations become unreliable.
  Bottom: gap between third and first entanglement energies. The singlet pattern $1 3 3$ is characteristic of a product state of consecutive singlets. At large $J_2$, the pattern becomes abruptly compatible with a gapless phase.}
    \label{fig:ent_spec_su3}
\end{figure}

The trimerization means that the SU(3) spins form singlets over three consecutive sites.
This can be immediately seen by considering the entanglement spectrum and more precisely, the degeneracy of its largest eigenvalue.
We define here the entanglement Hamiltonian at site $n$
\begin{equation}
H_\mathrm{ent}(n) = - \log \mathrm{Tr}_{s > n} \vert \Psi \rangle \langle \Psi \vert,\label{eq:defEntSpec}
\end{equation}
where $s$ counts sites to the right of $n$, and we denote $\varepsilon_{\alpha, j}$ its $j^\mathrm{th}$ eigenvalue in the irrep $\alpha$.
This is readily accessible for matrix product states.
In the gapless phase for $J_2 \leq 0.5$, the dominant eigenvalue is a singlet, with strict invariance by translation.
Conversely, in the trimerized phase, the dominant eigenvalue form a repeating pattern of a singlet at site $n$, then a triplet (the fundamental or conjugate irrep) at site $n+1$ and $n+2$.
This is the expected structure for a simple product state made of singlets on three consecutive sites.
This pattern is a clear marker of the trimerized phase, and, as seen in Fig.~\ref{fig:ent_spec_su3}, gives similar estimations for the phase boundaries.
Remarkably, close to the peak of trimerization, the states are close to such a product state.
At $J_2 = 1.2$ and $\chi = 8192$, for the singlet cut, the largest eigenvalue of the density matrix is about $0.745$ (singlet state).
The next ones are two quasi-degenerate octuplets (i.e. adjoint irrep) of total weight $\approx 0.125$ each.

%%%%%%%%%%%%%%%%%%%%

%%%%%%%%%%%%%%%%%%%

\begin{figure}[!htb]
\centering
    \includegraphics[width=\linewidth,clip]{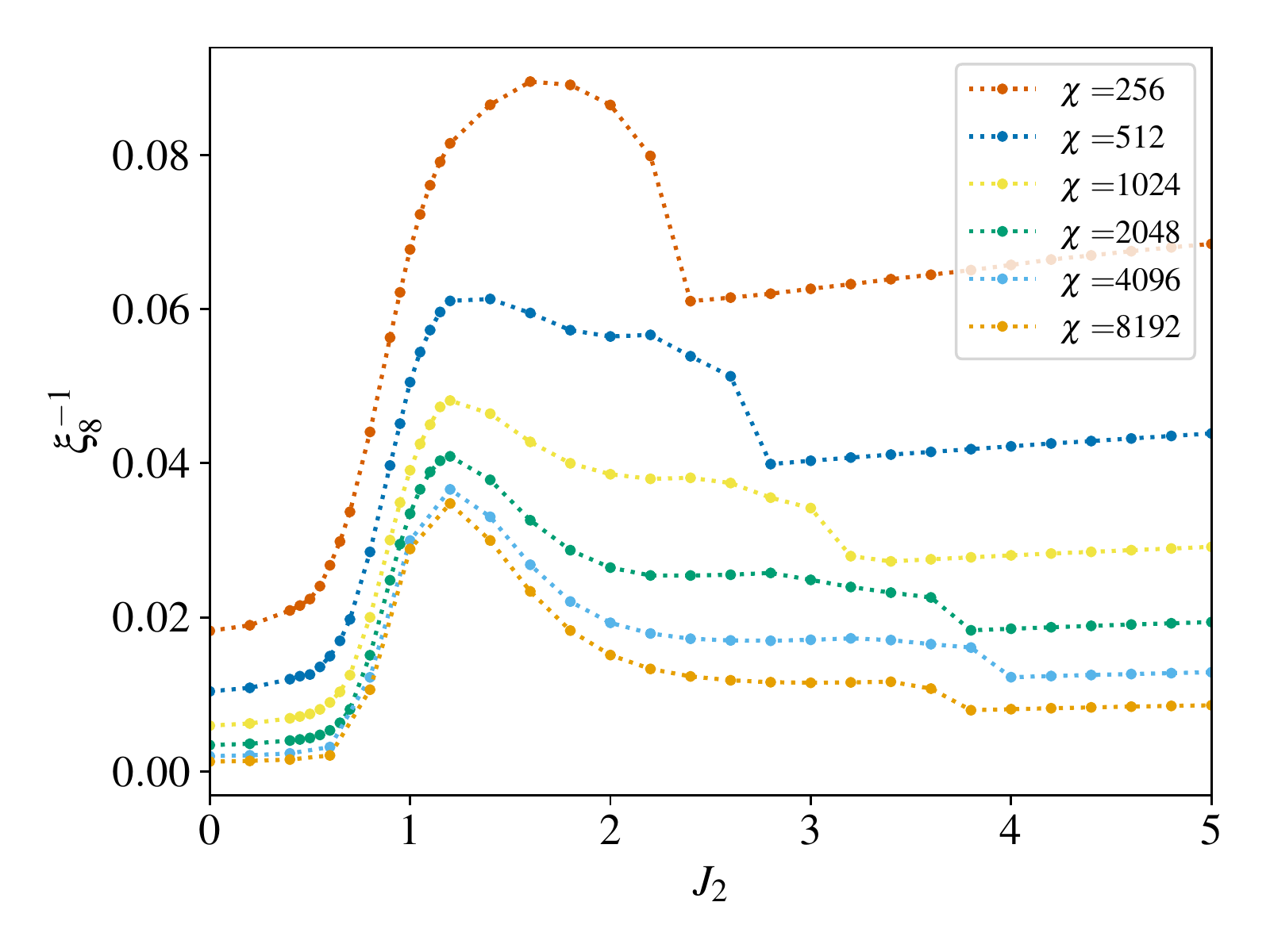}
    \caption{Inverse correlation length $\xi_8$ vs $J_2$ obtained by iDMRG in the SU(3) chain for various $\chi$. The gap opening at $J_2 \approx 0.5$ is immediately visible. The correlation length remains relatively large throughout the gapped phase. At large $J_2$, we observe also kinks in the correlation length, which marks the crossover due to finite $\chi$ that spoils our analysis.}
    \label{fig:su3_xi}
\end{figure}

\begin{figure}[!htb]
\centering
    \includegraphics[width=\linewidth,clip]{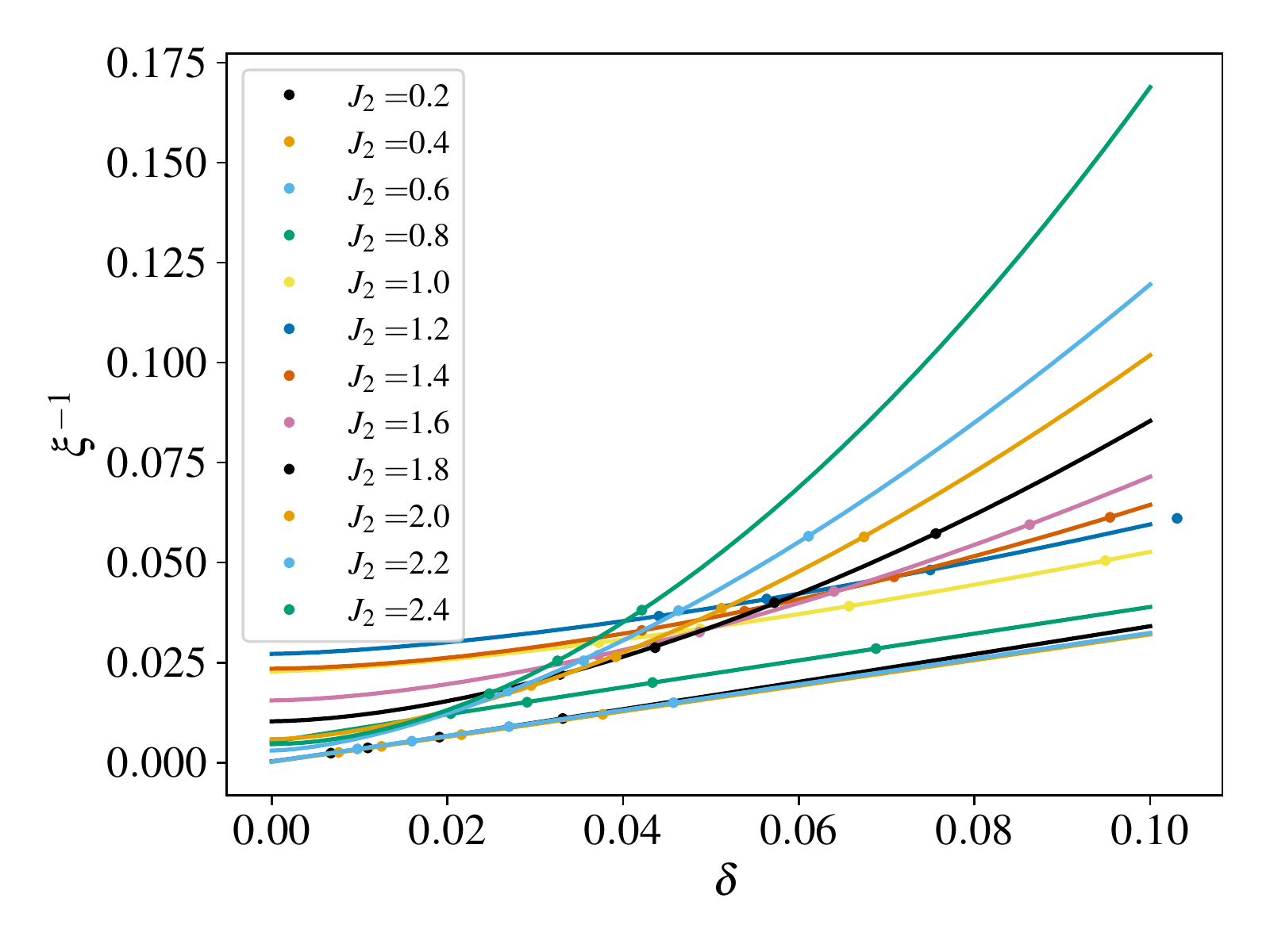}
    \caption{Scaling of the inverse correlation length $\xi$ vs $\delta$ obtained by iDMRG in the SU(3) case for various $\chi$, see text.}
    \label{fig:su3_xi_scaling}
\end{figure}

To complement these observables, we also investigated the correlation length $\xi$ (either induced from the finite $\chi$ in the gapless phases, or by the gap) obtained from the leading eigenvalues of the transfer matrix, see Fig.~\ref{fig:su3_xi}.
For a given irrep {\textbf{$\alpha$}}, we define the associated correlation length
\begin{equation}
\xi_\alpha = \frac{2N}{\log t_{\alpha, 1}},
\end{equation}
where $t_{\alpha, 1}$ is the first (nontrivial in the singlet case) eigenvalue of the transfer matrix.
For all $J_2$, the shortest correlation length appears to be in the adjoint irrep {\textbf{8}} and appears to be finite in the trimerized phase, see Fig.~\ref{fig:su3_xi_scaling}.
Even at the peak of trimerization, its value is about $30$ sites.
In order to limit the finite $\chi$ effects, we attempted to extrapolate $\xi$ using different gaps in the transfer matrix as scaling parameters~\cite{Vanhecke2019}.
In particular, we used the gap between the first and third {\textbf{8}} irreps in the eigenvalues
\begin{equation}
  \delta = \frac{1}{2N} \log\left(\frac{t_{8, 3}}{t_{8, 1}}\right),
\end{equation}
and the gap between the first irrep {\textbf{10}} and the first irrep {\textbf{8}}.
Both approaches led to similar results, suggestive of a finite region for the trimerized phase when $0.5 \lessapprox J_2 \lessapprox 3.8$, but significant level crossings with $\chi$ prevented us from getting a good approximation for the second transition point.

%%%%%%%%%%%%%%%%%%%%

%%%%%%%%%%%%%%%%%%%%

%%%%%%%%%%%%%%%%%%%%
\begin{figure}[!htb]
\centering
    \includegraphics[width=\linewidth,clip]{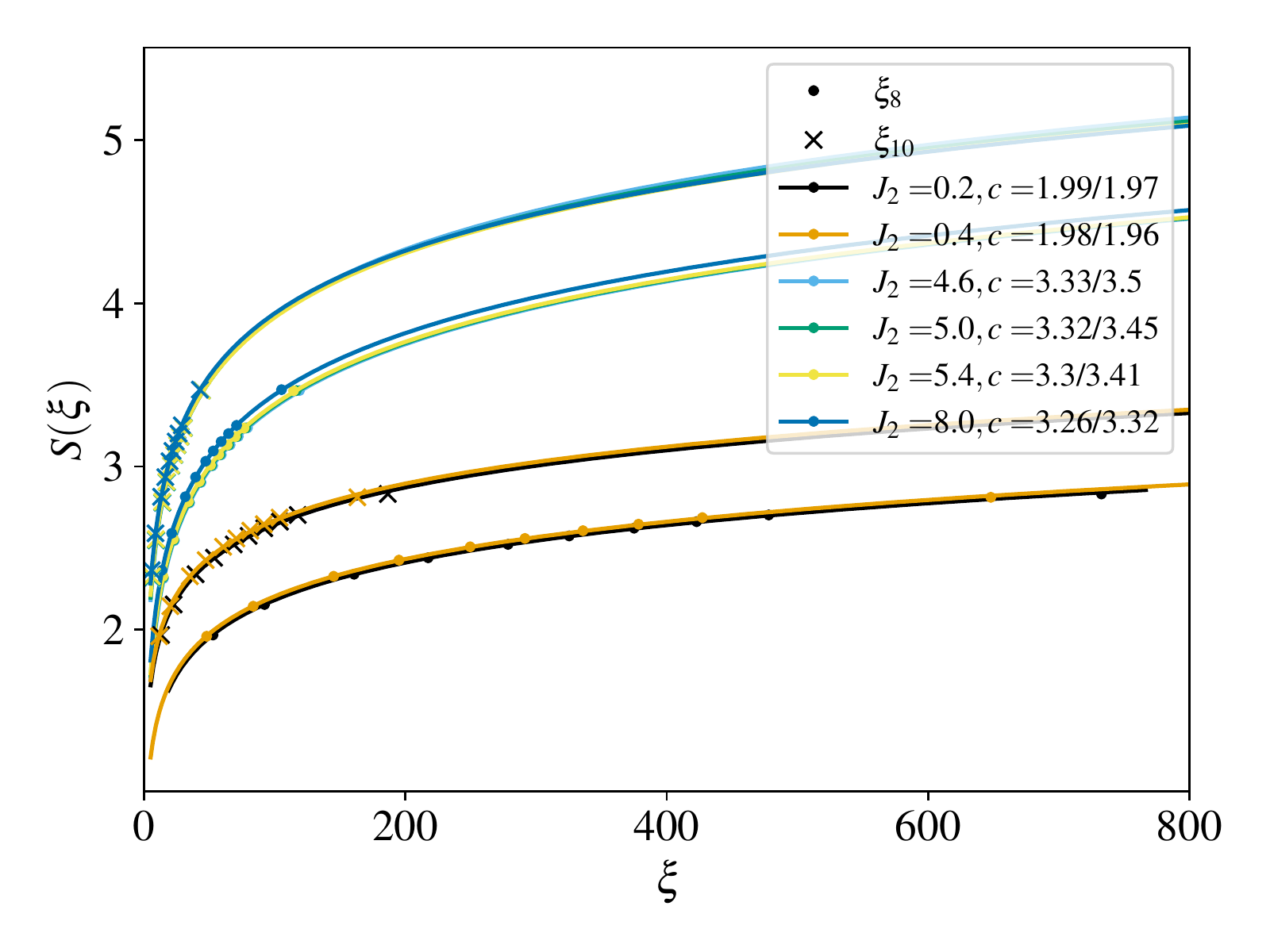}
    \caption{Entanglement entropy vs correlation length obtained by iDMRG in the SU(3) chain for various $\chi$, using the different correlation lengths we have access to. The lines are fit using the conformal formula in Eq.~\eqref{eq:centralCharge}. At low $J_2$, it matches the expected result of a SU(3)$_1$ critical theory. At large $J_2$, it reliably indicates a SU(3)$_2$ CFT, probably due to finite $\chi$, as discussed in the main text. The first indicated central charge is obtained using $\xi_8$ while the second one is obtained using $\xi_{10}$. They are in good agreement.
    }
    \label{fig:entropy_su3}
\end{figure}
%%%%%%%%%%%%%%%%%%%%

To further characterize the critical phases at small and large $J_2$, we have also computed the entanglement entropy $S$ in order to extract the central charge~\cite{Calabrese2009}
\begin{equation}
S \sim \frac{c}{6} \ln \xi + b,  \label{eq:centralCharge}
\end{equation}
when fitting with respect to the correlation length $\xi_8$ or $\xi_{10}$.
Data are shown in Fig.~\ref{fig:entropy_su3}. 
For $J_2< 0.5$, we obtain an excellent agreement with the expected value for SU(3)$_1$ criticality, namely $c=2$, despite using a very naive fit with only $c$ and $b$ as parameters. 
In the intermediate trimerized phase, the entanglement entropy $S$ saturates to a finite value. 
For large $J_2$, we obtain large values of $S$, which necessary limit the precision of our matrix product states.
We measure an apparent effective $c \simeq 3.2$, i.e. indicative of SU(3)$_2$ criticality~\footnote{We remind that the central charge for SU(N)$_k$ criticality is $c=k(N^2-1)/(k+N)$.}.
As discussed in Sec. \ref{low-energyoddN}, the low-energy approach starting from the zigzag regime, contains two different sectors, a first one described by Eq.~(\ref{effactionparaJperpF}) for the ${\mathbb{Z}}_3$ degrees of freedom which is fully gapped, and an SU(3)$_2$ one which enjoys a massless flow to SU(3)$_1$. 
The numerical observation of the SU(3)$_2$ criticality means that we resolve the  ${\mathbb{Z}}_3$ gap but not the SU(3)$_1$ criticality with $c=2$ obtained in the infrared regime. 
Related numerical problems were also noted in the context of the SU(3) Heisenberg spin chain in symmetric representation \cite{NatafMila-2016} and the correct quantum criticality has been obtained only by large-scale numerical simulations by exploiting the full non-Abelian SU(3) symmetry of the model~\cite{Nataf-G-M2021}. Our numerical approach exploits the color conservation but does not take into account the full SU(3) symmetry.
Hence, we expect that the observation of an apparent effective central charge $c \simeq 3.2$ is a crossover effect and $c$ should tend to 2 for large enough $\xi$.

As a side note, we find the scaling with $\xi_8$ more practical than a scaling directly with $\chi$: the effective exponents relating $\chi$ and $\xi$ differ from Ref.~\onlinecite{Pollmann2009}. We find numerically that 
\begin{equation}
 \xi_8 = \chi^\kappa 
 \end{equation} 
with $\kappa \approx 0.7$ for $J_2 \leq 0.5$ and $\kappa \approx 0.60$ at large $J_2$ (to compare with $0.87$ and $0.64$ using the standard formula). Data can be found in Appendix ~\ref{app:AddNumRes}.
This should not come as a surprise due to the large degeneracy of the entanglement spectrum enforced by the non-Abelian symmetry.

In the large $J_2$ limit for the SU(2) zigzag spin ladder, some incommensurability has been measured.~\cite{White-A-96}
This can be found using iDMRG by analyzing the imaginary part of the transfer matrix eigenvalues, see  Appendix \ref{app:su2}. 
We define the incommensuration associated to the eigenvalue $t_{\alpha, j}$ of the transfer matrix by
\begin{equation}
\theta_{\alpha, j} = \frac{1}{2N}\mathrm{Im} \log t_{\alpha, j}.\label{eq:defInc}
\end{equation}
Strictly speaking, nonzero $\theta_{\alpha, j}$ means that we expect some correlations $\langle \hat{O}(x) \hat{O}(y)\rangle$ to oscillate as $\cos ( 2N \theta_{\alpha, j} (X-Y))$ where $X$ (resp. $Y$) is the label of the unit-cell of $x$ (resp. $y$).
In practice, in our models, it is reasonable to expect that they will oscillate as $\cos \left( (\theta_{\alpha, j} + \frac{2m\pi}{L}) (x-y) \right)$, with $m$ an integer which cannot be determined directly from the transfer matrix. 
Our data for SU(3) shows no sign of finite incommensuration in the large $\chi$ limit, with nonzero $\theta$ remaining largely below $10^{-3}$ for eigenvalues of norm larger than $10^{-2}$, i.e. purely finite convergence and finite $\chi$ effects.

%%%%%%%%%%%%%%%%%%%%%%%%%%%%%%%%%%%%%%%%%%%%%%%%%%%%%%%%%
\subsection{SU(4) case}

\begin{figure}[!htb]
    \includegraphics[width=\linewidth,clip]{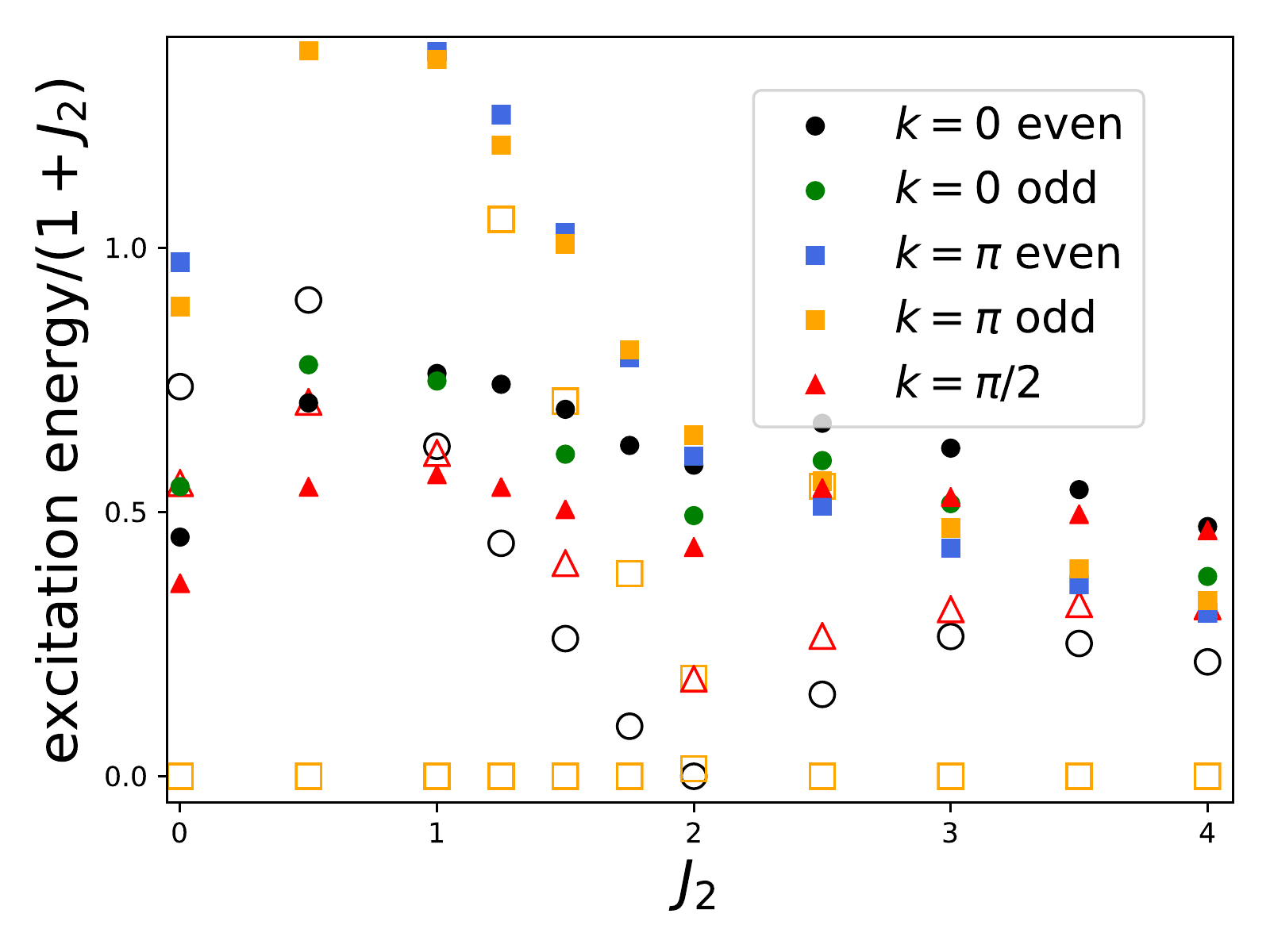}
    \caption{Low-energy spectrum vs $J_2$ obtained by ED in the SU(4) case on $L=20$ periodic chain. We have plotted only some particular momenta, relevant for the lowest excitations. Note that for $J_2 \geq 1$ the lowest excitation is a singlet with a momentum shift of $\pi$ with respect to the groundstate.}
    \label{fig:ed_su4}
\end{figure}

For the SU(4) case, we have performed extensive ED up to $L=20$ ($L=24$ for some parameters) periodic chains and the corresponding low-energy excitations are shown in Fig.~\ref{fig:ed_su4}. 
While the integrable $J_2=0$ case (corresponding to the critical Sutherland model) does exhibit a lowest excitation in the adjoint representation at momentum $\pm \pi/2$ as expected, we observe that for $J_2$ roughly larger than 1, the lowest excitation becomes a singlet with a $\pi$ momentum difference from the ground-state, possibly indicating a singlet phase that breaks translation symmetry. Quite interestingly, these two lowest singlets are quasi degenerate for $J_2\simeq 2$, although the model is quite simpler than the exact one showing a perfectly tetramerized VBS~\cite{Rachel-G-08}. For even larger $J_2$, the first excitation still remains a singlet with a $\pi$ momentum shift.

%%%%%%%%%%%%%%%%%%%%

%%%%%%%%%%%%%%%%%%%%

\begin{figure}[!htb]
    \includegraphics[width=\linewidth,clip]{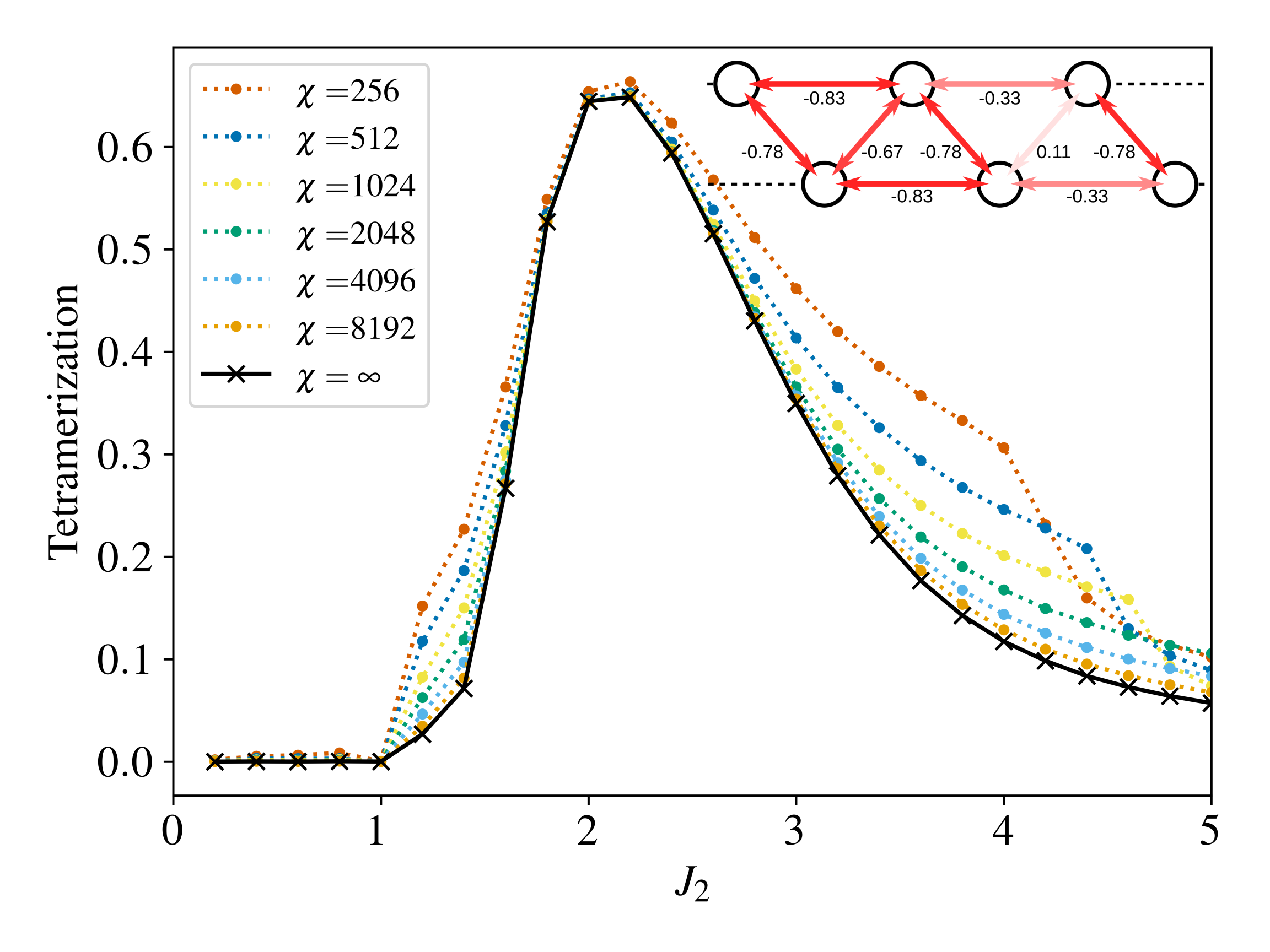}
    \caption{Tetramerization $T_{i, 4}$ vs $J_2$ obtained by iDMRG for the SU(4) chain for various $\chi$ as well as its extrapolation to infinite bond dimension  using a naive second degree polynomial fit. The tetramerized phase opens at $J_2 \approx 1$, and remains nonzero up to the largest $J_2 = 8.0$ we considered. Small non-analyticities are nonetheless visible, revealing the finite $\chi$ and convergence problems at large $J_2$.
In inset, we represented the numerically obtained bond energies $\langle \hat{\mathrm{P}}_{ij} \rangle$ for $J_2 = 2.0$, revealing the appearance of the singlet phase. 
    }
    \label{fig:dmrg_su4}
\end{figure}

In order to fully characterize this symmetry breaking, we have performed iDMRG simulations with a unit-cell of $2N=8$ sites and various maximal bond dimensions $\chi$. 
We define a tetramerization order parameter as
$$ T_{i, 4} = \langle \hat{\mathrm{P}}_{i,i+1} - \frac{1}{4}\sum\limits_{k = 0}^3 \hat{\mathrm{P}}_{i+k,i+k+1}\rangle$$
measured in the ground-state and choosing appropriate bonds so that it is positive and largest.
Data are shown in Fig.~\ref{fig:dmrg_su4} and are consistent with a critical value $J_2 \simeq 1$ for the transition between a critical phase and a tetramerized one which extends up to the large $J_2$ regime  in agreement with the field theory prediction.
 Note that the apparent anomaly at $J_2 \simeq 4$ becomes less pronounced when increasing the bond dimension $\chi$. We see no clear signs of the reopening of a uniform gapless phase at large $J_2$, with significantly better stability and convergence than for SU(3).
The $J_2$ bonds pattern exhibits a $\pi/2$ modulation (strong/strong/strong/weak), see e.g. the corresponding modulation in inset of Fig.~\ref{fig:dmrg_su4}. 
As in the SU(3) case, this tetramerized phase appears to be adiabatically connected to the limit of a tensor product of singlets over four consecutive sites.
The degeneracy of the largest eigenvalue of the entanglement spectrum, defined in Eq.~\eqref{eq:defEntSpec}, supports this picture, with a repeating pattern of ${\bf 1}$ (trivial irrep), ${\bf 4}$  (fundamental irrep), ${\bf 6}$ (the fully antisymmetric self-conjugate irrep) and  ${\bf \bar{4}}$ (the conjugate of the fundamental irrep).
Similarly to the SU(3) limit, for $J_2 = 2.0$ and $\chi = 8192$, the largest eigenvalue of the reduced density matrix reaches about $0.811$ in the singlet cut, while the second eigenvalue has a total weight of $0.104$ in the irrep {\bf 15} (adjoint irrep), revealing how close the system is from the singlet product state. 

%%%%%%%%%%%%%%%%%%%%
%%%%%%%%%%%%%%%%%%%%

\begin{figure}[!htb]
\centering
   \subfloat{\includegraphics[width=\linewidth,clip]{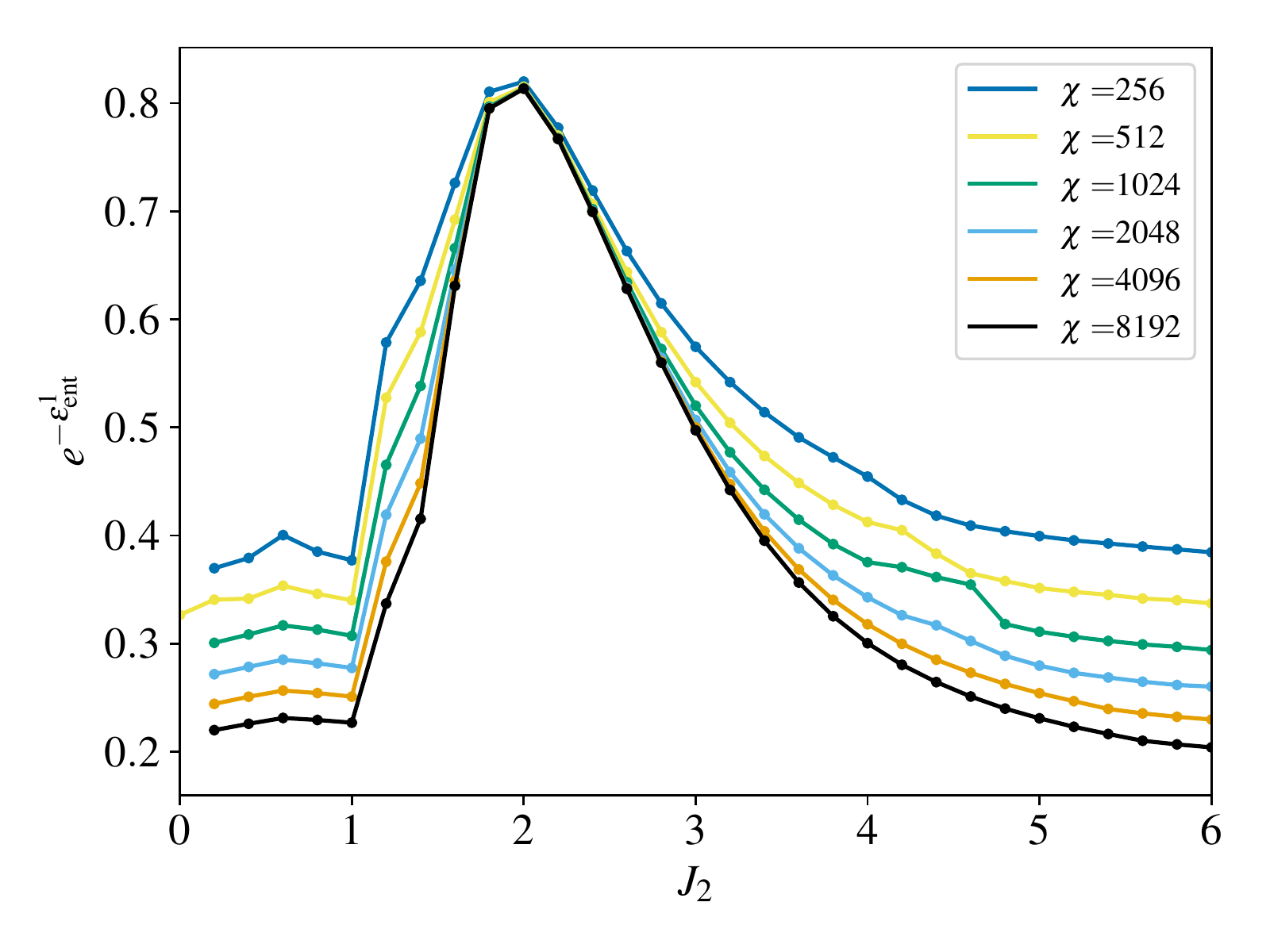}}\\
   \subfloat{\includegraphics[width=\linewidth,clip]{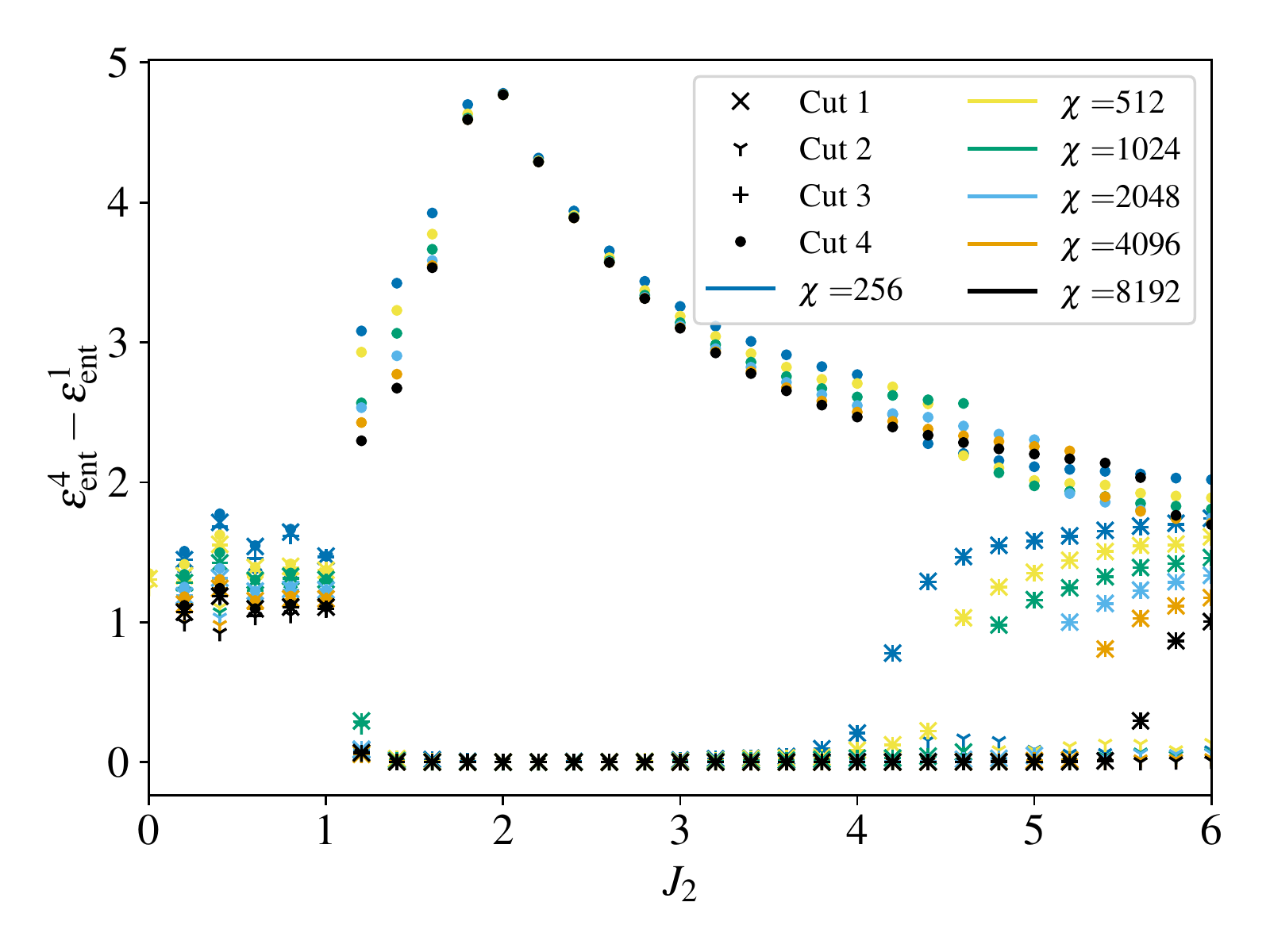}}
    \caption{Top: the largest eigenvalue of the "mid-chain" reduced density matrix on the weakest bond of the unit-cell for SU(4), as defined in Eq.~\eqref{eq:defEntSpec}. The opening of the tetramerized phase is characterized by a large peak where the matrix product state becomes close to a product state. Right: gap between sixth and first entanglement energies. The singlet pattern $1 4 6 4$ is characteristic of a product state of consecutive SU(4) singlets. We observe a breakdown of SU(4) at large $J_2$ due to finite $\chi$.}
    \label{fig:ent_spec_su4}
\end{figure}

%
At large $J_2 \geq 5.0$, we see a crossover towards a degeneracy pattern of $(1, 2, 4, 2)$, which is also visible at lower $J_2$ and $\chi$, and explain the small anomaly in the tetramerization for $J_2 \approx 5.0$, see Fig.~\ref{fig:ent_spec_su4}.
Note that this pattern is not SU(4) invariant, it is therefore a finite convergence and $\chi$ effect coming from the color conservation.
Note that the gapless phase for $J_2 \leq 1$ is again characterized by a single, nondegenerate largest eigenvalue.\\
%

%%%%%%%%%%%%%%%%%%%%

%%%%%%%%%%%%%%%%%%%

\begin{figure}[!htb]
    \includegraphics[width=\linewidth,clip]{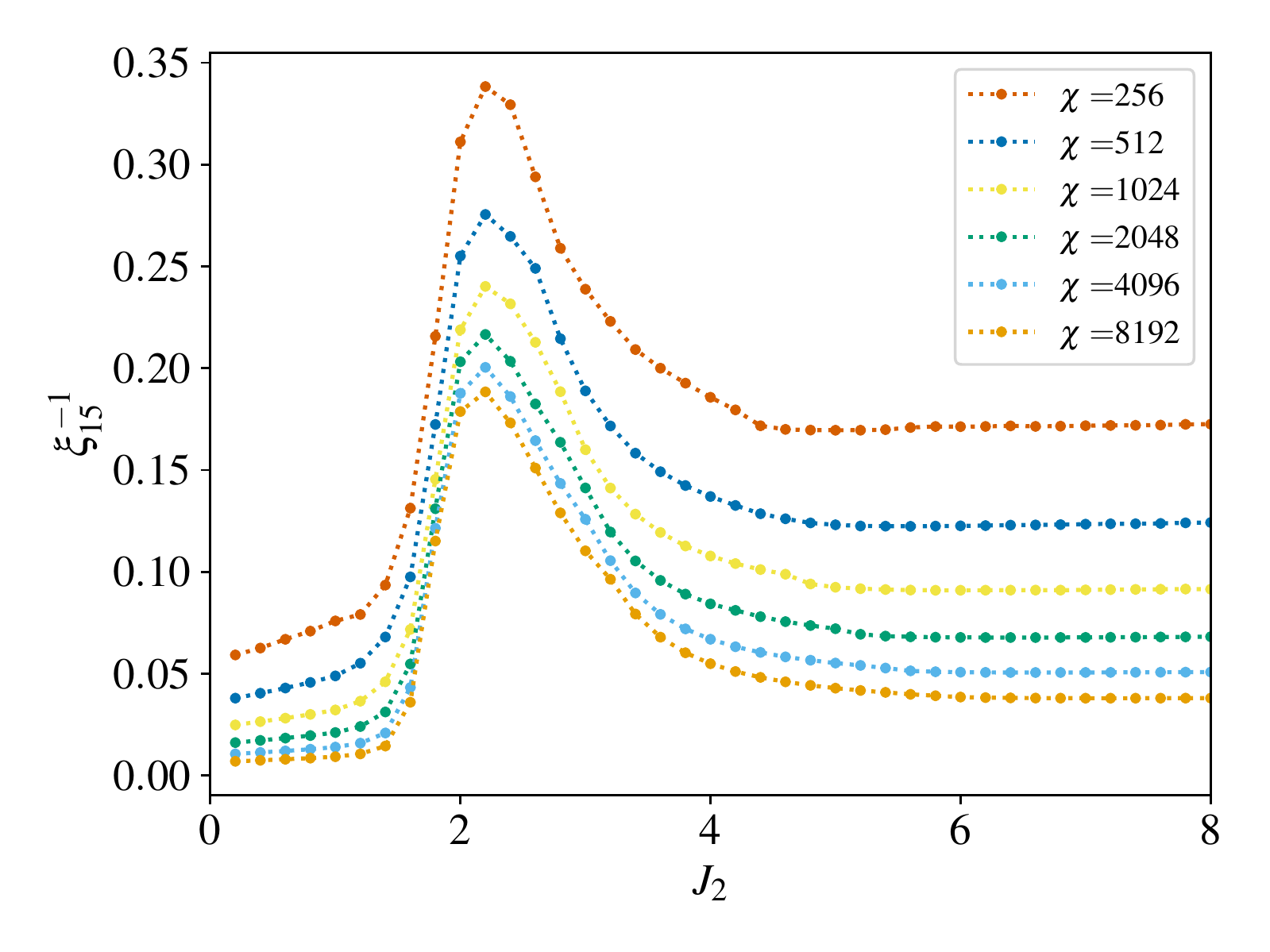}
    \caption{Inverse correlation length $\xi_{15}$ vs $J_2$ obtained by iDMRG in the SU(4) chain for various $\chi$. The gap opening at $J_2 \approx 0.5$ is immediately visible. The correlation length remains relatively large throughout the gapped phase. At large $J_2$, we observe also kinks in the correlation length, which marks the crossover due to finite $\chi$ that spoil our analysis.}
    \label{fig:su4_xi}
\end{figure}

The dominant correlation length comes from the adjoint representation {\bf 15}, shown in Fig.~\ref{fig:su4_xi}, and is of the order of the unit-cell close to the maximum of tetramerization.
We use as convergence parameters the gaps between the first and third adjoint irreps, and the gap $\xi^{-1}_{10,1}-\xi^{-1}_{8,1}$.
Both leads to similar results for $J_2 \leq 4.5$, beyond which the finite $\chi$ effects are too large.
The results are also coherent with a gapped phase remaining open for $J_2 \gg 1$.\\

%%%%%%%%%%%%%%%%%%%%

%%%%%%%%%%%%%%%%%%%%

\begin{figure}[!htb]
    \includegraphics[width=\linewidth,clip]{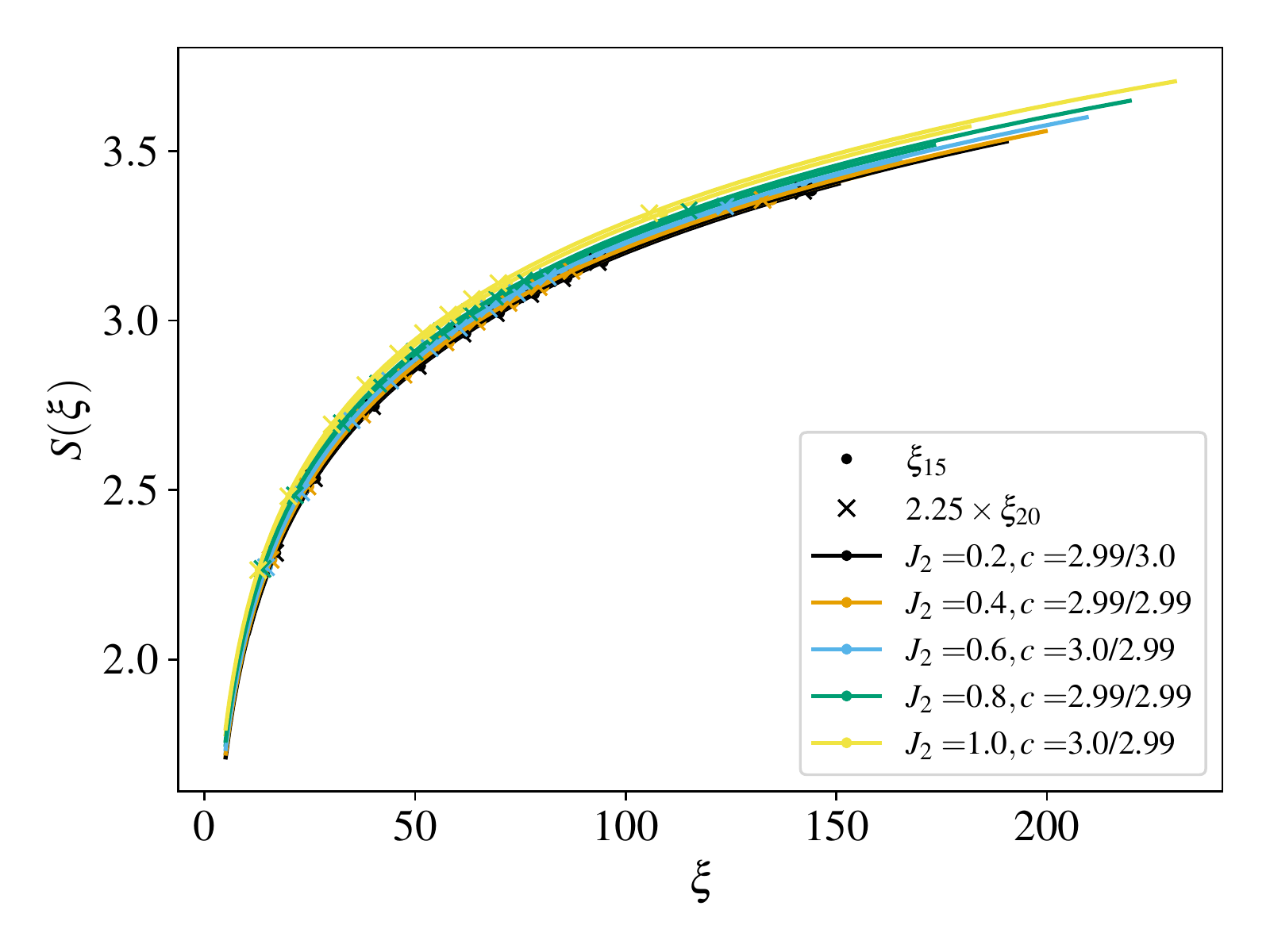}
    \caption{Entanglement entropy vs correlation length obtained by iDMRG in the SU(4) chain for various $\chi$ for the different correlation lengths we used. The lines are fit using the conformal formula in Eq.~\eqref{eq:centralCharge}. It matches the expected result of a SU(4)$_1$ critical theory for $J_2 \leq 1$. The first central charge is obtained using $\xi_{15}$ while the second one is obtained with $\xi_{20}$. }
    \label{fig:entropy_su4}
\end{figure}

We verified that the gapless phase for $J_2 \leq 1$ has central charge $c=3$, using the scaling with the correlation lengths (see Fig. \ref{fig:entropy_su4}).
We numerically found that $\xi_{15} \approx \chi^{0.6}$ instead of the naive $\chi^{2/3}$.
For $J_2>5$, our bond dimension is too low to resolve the gap, despite the finite tetramerization.
A naive fit of the entropy gives a central charge $c\approx 5.6$, only slightly reduced compared to two independent $c=3$ chains.
This estimate of the central charge seems to be consistent with the low-energy approach for $N=4$ since, in the large $J_2$ regime,  
we expect a first crossover from the $c=6$ decoupling regime to a critical regime with
$c=6- (1- 4/5) = 5.8$ which stems form the existence of the massless RG flow of model (\ref{SDSG}) from the ${\mathbb Z}_4$ $c=1$ UV fixed point to the 
${\mathbb Z}_3$ IR one with $c=4/5$. In the far-IR regime, a fully gapped tetramerized phase should eventually emerge.

%%%%%%%%%%%%%%%%%%%%
%%%%%%%%%%%%%%%%%%%%

\begin{figure}[!htb]
    \includegraphics[width=\linewidth,clip]{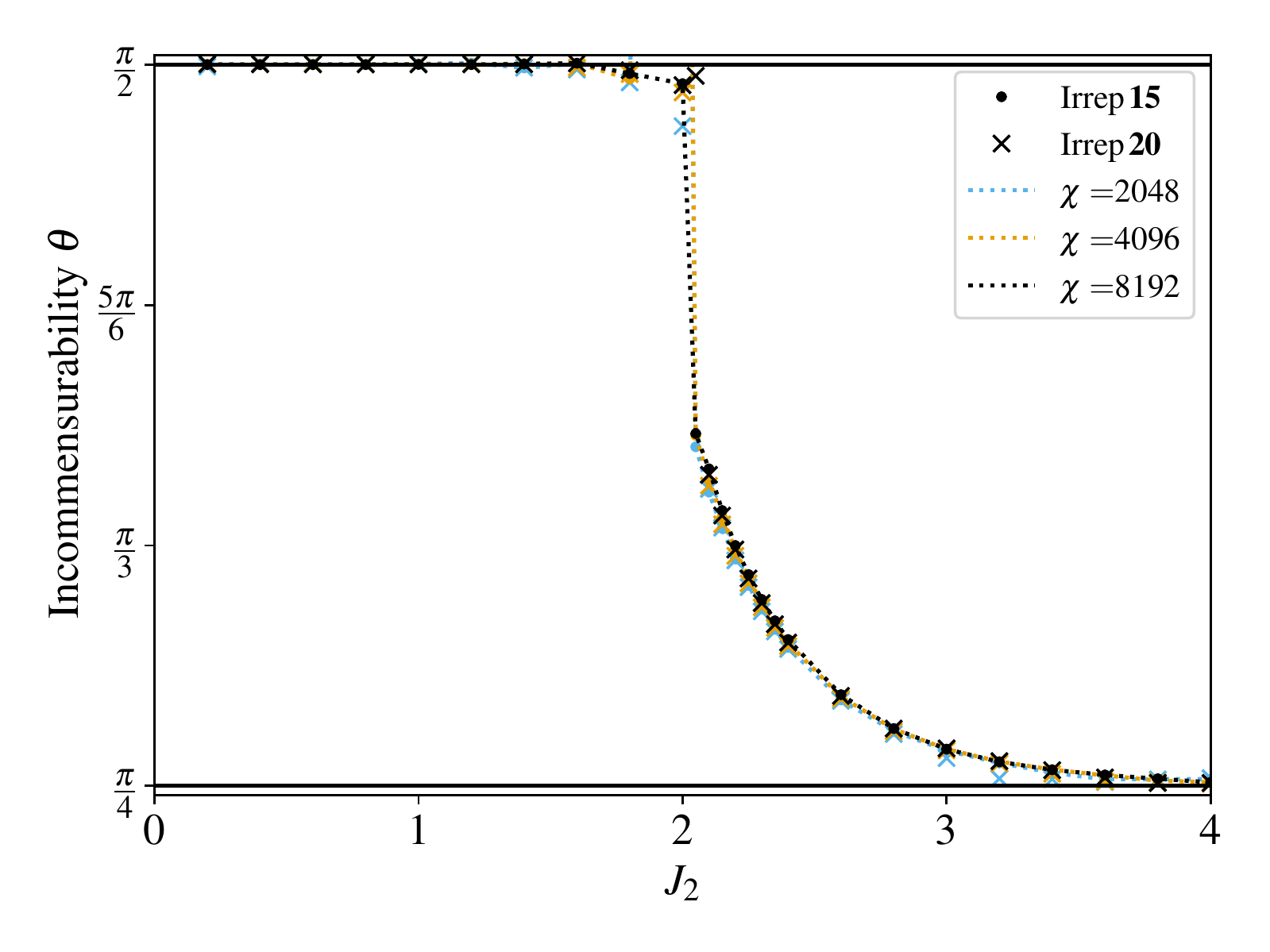}
    \caption{Incommensuration measured as in Eq.~\eqref{eq:defInc} for the SU(4) chain as a function of $J_2$ for various $\chi$. Unlike the SU(3) chain, incommensuration appears for $J_2 \approx 2.0$, close to the point with the shortest correlation length. The complex eigenvalues start one order of magnitude smaller than the dominant eigenvalues, and appear (at least) in the singlet and adjoint irrep. Note that with our choice of unit-cell, the incommensurability is only defined modulo $\pi/4$. We suggestively chose $\theta = \frac{\pi}{2}$ for $J_2 < 2$ as it matches the pattern induced by the four sites singlets in an open chain.}
    \label{fig:incoSU4}
\end{figure}

Finally, a careful investigation of the transfer matrix revealed a commensurate-incommensurate crossover at $J_2 \approx 2$, similar to the one observed in the SU(2) $J_1-J_2$ spin chain.
This incommensuration is not visible in the first nontrivial eigenvalues of the transfer matrix, even taking into account the different irreps.
This is actually also the case for SU(2) spin chains close to the Majumdar-Ghosh point (See Appendix E).
Unlike SU(2) though, the incommensurate eigenvalues never become dominant in the regime where we reliably measure incommensuration (for $J_2 \leqslant 4.0$).
As an example, for $J_2 = 2.2$ and $\chi = 8192$, the first non trivial values are $t_{15, 1} \approx 0.221$ and we see incommensuration in this irrep only for a smaller eigenvalue $t^\mathrm{inc}_{15, j} \approx -0.023 + 0.040i$.
Note that for this $J_2$ and $\chi$, the entropy is converged at $10^{-9}$ and the absolute value of $t^{inc}_{15, j}$  slowly increases with $\chi$.
At larger $J_2 = 4.0$ and $\chi = 8192$, we have  $t_{15, 1} \approx 0.645$ and the first incommensuration for $t^\mathrm{inc}_{15, 4} \approx 0.456 - 0.014 i$.
Therefore, given the clear converged signal shown in Fig.~\ref{fig:incoSU4}, we are confident that this incommensuration is not a numerical artifact.
Within our precision, the commensurate-incommensurate crossover coincides with the point where the singlets are maximally localized, i.e., when the entropy between the $4$th and $5$th sites of the unit-cell is minimal.
The natural interpretation of our data is that the incommensuration opens only in a specific excitation channel with a significantly larger gap.
We verified that the corresponding right eigenvector has nonzero overlap with the tensor
\begin{center}
\includegraphics[width=0.1\linewidth]{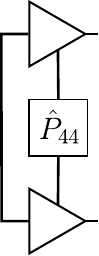}
\end{center}
where $\hat{P}_{44}$ is the local projector on $(0, 0, 0, 1)$ and the triangles denote the left canonical matrix product state.
Hence, the incommensuration will be observable in the spin-spin correlations in the form of some $\cos(\theta (x-y))$ modulations, although they are subdominant (see Appendix \ref{app:AddNumRes}).

It is tempting to explain the numerical evidence of this incommensuration for $N=4$ by the existence of a nonzero conformal spin operator, a twist term, in  the low-energy approach of the two-leg zigzag spin ladder (\ref{eqn:def-2leg-ladder}). Such spin-1 conformal spin operator, $N_1^{A} \partial_x N_2^{A}$, with scaling dimension $3 - 2/N$ has been conjectured in Ref. \onlinecite{Nersesyan1998} to be the source of the incommensuration of the SU(2) $J_1-J_2$ Heisenberg spin chain. Such an operator is marginal in the $N=2$ whereas slightly irrelevant when $N=4$, a fact  which is  consistent with the subdominant incommensurate behavior observed for $N=4$.

%%%%%%%%%%%%%%%%%%%%%%%%%%%%%%%%%%%%%%%%%%%%%%%%%%%%%%%
\section{Conclusion}
\label{sec:conclusion}
Using complementary analytical and numerical techniques, we have investigated the ground-state phase diagram of the $J_1-J_2$ SU($N$) antiferromagnetic Heisenberg spin chain, in the fundamental representation, which is a minimal model to study frustration effect in SU($N$) chains. By varying the sole parameter $J_2/J_1$, we have shown that different phases can emerge. First, the well-known (integrable) case at $J_2=0$ corresponds to an SU($N$)$_1$ critical phase (Luttinger liquid) with $c=N-1$ gapless modes, which has a finite extension for all $N$. Then, for $J_2 > J_2^{c, 1}$, the system enters an $N$-merized phase which breaks spontaneously lattice translation symmetry (also known as a valence bond solid or VBS) and which can be understood as being adiabatically connected to a product of SU($N$) singlets (made of $N$ consecutive sites). Such a perfect  SU($N$) VBS is the exact ground-state of parent Hamiltonians with multi-site interactions~\cite{Rachel-G-08} and fine tuned parameters, while our microscopic model appears to be more realistic.
In the $N$-merized phase, the low-lying excitations are made of SU($N$) spinons with fractional quantum numbers which have been shown, within the low-energy approach, to belong to the ${\bar {\bf N}}$ irrep of the SU($N$) group. As in the $N=2$ case, these spinons can be viewed as domain-wall configurations between the $N$-degenerate ground states of the $N$-merized phase.

 For very strong $J_2$, the nature of the ground-state crucially depends on the parity of $N$: for even $N$, there is no other phase transition and the chain remains in a gapped $N$-merized phase that breaks translation symmetry; conversely, for odd $N$, the $N$-merized phase has only a finite extension and gives rise to another SU($N$)$_1$ gapless phase which persists for $J_2/J_1\rightarrow \infty$.
On top of this quantitative difference between even and odd $N$, we have numerically observed a much subtle effect with the emergence of an incommensurate behavior when $N=4$. Our iDMRG simulations have revealed a nonzero  imaginary part of some transfer matrix eigenvalues, leading to incommensurate spin-spin correlations for $N=2,4$ whereas no incommensuration is found for $N=3$. The incommensurate behavior for $N=4$ is reminiscent of the  behavior found for SU(2) in Ref. \onlinecite{White-A-96} with incommensurate correlations above the Majumdar-Ghosh point, the main difference being that the incommensuration is subdominant in the spin-spin correlation function for SU(4).
 
As perspectives, it would be worth investigating the physical origin of this incommensuration phenomenon, as well as checking if it exists for all even $N$.  From a numerical point of view, we have discussed the difficulty of simulating the regime $J_2 \gg J_1$, in particular to get the correct criticality for $N=3$ or to get an accurate second critical value, which could be overcome using full SU($N$) symmetry in the simulations.~\cite{NatafMila-2016,Devos2022} 
Finally, due to the experimental realizations of SU($N$)-symmetric interactions in ultracold atomic gases, it will be very interesting to investigate experimentally the competition, revealed here in a simple model, between magnetic ordering and valence-bond solids.

%%%%%%%%%%%%%%%%%%%%%%%%%%%%%%%%%%%%%%%%%%%%%%%%%%%%%%%
%%%%%%%%%%%%%%%%%%%%%%%%%%%%%%%%%%%%%%%%%%%%%%%%%%%%%%%%%%%%%
\begin{acknowledgments}
We would like to thank Matt Fishman, Olivier Gauth\'e, Fr\'ed\'eric Mila, and Keisuke Totsuka for useful discussions.
S. C. acknowledges the use of HPC resources from CALMIP (Grant No. 2022-P0677) and GENCI (Project No. A0110500225).  L. H. has been supported by the Swiss National Science Foundation (FM) Grant No. 182179. Part of the simulations have been performed on the facilities of the Scientific IT and Application Support Center of EPFL. S.C. and P.L. would like to thank CNRS for constant support over the years and acknowledge the financial support (grant CNRS IRP EXQMS).
\end{acknowledgments}

%%%%%%%%%%%%%%%%%%%%%%%%%%%%%%%%%%%%%%%%%%%%%%%%%%%%
\appendix
%%%%%%%%%%%%%%%%%%%%%%%%%%%%%%%%%%%%%%%%%%%%%%%%%%%%

%%%%%%%%%%%%%%%%%%%%%%%%%%%%%%%%%%%%%%%%%%%%%%%%%%%%
\section{Abelian-bosonization description of the $N$-merized phase}
\label{sec:AppendixA}

In this Appendix, we use the Abelian-bosonization approach to model \eqref{hamJ1J2} in the weak-coupling limit $J_2 \ll J_1$ to describe the low-energy properties of the $N$-merized phase.

As is well-known, the physical properties of the SU($N$) Sutherland model, e.g. model \eqref{hamJ1J2} with $J_2=0$, can be obtained from the large 
repulsive $U$ limit of the U($N$) Hubbard chain at $1/N$-filling with Hamiltonian:\cite{Affleck-NP86,Affleck-88,James-K-L-R-T-18}
\begin{eqnarray}
H_{\rm U} &=& - t \sum_i \sum_{\alpha=1}^{N} \left( c^{\dagger}_{\alpha,i+1} c_{\alpha,i} + \mathrm{H.c.}\right)
\nonumber \\
&+& \frac{U}{2} \sum_{i, \alpha,\beta} n_{\alpha,i}   n_{\beta,i} \left( 1 - \delta_{\alpha\beta} \right),
\label{hubbardSUN}
\end{eqnarray}
where $c^{\dagger}_{\alpha,i}$ creates a  fermion with color $\alpha=1, \ldots N$ of the site $i$ and $
n_{\alpha,i} = c^{\dagger}_{\alpha,i} c_{\alpha,i}$ is the occupation number. 
When $U$ is sufficiently large, the system becomes a Mott insulator and 
the charge degrees of freedom gets decoupled from the low-energy physics. \cite{Affleck-88,Assaraf-A-C-L-99}
The latter is described by the SU($N$) Sutherland model with  an SU($N$) spin operator on site $i$  which reads as follows in terms of the lattice fermions:
\begin{equation}
 S^{A}_i = c^{\dagger}_{\alpha,i}  T^{A}_{\alpha \beta}   c_{\beta,i} ,
 \label{spinopfermion}
 \end{equation}
 where $T^A$ are the generators of the $\bolN$-representation of SU($N$)  normalized such that $\text{Tr}(T^{A} T^{B})=\delta^{AB}/2$.
Using the continuum limit of the lattice fermion operators in terms of $N$ left-right moving Dirac fermions: $c_{\alpha,i}/\sqrt{a_0} \rightarrow e^{-i k_F x } L_{\alpha} (x)
+ e^{i k_F x }  R_{\alpha}(x) $ (with $k_{F} = \pi/(Na_0)$), one gets the low-energy identification as in Eq. (\ref{spinopappen}):
\begin{equation}
S^{A}_i/ a_0 \rightarrow J^{A}_{R} + J^{A}_{L} + e^{i 2k_F x }  N^{A} + \mathrm{H.c.} +...,
\label{spinopapp}
\end{equation}
where $J^{A}_{L} = L^{\dagger}_{\alpha}T^{A}_{\alpha \beta}  L_{\beta}$ is the left SU($N$)$_1$ current with a similar definition for the right one. 
In Eq. (\ref{spinopapp}), the 2$k_F$ SU($N$) spin density   is 
$N^{A} = \langle  L^{\dagger}_{\alpha}T^{A}_{\alpha \beta}  R_{\beta} \rangle_c$, the average being over the fully gapped
charge mode. 
The Hamiltonian density  (\ref{HcontJ1J2}) which captures the IR physical properties of the $J_1-J_2$ SU($N$) Heisenberg spin chain in the weak-coupling $J_2 \ll J_1$ can be expressed in terms of the Dirac fermions:
\begin{eqnarray}
  \mathcal{H}_{J_1-J_2} &=& \frac{2\pi v}{N + 1} \left( : J^A_{R} J^A_{R}: + : J^A_{L} J^A_{L}: 
\right)  + \lambda  J_R^{A} J_L^{A}  \nonumber \\ 
&=&- i v \left (: R^{\dagger}_{\alpha} \partial_x R_{\alpha}: - : L^{\dagger}_{\alpha} \partial_x L_{\alpha}: \right)
+ \frac{\lambda }{2} R^{\dagger}_{\alpha} R_{\beta} L^{\dagger}_{\beta} L_{\alpha} \nonumber \\
&-& \frac{\lambda }{2N} : R^{\dagger}_{\alpha} R_{\alpha}:: L^{\dagger}_{\beta} L_{\beta}:,
\label{HcontJ1J2fer}
\end{eqnarray}
where we have used the identity: 
\begin{eqnarray}
T^{A}_{\alpha\beta} T^{A}_{\gamma \delta} = \frac{1}{2} \left( \delta_{\alpha \delta} \delta_{ \beta \gamma} - \frac{1}{N} \delta_{\alpha \beta} \delta_{ \ \delta\gamma} \right).
\end{eqnarray}

We introduce now  $N$ left-right moving bosons $\varphi_{\alpha L,R}$  to bosonize the Dirac fermions in Eq. (\ref{HcontJ1J2fer}): \cite{James-K-L-R-T-18,Gogolin-N-T-book}
\begin{eqnarray}
 L_{\alpha} &=& \frac{\kappa_{\alpha}}{\sqrt{2 \pi a_0}} \; e^{ - i \sqrt{4 \pi}\varphi_{\alpha L}}, \nonumber \\
 R_{\alpha} &=& \frac{\kappa_{\alpha}}{\sqrt{2 \pi a_0}} \; e^{ i \sqrt{4 \pi}\varphi_{\alpha R}}  ,
\label{bosoabeleq}
\end{eqnarray}
where $[\varphi_{\alpha R}, \varphi_{\beta L} ] = i \delta_{\alpha\beta}/4$ and $\kappa_{\alpha}$ are Klein factors to ensure 
the anticommutation of fermions with different colors: $\{\kappa_{\alpha}, \kappa_{\beta} \} = 2   \delta_{\alpha\beta} $,
$\kappa^{\dagger}_{\alpha} = \kappa_{\alpha}$. The 2$k_F$ SU($N$) spin density  can thus be expressed in terms
of these bosonic fields:
\begin{equation}
N^{A} =  \frac{\kappa_{\alpha} \kappa_{\beta} i^{\delta_{\alpha\beta}}}{2 \pi a_0}  T^{A}_{\alpha \beta}    
\langle e^{  i \sqrt{4 \pi}\varphi_{\alpha L} + i \sqrt{4 \pi}\varphi_{\beta R} } \rangle_c .
\label{BoseNA}
\end{equation}
It is then helpful to introduce a new basis with one charge bosonic field $\Phi_{cR,L}$ and 
$N-1$ spin fields $\Phi_{smR,L}$ ($ m=1, \ldots N-1$) to perform the charge average:\cite{Assaraf-A-C-L-99}
\begin{eqnarray}
&& \Phi_{cR,L} = \frac{1}{\sqrt{N}} \sum_{\alpha =1}^{N} \varphi_{\alpha R,L } \label{SUNbasis} \\
&& \Phi_{sm R,L} =  \frac{1}{\sqrt{m(m+1)}} \left( \sum_{p=1}^{m} \varphi_{pR,L} - m
\varphi_{m+1 R,L} \right)  \nonumber,
\end{eqnarray}
the inverse transformation being:
\begin{eqnarray}
 \varphi_{\alpha R,L }  &=& \frac{\Phi_{cR,L}}{\sqrt{N}} + \sum_{m=1}^{N-1} e^{m}_{\alpha} \Phi_{sm R,L} 
 \nonumber \\
 &=&   \frac{\Phi_{cR,L}}{\sqrt{N}} + {\vec  e}_{\alpha} \cdot {\vec  \Phi}_{sR,L}  ,
\label{invSUN}
\end{eqnarray}
where ${\vec  e}_{\alpha}$ ($\alpha=1, \ldots, N$)  are $N-1$-dimensional vectors  which satisfy:
\begin{subequations}
\begin{align}
\sum_{\alpha=1}^N \vec{e}_\alpha &= {\vec 0}, \\
\sum_{\alpha=1}^N [\vec{e}_\alpha]^m [\vec{e}_\alpha]^{m'} &= \delta_{mm'}, \\
 \vec{e}_\alpha {\cdot} \vec{e}_\beta
&= \delta_{\alpha \beta} -\frac{1}{N},
\end{align}
\label{weightSUN}
\end{subequations}
where $m=1, \ldots, N -1$ describes the components of the ${\vec  e}_{\alpha}$ vectors. An explicit choice is:
\begin{align} \label{eq:SpecificWeight}
[\vec{e}_\alpha]^m = \begin{cases} \frac{1}{\sqrt{m(m+1)}} & (m \geq \alpha) \\ -\sqrt{\frac{m}{m+1}} & (m=\alpha-1) \\ 0 & (m<\alpha-1) \end{cases}. 
\end{align}

 A free-field representation of  the SU($N$)$_1$ WZNW $g$ primary field can then be obtained using 
Eq. (\ref{BoseNA}) and the identity (\ref{WZNWcont}):
\begin{eqnarray}
g_{\beta \alpha} &= \frac{\kappa_{\alpha} \kappa_{\beta} i^{\delta_{\alpha\beta}-1}}{\sqrt{N}}
: e^{  i \sqrt{4 \pi} {\vec  e}_{\alpha} \cdot {\vec  \Phi}_{sL}
+ i \sqrt{4 \pi}  {\vec  e}_{\beta} \cdot {\vec  \Phi}_{sR}  }: .
\label{freefieldrepWZWGfield}
\end{eqnarray}
A more rigorous free-field representation of the $g$ WZNW  field can be found 
in Ref. \onlinecite{Fuji-L-17}  (see Appendix B) where the Klein factors are constructed out of the zero mode operators of the
spin bosonic fields.  From the correspondence (\ref{freefieldrepWZWGfield}), we deduce an Abelian bosonization of the $N$-merization operator (\ref{Nmerisation}):
\begin{eqnarray}
e^{- i \frac{2 \pi n}{N}} S^{A}_{n}  S^{A}_{n+1}  \sim {\rm Tr} g  =   \frac{1}{\sqrt{N}} \sum_{\alpha=1}^{N} e^{i \sqrt{4 \pi}   {\vec  e}_{\alpha} \cdot {\vec  \Phi}_{s}}, 
\label{Nmerisationboso}
\end{eqnarray}
where ${\vec  \Phi}_{s} = {\vec  \Phi}_{sL} + {\vec  \Phi}_{sR}$. Since the WZNW $g$  field is the elementary primary field of the SU($N$)$_1$ CFT, the bosonic field ${\vec  \Phi}_{s}$ is compactified with the following redundancy which leaves invariant (\ref{Nmerisationboso}):
\begin{eqnarray}
 {\vec  \Phi}_{s} \sim  {\vec  \Phi}_{s} + \sqrt{\pi} \sum_{i=1}^{N-1} n_i  \; {\vec  \alpha_i },   
\label{redundancy}
\end{eqnarray}
$n_i$ being integers.  In Eq. (\ref{redundancy}), ${\vec  \alpha_i }$ ($i=1, \ldots, N-1$) are the simple roots ($ {\vec  \alpha_i } = {\vec  e}_{i} - {\vec  e}_{i+1} $) which generate the root lattice $Q$ of the Lie algebra of SU($N$). 

Model (\ref{HcontJ1J2fer}) can then be bosonized using Eqs. (\ref{bosoabeleq}, \ref{invSUN}):
\begin{eqnarray}
 \mathcal{H}_{J_1-J_2} &=& v    \left( \left(\partial_x  {\vec  \Phi}_{sR} \right)^2 + 
\left(\partial_x  {\vec  \Phi}_{sL} \right)^2 \right)  +  \frac{\lambda }{2\pi}  \partial_x {\vec  \Phi}_{sR} \cdot  \partial_x  {\vec  \Phi}_{sL}  \nonumber \\
&-&  \frac{\lambda }{4\pi^2 a^2_0} \sum_{1 \le \alpha < \beta \le N} \cos \left( \sqrt{4\pi} \left( {\vec  e}_{\alpha} - {\vec  e}_{\beta}\right) \cdot {\vec  \Phi}_{s}\right).
\label{HcontJ1J2boso}
\end{eqnarray}
A similar expression has been obtained before in the bosonization approach of the U($N$) Thirring model \cite{Ha-84} whose low-energy properties are governed by model (\ref{HcontJ1J2fer}). 
We introduce the set of positive roots $\Delta_+$ of the Lie algebra of SU($N$):
\begin{align}
\Delta_+ = \{ \vec  e_\alpha-\vec  e_{\beta} \ | \ 1 \leq \alpha < \beta \leq N \},
\end{align}
to get:
\begin{eqnarray}
 \mathcal{H}_{J_1-J_2} &=& v    \left( \left(\partial_x  {\vec  \Phi}_{sR} \right)^2 + 
\left(\partial_x  {\vec  \Phi}_{sL} \right)^2  \right)  +  \frac{\lambda }{2\pi}  \partial_x {\vec  \Phi}_{sR} \cdot  \partial_x  {\vec  \Phi}_{sL}  \nonumber \\
&-&  \frac{\lambda }{4\pi^2 a^2_0} \sum_{\vec{\alpha} \in \Delta_+} \cos \left( \sqrt{4\pi}  \; {\vec  \alpha}  \cdot {\vec  \Phi}_{s}\right).
\label{HcontJ1J2bosofin}
\end{eqnarray}

When $\lambda >0$, the interacting part of model (\ref{HcontJ1J2fer}) is a marginal relevant operator and a perturbative spectral gap is formed. Its bosonized description (\ref{HcontJ1J2bosofin}) takes the form of a sum of sine-Gordon models. In the fully gapped phase, the bosonic fields ${\vec  \Phi}_{s}$ are pinned in the degenerate minima of the cosine potential of Eq. (\ref{HcontJ1J2bosofin}):
\begin{eqnarray}
\langle {\vec  \Phi}_{s} \rangle = \sqrt{\pi} \sum_{i=1}^{N-1} p_i  \; {\vec  \omega_i },   
\label{pinning}
\end{eqnarray}
$p_i$ being integers. In Eq. (\ref{pinning}), ${\vec  \omega_i }$ ($i=1, \ldots, N-1$) are the fundamental weights which generate the weight lattice $P$ of the Lie algebra of SU($N$). They are defined as follows in terms of the ${\vec  e}_{\alpha}$ vectors (\ref{weightSUN}):
\begin{eqnarray}
 {\vec  \omega_i } = \sum_{j=1}^{i}{\vec  e}_{j} ,  
\label{fundamentalweigths}
\end{eqnarray}
so that ${\vec  \omega_i }  \cdot {\vec  \alpha_j }   = \delta_{ij} $. Using the identification (\ref{redundancy}), we find that the ground state of model (\ref{HcontJ1J2bosofin}) is $N$-fold degenerate since the ratio $P/Q$ of the SU($N$) lattices is isomorphic to the center of SU($N$): $P/Q \sim {\mathbb Z}_N$. \cite{DiFrancesco-M-S-book}
The $N$ inequivalent ground states ($|  k \rangle, k=0, \ldots, N -1$) of the sine-Gordon models  (\ref{HcontJ1J2bosofin}) are chosen such that: 
\begin{eqnarray}
\langle {\vec  \Phi}_{s} \rangle = - \sqrt{\pi}  \; {\vec  \omega_k},   
\label{Npinning}
\end{eqnarray}
with the convention ${\vec  \omega_0} = {\vec  0}$. In these ground states, the $N$-merization operator (\ref{Nmerisationboso}) takes $N$ different values since
\begin{eqnarray}
\langle {\rm Tr} g  \rangle = \sqrt{N} e^{ i  \frac{2 k \pi}{N}}.
\label{N-merisationGSg}
\end{eqnarray}
This signals the spontaneous breaking of the one-step translation symmetry T$_{a_0}$ in the $N$-merized phase since under T$_{a_0}$ 
the SU($N$)$_1$ WZNW $g$ field transforms as $g \rightarrow e^{ i  \frac{2  \pi}{N}} g$ (see Eq. (\ref{Tao})).

The low-lying excitations of the $N$-merized phase are massive solitons and anti-solitons of the sine-Gordon models  (\ref{HcontJ1J2bosofin})  which represent domain walls between different ground states (\ref{Npinning}). There are $N$ solitons between two consecutive ground states 
$|  k \rangle$ and $|  k +1 \rangle$, ($k=0, \ldots, N -1$) of the $N$-merized phase (with the identification $ |  N \rangle =  |  0 \rangle $). From Eq. (\ref{Npinning}), we deduce that the field configurations related to such solitons are 
\begin{eqnarray}
 {\vec  \Phi}_{s} \left(\infty\right) -  {\vec  \Phi}_{s} \left(- \infty\right)= -  \sqrt{\pi}  \; \left( {\vec  \omega_{k+1}} - {\vec  \omega_k}\right),
\label{kink}
\end{eqnarray}
with the identification ${\vec  \omega_{N}}= {\vec  \omega_{0}} = {\vec  0}$.
The quantum numbers of these excitations can be determined by considering the $N-1$ mutual commuting conserved charges $Q^m$ ($m=1, \ldots, N-1$) of model (\ref{HcontJ1J2fer}).  They correspond to the $N-1$ diagonal Cartan generators $H^{m}$  in the defining representation $\bolN$:
\begin{equation}
[H^{m}]_{\alpha \beta} = \frac{1}{\sqrt{2 m(m+1)}} \delta_{\alpha \beta} \left( \sum_{l=1}^{m} \delta_{ \beta l} - m \delta_{\beta, m+1} \right)  \; .
\label{eqn:Cartan-in-GellMann}
\end{equation}
Using Eqs. (\ref{spinopfermion}, \ref{spinopapp}) and the bosonized descriptions (\ref{bosoabeleq}, \ref{SUNbasis}), the $Q^m$ charges are given by:
\begin{equation}
{\vec  Q} =  \frac{1}{\sqrt{2 \pi}} \int_{-\infty}^{\infty} d x \; \partial_x {\vec  \Phi}_{s}  .
\label{conservedcharges}
\end{equation}
We deduce that the quantum numbers ${\vec  Q}_{k+1} $ of the  excitation (\ref{kink}) between the two consecutive ground states 
$|  k \rangle$ and $|  k +1 \rangle$ ($k=0, \ldots, N -1$) are 
\begin{equation}
{\vec  Q}_{k+1}  =  - \frac{ {\vec  \omega_{k+1}} - {\vec  \omega_k}}{\sqrt{2}} =  - \frac{\vec{e}_{k+1}}{\sqrt{2}} .
\label{chargeskinkk}
\end{equation}
From Eq. (\ref{eq:SpecificWeight}), we observe that the $N$ quantum numbers (\ref{chargeskinkk})  correspond to the $N$  weight vectors of the conjugate ${\bar \bolN}$-representation of the SU($N$) group which are the opposite of the weights of the $ \bolN$-representation.
For instance, in the $N=3$ case, we have from Eqs. (\ref{chargeskinkk}, \ref{eq:SpecificWeight}):
\begin{eqnarray}
{\vec Q}_1 &=& - \begin{pmatrix} \frac{1}{2},  \frac{1}{2 \sqrt{3}} \end{pmatrix} \, , \; 
{\vec Q}_2 = - \begin{pmatrix} -\frac{1}{2},  \frac{1}{2 \sqrt{3}} \end{pmatrix} \, , \; \; \nonumber \\
{\vec Q}_3 &=&  \begin{pmatrix} 0,  \frac{1}{\sqrt{3}}  \end{pmatrix} \; ,
\label{weightSU3}
\end{eqnarray}
and in the $N=4$ case, we have:
\begin{eqnarray}
  {\vec Q}_1 = - \begin{pmatrix} \frac{1}{2},  \frac{1}{2 \sqrt{3}},  \frac{1}{2 \sqrt{6}} \end{pmatrix} \, & , & \;
  {\vec Q}_2 = - \begin{pmatrix} -\frac{1}{2},  \frac{1}{2 \sqrt{3}},  \frac{1}{2 \sqrt{6}} \end{pmatrix} \, , \; \nonumber \\
{\vec Q}_3 = - \begin{pmatrix} 0,  - \frac{1}{\sqrt{3}},  \frac{1}{2 \sqrt{6}}  \end{pmatrix} \, & , & \;
{\vec Q}_4 = \begin{pmatrix} 0,  0,  \frac{1}{2} \sqrt{\frac{3}{2}}  \end{pmatrix} \; .
\label{weightSU4}
\end{eqnarray}

The massive solitons of the $N$-merized phase share thus the same quantum numbers as the original gapless spinons of the Sutherland model which transform in the ${\bar \bolN}$-representation of  SU($N$). \cite{Bouwknegt-S-96, SchurichtGreiterPRB06} The latter become massive in the $N$-merized phase but still deconfined excitations that correspond to the domain-wall excitations between the $N$-degenerate ground states. Similarly, the anti-kinks of the $N$-merized phase transform in the $ \bolN$-representation of  SU($N$) and can be viewed as massive SU($N$) anti-spinons.

%%%%%%%%%%%%%%%%%%%%%%%%%%%%%%%%%%%%%%%%%%%%%%%%%%%%
\section{Identification of fields}
\label{AppCFT}

In this Appendix, we exploit the conformal embedding  (\ref{embedding}) to rewrite  the interacting part of the Hamiltonian density (\ref{cft}) in terms of the fields in the SU($N$)$_2$ $\times$ ${\mathbb{Z}}_N$ basis. We use a different approach than the one presented in Ref. \onlinecite{Lecheminant-T-15}.

We introduce two operators written in terms of the WZNW fields of the SU($N$)$_1$ $\times$ SU($N$)$_1$ CFT
\begin{eqnarray}
 {\cal O}_1 &=& \mbox{Tr} (g_1  g^{\dagger}_2)  \nonumber \\
  {\cal O}_2 &=&   \mbox{Tr}g_1  \mbox{Tr}g^{\dagger}_2 - \frac{1}{N} \mbox{Tr} (g_1  g^{\dagger}_2)  .
\label{SUNprimaries}
\end{eqnarray}
In this Appendix, we find the expression of these fields in the SU($N$)$_2$ $\times$ ${\mathbb Z}_N$  basis.
In particular, we show that ${\cal O}_1$  does not depend on the SU($N$)$_2$ degrees of freedom and is a primary field of  the 
${\mathbb Z}_N$  CFT  with conformal weights $h,\bar h =(N-1)/N$:
\begin{eqnarray}
T_{{\mathbb Z}_N} (z)  {\cal O}_1(0,0)  &\sim&  \frac{N-1}{Nz^2}  {\cal O}_1 (0,0) +  \frac{1}{z}  \partial {\cal O}_1 (0,0) 
\nonumber \\
T_{SU(N)_2} (z)  {\cal O}_1(0,0)  &\sim& 0,
\label{primarySUN2sym}
\end{eqnarray}
where $T_{{\mathbb Z}_N}$ (respectively $T_{SU(N)_2}$) is the stress-energy tensor of the ${\mathbb Z}_N$ (respectively SU($N$)$_2$)  CFT.
We have similar equations for the antiholomorphic sector that we will not consider here for simplicity.
It is also shown that the operator ${\cal O}_2$  is a primary field of the  ${\mathbb Z}_N \times$ SU($N$)$_2$ CFT:
 \begin{eqnarray}
\left( T_{{\mathbb Z}_N} + T_{SU(N)_2}\right) (z)  {\cal O}_2(0,0)  &\sim&  \frac{N-1}{N z^2}  {\cal O}_2 (0,0) 
\nonumber \\
&+&  \frac{1}{z}  \partial {\cal O}_2 (0,0) ,
\label{primarySUN2antisym}
\end{eqnarray}
with holomorphic weight $h = h_{\sigma_2}  + h_{\rm adj} = (N-2)/2 N(N+2) + N/(N+2)  = (N-1)/N$, $h_{\sigma_2}$  (respectively $h_{\rm adj}$) being the holomorphic weight of the ${\mathbb Z}_N$ (respectively SU($N$)$_2$) primary field $\sigma_2$  (respectively $\Phi_{\rm adj}$).

\subsection{Preliminaries}

We first present our normalization for the  SU($N$)$_1$ current which is defined by the following operator product expansion
(OPE):
\begin{equation}
J^{A}_{L} (z)  J^{B}_{L}(0) \sim \frac{\delta^{AB}}{8 \pi^2 z^2} +   \frac{i f^{ABC}  J^{C}_{L}(0) }{2 \pi z} ,
 \label{OPESUN1curr}
\end{equation}
where $f^{ABC}$ is the structure constants of the SU($N$) Lie algebra: $\left[ T^{A}, T^{B} \right] = i f^{ABC} T^{C} $.
The stress-energy tensor of the SU($N$)$_1$ CFT in our normalization is:
\begin{equation}
T_{SU(N)_1} = \frac{4\pi^2}{N+1}  : J^{A}_{L} J^{A}_{L}: .
\label{stresstensorSUN1}
\end{equation}
The defining OPE of the SU($N$)$_1$ WZNW $g$ field reads as:\cite{DiFrancesco-M-S-book}
\begin{eqnarray}
J_{ L}^A\left(z\right)  g_{\alpha \beta} (0,0) &\sim& - \frac{1}{2 \pi z} \; T^{A}_{\alpha \gamma}
g_{\gamma \beta} (0,0) \nonumber \\
J_{L}^A\left(z\right)  g^{\dagger}_{\beta \alpha} (0,0) &\sim& \frac{1}{2 \pi z}  \;
 g^{\dagger}_{\beta  \gamma} (0,0) T^{A}_{\gamma \alpha} ,
\label{OPEWZW}
\end{eqnarray}
$ T^{A}$ being the generator which transforms in the fundamental representation of the 
SU($N$) group. Actually, we need the subleading contribution of the OPE (\ref{OPEWZW}) to establish 
the relations (\ref{primarySUN2sym}, \ref{primarySUN2antisym}).
In this respect, let us now show that for an SU($N$)$_1$ CFT, we have:
\begin{eqnarray}
J_{L}^A\left(z\right)  g_{\alpha \beta} (0,0) &\sim& - \frac{1}{2 \pi z} \; T^{A}_{\alpha \gamma}
g_{\gamma \beta} (0,0)    \nonumber \\
&+& \frac{C}{2\pi} \; T^{A}_{\alpha \gamma}  \partial g_{\gamma \beta} (0,0)  \label{OPEWZWbis} \\
J_{L}^A\left(z\right)  g^{\dagger}_{\beta \alpha} (0,0) &\sim& \frac{1}{2 \pi z}  \;
 g^{\dagger}_{\beta \gamma} (0,0) T^{A}_{\gamma \alpha} \nonumber \\
& -& \frac{C}{2\pi} \;  \partial g^{\dagger}_{\beta \gamma} (0,0) T^{A}_{\gamma \alpha},
\label{OPEWZWbiss}
\end{eqnarray}
with $C = - N/ (N-1)$.

Indeed, from Eq. (\ref{OPEWZWbis}), we get:
\begin{eqnarray}
 g_{\alpha \beta} (x, \bar w)  J_{L}^A\left(w\right) &\sim& 
- \frac{1}{2 \pi (w-x) } \; T^{A}_{\alpha \gamma}
g_{\gamma \beta} (x, \bar w)  \nonumber \\
&+& \frac{C}{2\pi}  \; T^{A}_{\alpha \gamma}  \partial g_{\gamma \beta} (w,\bar w)  \nonumber \\
&\sim& \frac{1}{2 \pi (x-w)} \; T^{A}_{
   \alpha \gamma}
g_{\gamma \beta} (w, \bar w)  \nonumber \\
&+& \frac{\left(C+1\right)}{2\pi}  \; T^{A}_{\alpha \gamma}  \partial g_{\gamma \beta} (w,\bar w)  .
 \label{gJ}
\end{eqnarray}

We then consider the following OPE:
\begin{eqnarray}
&& \contraction{}{g_{\alpha \beta}}{(z, \bar w)}{:J_{L}^A J_{L}^A :} 
 g_{\alpha \beta}(z, \bar w) :J_{L}^A J_{L}^A :\left(w\right) = 
\frac{1}{2 \pi i} \oint_{w} \frac{dx}{x-w} \; \left[
\contraction{}{g_{\alpha \beta}}{(z, \bar w)}{J_{L}^A (x)} g_{\alpha \beta}(z, \bar w)J_{L}^A (x)
\right. \nonumber \\
&\times&  \left. J_{L}^A (w) + J_{L}^A (x) \contraction{}{g_{\alpha \beta}}{(z, \bar w)}{J_{L}^A (w)} g_{\alpha \beta}(z, \bar w)J_{L}^A (w)
\right]  \nonumber \\
&=&
\frac{1}{2 \pi i} \oint_{w} \frac{dx}{x-w} \; \left[\frac{1}{2 \pi (z-x)} \; T^{A}_{\alpha \gamma} g_{\gamma \beta}(x, \bar w)J_{L}^A (w)
\right. \nonumber \\
&+&  \left. \frac{1}{2 \pi (z-w)} \; J_{L}^A (x) T^{A}_{\alpha \gamma}  g_{\gamma \beta}(w, \bar w)
\right].
 \label{gJOPE}
\end{eqnarray}
By using the results (\ref{OPEWZWbis},\ref{gJ}), we deduce:
\begin{eqnarray} 
 &&g_{\alpha \beta}(z, \bar w) :J_{L}^A J_{L}^A :\left(w\right) \sim 
\frac{N^2-1}{8 \pi ^2 N(z-w)^2} \; g_{\alpha \beta} (w, \bar w)  \nonumber  \\
&+& \frac{N^2-1}{4 \pi^2 N} \frac{C+1/2}{z-w} \;{\partial g}_{\alpha \beta} (w, \bar w),
 \label{gJOPEfin}
\end{eqnarray}
and thus using the definition (\ref{stresstensorSUN1})
\begin{eqnarray} 
&&g_{\alpha \beta}(z, \bar w) T_{SU(N)_1} (w) \sim
\frac{N-1}{2N(z-w)^2} \; g_{\alpha \beta} (w, \bar w)\nonumber \\
&+& \frac{N-1}{N} \frac{C+1/2}{z-w} \;{\partial g}_{\alpha \beta} (w, \bar w).
\label{gTOPEfin}
\end{eqnarray}

Finally, we find
\begin{eqnarray} 
&&T_{SU(N)_1} (z)  g_{\alpha \beta}(w, \bar w)  \sim
\frac{N-1}{2N(z-w)^2} \; g_{\alpha \beta} (w, \bar w) \nonumber \\
&-& \frac{N-1}{N} \frac{C}{z-w} \;{\partial g}_{\alpha \beta} (w, \bar w) .
\label{priOPEfin}
\end{eqnarray}
Since $g_{\alpha \beta}$ is an SU($N$)$_1$ primary field with holomorphic weight $h = (N-1)/2N$, we have by definition
\begin{eqnarray} 
&&T_{SU(N)_1} (z)  g_{\alpha \beta}(w, \bar w)  \sim
\frac{N-1}{2N(z-w)^2} \; g_{\alpha \beta} (w, \bar w) \nonumber \\
&+&  \frac{\;{\partial g}_{\alpha \beta} (w, \bar w)}{z-w},
\label{priOPEfdef}
\end{eqnarray}
and we thus obtain $C = - N/(N-1)$.

\subsection{Identifications of ${\bf {\cal O}_{1,2}}$ operators}

We are now in position to show the identities (\ref{primarySUN2sym}, \ref{primarySUN2antisym}).
The holomorphic stress-energy tensors of two decoupled SU($N$)$_1$ CFTs read as follows in terms of their currents:
\begin{eqnarray}
T_{SU(N)_1\times SU(N)_1} &=&  T_{SU(N)_1} + T_{SU(N)_1} \nonumber \\ 
&=& \frac{4 \pi^2}{N+1} \sum_{l=1}^{2} : J^{A}_{lL} J^{A}_{lL}:,
\label{stresstensor}
\end{eqnarray}
while the SU($N$)$_2$ one is in terms of the diagonal current $I^{A}_{L} = J^{A}_{1L}  + J^{A}_{2L}$:
 \begin{equation}
T_{SU(N)_2} = \frac{4\pi^2}{N+2}  : I^{A}_{L} I^{A}_{L}: .
\label{stresstensorSUN2}
\end{equation}
From the coset construction of the embedding  (\ref{embedding}), we deduce the stress-energy tensor of the ${\mathbb Z}_N$
parafermionic CFT:
\begin{eqnarray}
T_{{\mathbb Z}_N} &=&   \frac{4\pi^2}{(N+1)(N+2)}  \left( \sum_{l=1}^{2} : J^{A}_{lL} J^{A}_{lL}:
\right. \nonumber \\ 
&-& \left. 2(N+1) J^{A}_{1L} J^{A}_{2L} \right) .
\label{stresstensorZN}
\end{eqnarray}

Using the fact that $g_1$ and $g^{\dagger}_2$ are SU($N$)$_1$ primaries field with holomorphic weight $h= (N-1)/2N$, we have
\begin{eqnarray} 
4 \pi^2 \sum_{l=1}^{2} :J^{A}_{lL} J^{A}_{lL}:&(z)&  {\cal O}_{1,2} (0,0) \sim \frac{N^2-1}{Nz^2} {\cal O}_{1,2} (0,0) 
\nonumber \\
&+& \frac{N+1}{z}  \partial {\cal O}_{1,2} (0,0) .
\label{curcurO12}
\end{eqnarray}
We need also the following OPEs which can be deduced from the results (\ref{OPEWZWbis}, \ref{OPEWZWbiss}):
\begin{eqnarray} 
4 \pi^2 J^{A}_{1L} J^{A}_{2L}:(z)  {\cal O}_{1} (0,0) &\sim& - \frac{N^2-1}{2Nz^2} {\cal O}_{1} (0,0) 
\nonumber \\
&-& \frac{N+1}{2z}  \partial {\cal O}_{1} (0,0)  \label{cur1cur2O12}   \\
4 \pi^2 J^{A}_{1L} J^{A}_{2L}:(z)  {\cal O}_{2} (0,0) &\sim&  \frac{1}{2Nz^2} {\cal O}_{2} (0,0) 
\nonumber \\
&+& \frac{1}{2(N-1)z}  \partial {\cal O}_{2} (0,0)  .
\nonumber
\end{eqnarray}
It is now straightforward to check that the ${\cal O}_{1}$ field satisfies the OPEs (\ref{primarySUN2sym}) which 
means that it is a singlet under the SU($N$)$_2$ CFT and a ${\mathbb Z}_N$ primary field with 
conformal weights $((N-1)/N, N-1)/N)$ as $ \Psi_{1L}  \Psi_{1R}$:
\begin{equation}
 {\cal O}_1 =  \mbox{Tr} (g_1  g^{\dagger}_2)
 \sim   \Psi_{1L}  \Psi_{1R} .
 \label{identO1fin}
\end{equation}

Finally, we find that the ${\cal O}_{2}$ field satisfies the following OPEs from Eqs. (\ref{curcurO12}, \ref{cur1cur2O12}):
\begin{eqnarray}
T_{{\mathbb Z}_N} (z)  {\cal O}_2 (0,0)  &\sim& \frac{N - 2}{N(N+2)z^2}  {\cal O}_2 (0,0)  \nonumber  \\
&+&  \frac{N - 2}{(N-1)(N+2)z}  \partial {\cal O}_2 (0,0)  \nonumber  \\
T_{SU(N)_2} (z)  {\cal O}_2 (0,0)  &\sim&  \frac{N}{(N+2)z^2}  {\cal O}_2 (0,0) 
\label{O2Ope} \\
&+&  \frac{N^2}{(N-1)(N+2)z}  \partial {\cal O}_2 (0,0) 
\nonumber .
\end{eqnarray}
We thus deduce that the ${\cal O}_{2}$ field satisfies the OPE (\ref{primarySUN2antisym}).
It is not a singlet of ${\mathbb Z}_N $ or SU($N$)$_2$ CFTs but a primary field of the ${\mathbb Z}_N \times$ SU($N$)$_2$ CFT with holomorphic weight $h =  h_{\sigma_2}  + h_{\rm adj} = (N-2)/N(N+2) + N/(N+2)  = (N-1)/N$. One then obtains the correspondence:
\begin{equation}
 {\cal O}_2 = \mbox{Tr}g_1  \mbox{Tr}g^{\dagger}_2 - \frac{1}{N} \mbox{Tr} (g_1  g^{\dagger}_2)
 \sim \sigma_2 \mbox{Tr} \Phi_{\rm adj} .
 \label{identO2fin}
\end{equation}

\subsection{Sign ambiguity}

The precise value of the coefficient in the identification (\ref{identO1fin}) is not important but its sign turns out to be crucial for the determination of the ground state of the two-leg SU($N$) zigzag ladder (\ref{eqn:def-2leg-ladder}) in the limit $J_1 \ll J_2$. The approach, presented here, cannot fix the sign. We give here an additional argument that
\begin{equation}
 \mbox{Tr} (g_1  g^{\dagger}_2)
 = a    \Psi_{1L}  \Psi_{1R} ,
 \label{identO1final}
\end{equation}
where $a$ is necessarily positive. 

In this respect, let us consider the following action which couples two SU($N$)$_1$ CFTs:
\begin{eqnarray}
{\cal S} &=& {\cal S}[SU(N)_1; g_1] +   {\cal S}[SU(N)_1; g_2]
\nonumber \\
&-&    \int d^2x \; \Big[  \mbox{Tr}(g_1g_2^+) + \mathrm{H.c.}\Big] ,
\label{actionSUNcoupled}
\end{eqnarray}
where the action of the SU($N$)$_k$ WZNW model is given by:\cite{Witten-84,Knizhnik-Z-84}
\begin{eqnarray}
 && {\cal S}[SU(N)_k; g] =  \frac{k}{8\pi} \int d^2 x \; {\rm Tr} \; (\partial^{\mu} g^{+} \partial_{\mu} g) 
+ \Gamma(g) \nonumber \\
\Gamma(g) &=& \frac{i k }{12\pi}   \int_B d^3 y \; \epsilon^{\alpha \beta \gamma} 
 {\rm Tr} \; (g^{+} \partial_{\alpha} g g^{+} \partial_{\beta} g g^{+} \partial_{\gamma} g),
\label{WZWaccc}
\end{eqnarray}
$g$ being an SU($N$) matrix field and $\Gamma(g)$ is the WZNW topological term.
Using the conformal embedding (\ref{embedding}) and the identification (\ref{identO1final}), we get
\begin{eqnarray}
{\cal S} &=& {\cal S}[SU(N)_2 ; g] +   {\cal S}_{{\mathbb Z}_N}
\nonumber \\
&-a&    \int d^2x \; \Big[     \Psi_{1L}  \Psi_{1R} + \mathrm{H.c.}\Big] .
\label{actionSUNcoupledemb}
\end{eqnarray}
The interacting part acts only in the ${\mathbb Z}_N$ sector and takes the form of the integrable perturbation (\ref{effactionparaJperpFtrans}) of the ${\mathbb Z}_N$ parafermions. As discussed in the main text, a perturbative mass opens when $a$ is positive and the action (\ref{actionSUNcoupled}) displays critical properties in the SU($N$)$_2$ universality class with central charge $c= 2(N^2-1)/(N+2)$. If $a <0$,  the ${\mathbb Z}_N$  parafermions exhibit a massless RG flow and action (\ref{actionSUNcoupled}) has a central charge larger than $c= 2(N^2-1)/(N+2)$.

We focus now directly on action (\ref{actionSUNcoupled}) and introduce the matrix $G = g_1g_2^+$ which is still an SU($N$) matrix.  Using the Polyakov-Wiegmann formula \cite{Polyakov-W-84}: 
\begin{eqnarray}
&& {\cal S}[SU(N)_1;  g_1] = {\cal S}[SU(N)_1; G g_2] = {\cal S}[SU(N)_1; G] \nonumber \\
 &+& {\cal S}[SU(N)_1; g_2] 
  - \frac{1}{4 \pi} \int d^2 x \; {\rm Tr} \; (g_2^{+} \partial  g_2 G  {\bar \partial} G^{+} ) ,
\label{PolyakovFormula}
\end{eqnarray}
action (\ref{actionSUNcoupled}) simplifies as
\begin{eqnarray}
{\cal S} = {\cal S}[SU(N)_1; G] &+&  2 {\cal S}[SU(N)_1; g_2] - \int d^2x \;   \mbox{Tr}(G) + \mathrm{H.c.}
\nonumber \\
&-&   \frac{1}{4 \pi} \int d^2 x \; {\rm Tr} \; (g_2^{+} \partial  g_2 G  {\bar \partial} G^{+} )  .
\label{actionSUNcoupled2}
\end{eqnarray}
In the far-IR limit, the $G$ field is pinned to the configuration $G = I$ and the fluctuations are massive. Integrating out these degrees of freedom, we get in the low-energy regime: 
\begin{eqnarray}
{\cal S} =   2 {\cal S}[SU(N)_1; g_2] = {\cal S}[SU(N)_2; g_2] ,
\label{actionSUNcoupled1}
\end{eqnarray}
and thus an SU($N$)$_2$ critical behavior as it should be. This gives a strong argument that the identification (\ref{identO1final})  with $a>0$ is correct.

\section{Semiclassical approach}
\label{Semiclassical}

In this Appendix, in the $N=3$ case, we provide an alternative approach  to the conformal embedding one (\ref{embedding}) to show the emergence of the 
SU(3)$_1$ critical phase in the phase diagram of the SU(3) two-leg zigzag spin ladder (\ref{eqn:def-2leg-ladder})  when $J_1 \ll J_2$.

Let us first consider the Euclidean action corresponding to Eq. (\ref{cft}): 
\begin{eqnarray}
{\cal S} &=& {\cal S}[SU(3)_1; g_1] +   {\cal S}[SU(3)_1; g_2]
\nonumber \\
&+&  \lambda_1  \int d^2x \; \Big[ e^{i \pi/3}  \mbox{Tr}(g_1g_2^+) + \mathrm{H.c.}\Big] 
\nonumber \\
  &+& \lambda_2  \int d^2x  \;  \Big[  e^{i \pi/3}  \mbox{Tr}g_1\mbox{Tr}g_2^+ + \mathrm{H.c.}\Big].
\label{cftaction}
\end{eqnarray}

As already emphasized in section II B,  the first term of Eq. (\ref{cftaction}) with coupling constant $\lambda_1$ has  a nonperturbative mass gap $\Delta \sim J_1^{N/2}$.  The minimization of this  term ($ \lambda_1  > 0$)  leads to  $g_1 = - e^{-i \pi/3}  g_2$, which is an SU(3) matrix. We integrate out the $g_1$ degrees of freedom to  deduce the effective action for $g_2 =G$:
\begin{equation}
{\cal S}_{\rm eff} =  {\cal S}[SU(3)_2; G] - 2 \lambda_2  \int d^2 x \;  |\mbox{Tr}  \; G|^2,
\label{seffstrongcoup}
\end{equation}
with $ \lambda_2  < 0$ since $J_1> 0$. The minimization of the action leads to the condition: $\mbox{Tr}  \; G = 0$.
The eigenvalues of the $G$ matrix are  the $3$-rd roots of unity and the fundamental WZNW SU(3) $G$  field can be written as:
\begin{eqnarray}
G &=& U \Omega U^{\dagger} \nonumber \\
\Omega &=&  
\begin{pmatrix}
1  & 0  & 0 \\
 0 & \omega & 0\\
0  & 0 & \omega^{2} 
\end{pmatrix} , \label{wzwfieldmanifold}\\
\nonumber  
\end{eqnarray}
$U$ being a general U(3) matrix and $\omega = e^{i 2\pi/3}$. We then introduce 9 complex scalar fields $\Phi_{i j}$ ($i,j = 1, \ldots, 3$) 
such that $ U_{i j} = \Phi_{i j} = (\vec \Phi_j)_i$. These fields are constraint to be orthonormal complex vectors: $\vec \Phi^{*}_i \cdot \vec \Phi_j = \delta_{ij}$ to enforce the U(3) property: $ U^{\dagger} U = I$. The identification (\ref{wzwfieldmanifold}) reads thus as follows in terms of the scalar fields: 
\begin{equation}
G_{i j} = \sum_{a=1}^{3}  \Phi^{*}_{j a}   \Omega_{aa}  \Phi_{i a} .
\label{Gfieldident}
\end{equation}
A U(1)$^3$ redundancy in the description (\ref{Gfieldident})  is manifest since the transformation $\vec \Phi_a \rightarrow e^{ i \theta_a} \vec \Phi_a$ gives the same $G_{i j}$ for all $\theta_a$ ($a =1, 2, 3$). Distinct scalar fields take thus value in the manifold U(3)/U(1)$^3$ $\sim$ SU(3)/U(1)$^{2}$, which is the flag manifold. \cite{Affleck-B-W2022}

The next step of the approach is to replace the identification (\ref{Gfieldident}) into the action (\ref{seffstrongcoup}) to derive the low-energy effective field theory for the complex fields $\vec \Phi_i$. The action  takes the form of a nonlinear sigma model on the flag manifold  SU(3)/U(1)$^{2}$ with topological $\theta$ terms with a Lagrangian density: \cite{Ohmori-S-S-19, Tanizaki-S-18}  
\begin{eqnarray}
&& {\cal L} = \frac{3}{4\pi} \sum_{a=1}^{3} \left( |\partial_{\mu}  {\vec \Phi_{a}} |^2
-  |{\vec \Phi^{*}}_{a} \cdot \partial_{\mu}  {\vec \Phi_{a}} |^2  \right)   \nonumber \\
&& \; \; \; \; \; \; +  \sum_{a=1}^{3}  \frac{\theta_a}{2\pi} \epsilon^{\mu \nu} 
\partial_{\mu}  {\vec \Phi^{*}}_{a} \cdot \partial_{\nu}  {\vec \Phi_{a}}
\label{flagsigma}  \\
&+&  \sum_{1 \le a < b \le 3}  \left(g_{ab} \delta^{\mu \nu}  +  b_{ab} \epsilon^{\mu \nu} \right) \left({\vec \Phi^{*}}_{a} \cdot \partial_{\mu}  {\vec \Phi_{b}}  \right) \left( {\vec \Phi^{*}}_{b} \cdot \partial_{\nu}  {\vec \Phi_{a}} \right),
\nonumber 
\end{eqnarray}
with $\theta_a = 4\pi a/3$ ($a = 1, 2, 3$), $g_{ab} = 3 \cos( 2\pi (a-b)/3)/2\pi$ and $b_{ab}  
= 3 \sin( 2\pi (a-b)/3)/2\pi$. Model (\ref{flagsigma}) contains three topological angles $\theta_a$ with topological charges: 
\begin{equation}
q_{a} = \frac{i}{2\pi}   \int d^2 x  \epsilon^{\mu \nu} 
\partial_{\mu}  {\vec \Phi^{*}}_{a} \cdot \partial_{\nu}  {\vec \Phi_{a}} 
\label{topocharges}
\end{equation}
which are integers. However, the topological charges are not all independent due to the orthormalization constraint: 
$\vec \Phi^{*}_i \cdot \vec \Phi_j = \delta_{ij}$ and satisfy: $\sum_{a=1}^{3} q_{a} = 0$ since it can be shown 
$\sum_{a=1}^{3} \vec \Phi^{*}_{a} \cdot \partial_{\mu}  {\vec \Phi_{a}}  = 0$.\cite{Tanizaki-S-18,Lajko-W-M-A-17}  
It implies that model (\ref{flagsigma}) is left invariant by shifting all topological angles by a same amount: $\theta_a \rightarrow \theta_a + \theta$ for all $a$.
There are thus two independent topological angles $\theta_a = 4\pi a/3$ ($a = 1, 2$) in model (\ref{flagsigma})
in full agreement with the value of the second homotopy group for the flag manifold: $\Pi_2$ (SU(3)/U(1)$^{2}) = \mathbb{Z} \times \mathbb{Z} $. \cite{Bykov2012}

It has been shown recently that the flag sigma model (\ref{flagsigma}) with topological angles  $\theta_a = 2\pi p a/3$ control the IR properties of  SU(3) Heisenberg spin chain in symmetric rank-$p$ tensor representation in the large $p$ limit.\cite{Lajko-W-M-A-17} 
A gapless phase in the SU(3)$_1$ universality class has been predicted for model (\ref{flagsigma}) when $p$ and 3 are coprime.\cite{Ohmori-S-S-19, Tanizaki-S-18,Warmer2020, Warmer-L-M-A2020} In the particular $p=2$ case, which share the same topological angles as in (\ref{flagsigma}), a large-scale  DMRG calculation has shown the existence of a gapless SU(3)$_1$ behavior with central charge $c=2$. \cite{Nataf-G-M2021}
We thus conclude that the action (\ref{cftaction}) describes a gapless
behavior in  the SU(3)$_1$ universality class. This leads to a second argument which shows that  the two-leg SU(3) zigzag spin ladder (\ref{eqn:def-2leg-ladder}) 
displays a critical SU(3)$_1$  phase in the regime $J_1 \ll J_2$. 

\section{Additional numerical results}\label{app:AddNumRes}
\begin{figure}[!htb]
\begin{center}
    \includegraphics[width=0.8\linewidth,clip]{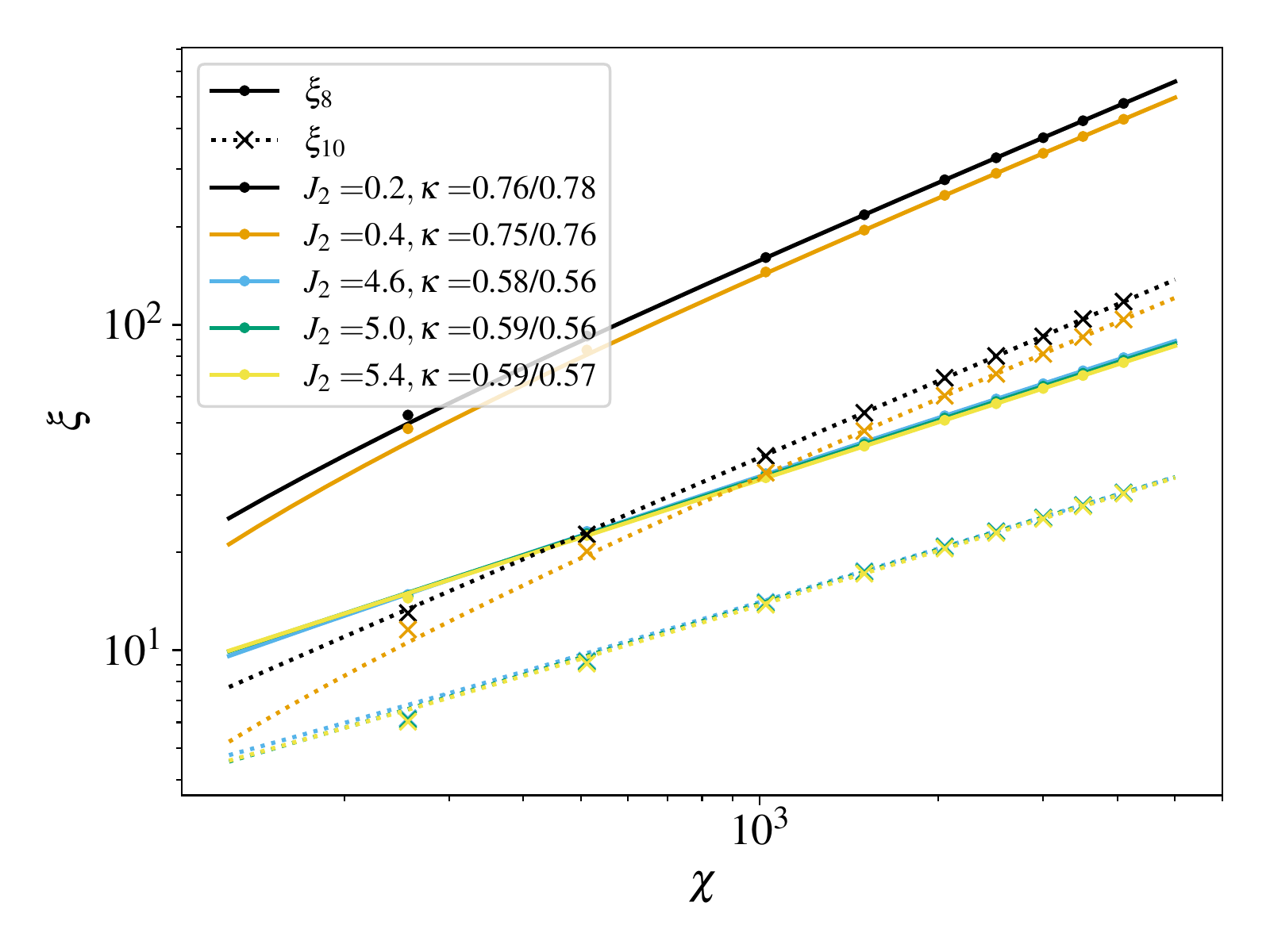}
    \includegraphics[width=0.8\linewidth,clip]{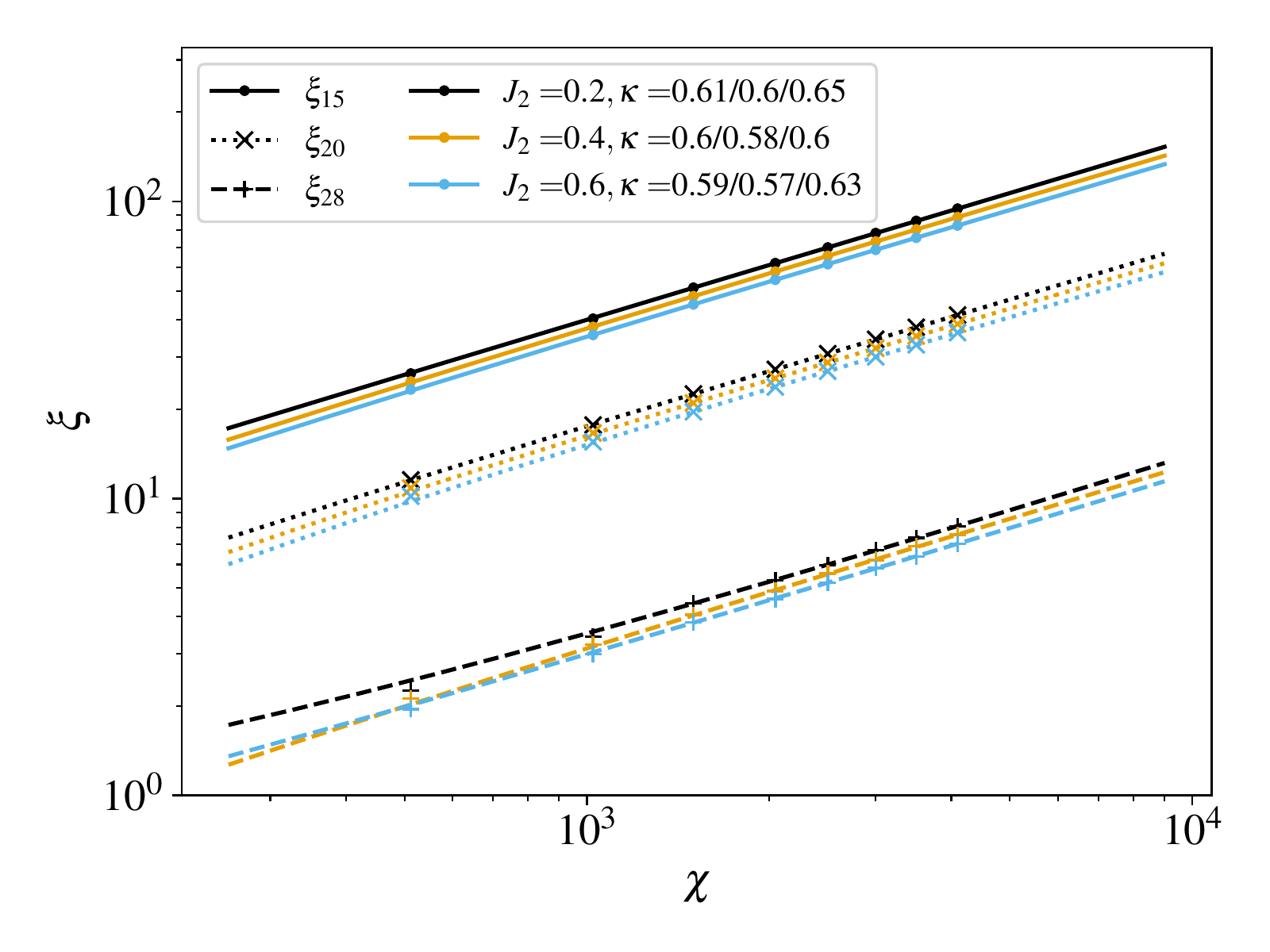}
\end{center}
    \caption{Scaling of the dominant correlation length with the bond dimension for SU(3) (top) and SU(4) (bottom). In both cases, the exponents slightly deviate from the usual predictions~\citep{Pollmann2009}.
    }
    \label{fig:corrScaling}
\end{figure}

We discussed in the main text the scaling of the correlation length with the bond dimension in the different gapless phases.
Fig.~\ref{fig:corrScaling} presents the relevant data.
We find that $\xi$ does scale as $\chi^\kappa$, but $\kappa$ deviates from the predicted value~\cite{Pollmann2009}
\begin{equation}
\kappa=\frac{6}{c (\sqrt{\frac{12}{c}}+1)}.
\end{equation}

\begin{figure}[!htb]
\includegraphics[width=\linewidth,clip]{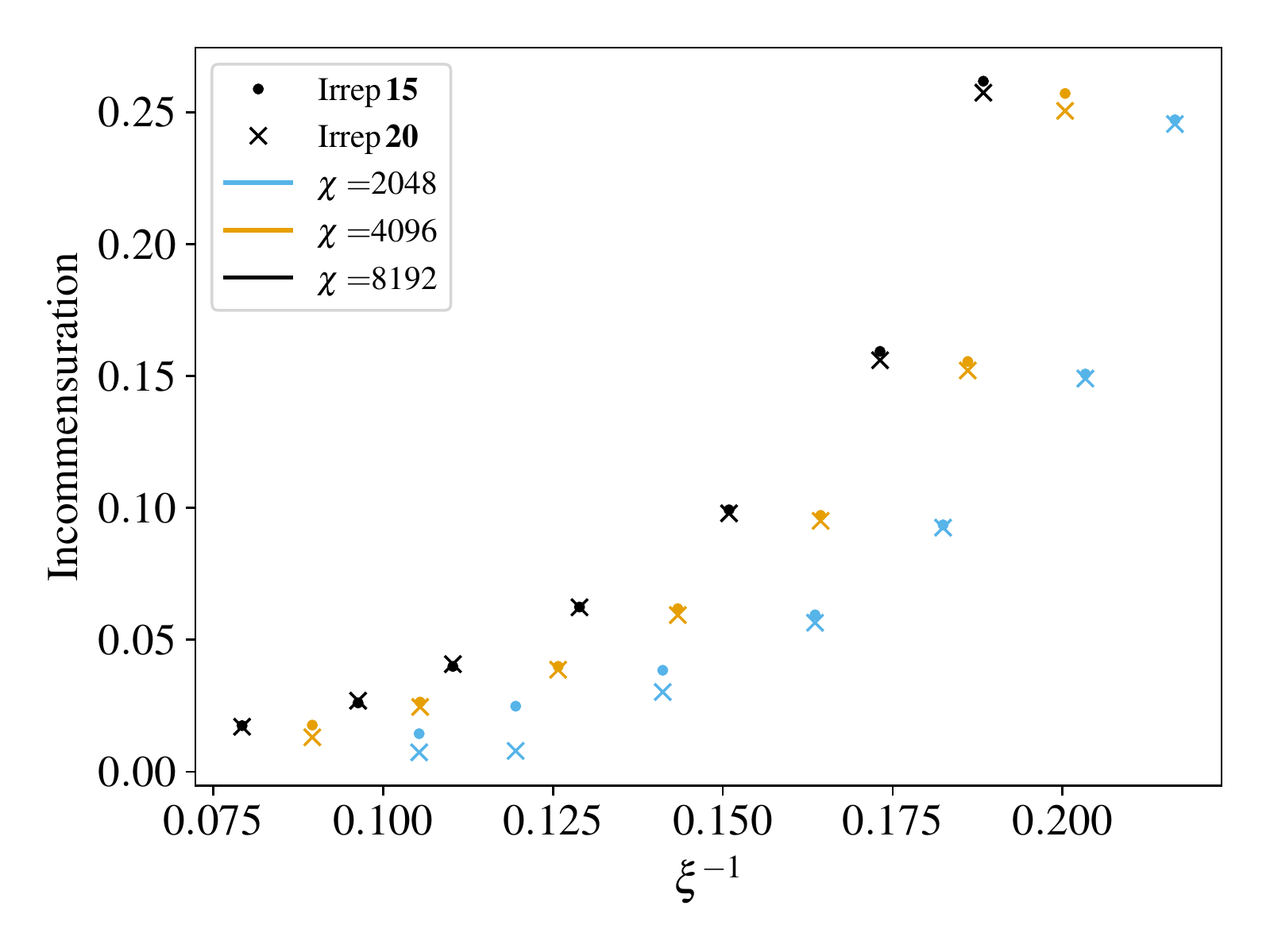}
    \caption{Scaling of the incommensuration with the correlation length. We do not observe the linear behavior seen in the SU(2) spin chain, though it could be due only to finite convergence. }
    \label{fig:incScaling}
\end{figure}

In Fig.~\ref{fig:incScaling}, we show the scaling of the incommensuration with the correlation length for the SU(4) chain above $J_2 = 2$. We have not yet reached the scaling regime for the coherence length, though the incommensuration is largely converged.

\begin{figure}[!htb]
\includegraphics[width=\linewidth,clip]{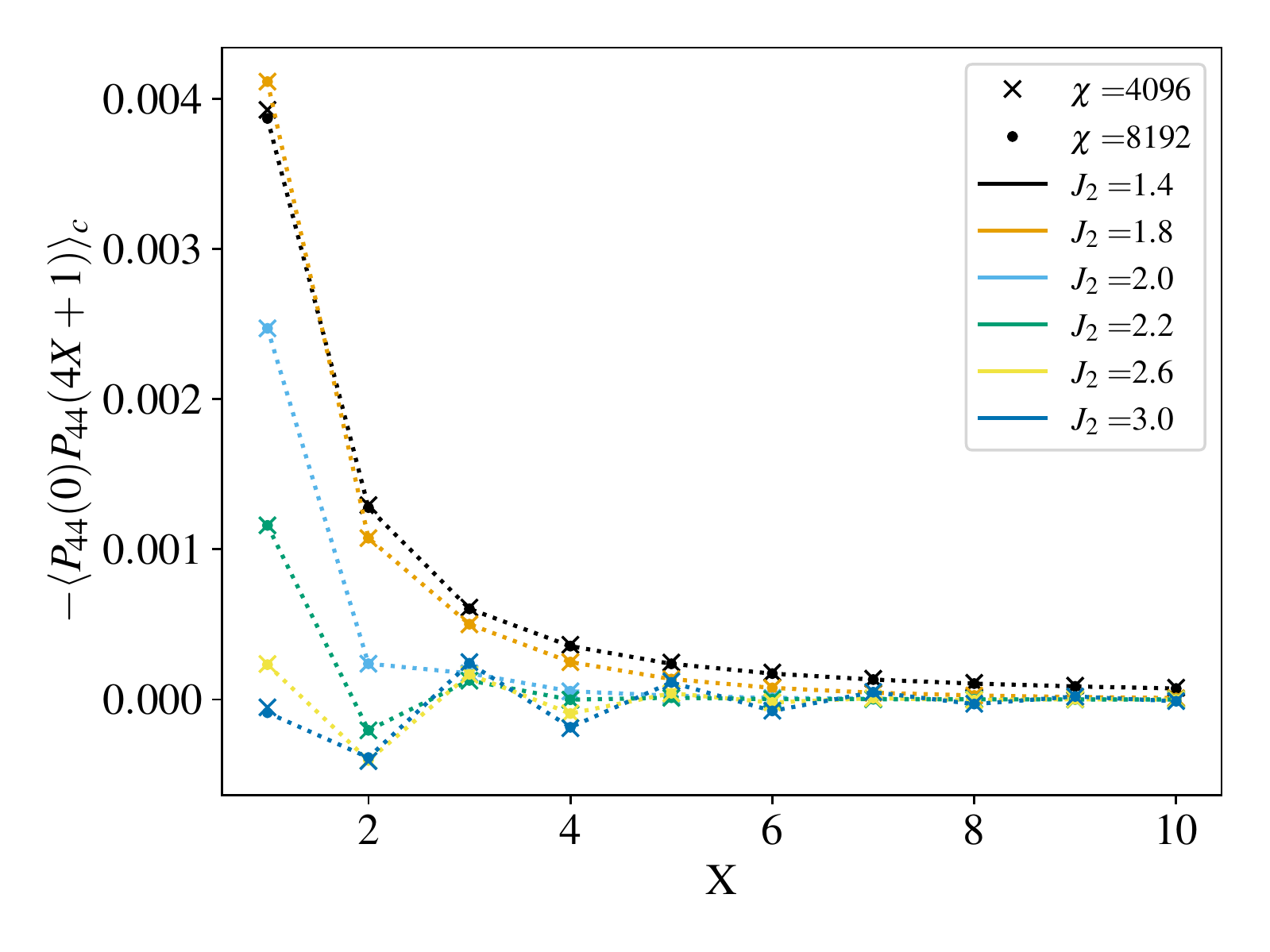}
    \caption{Correlations of the projector $\hat{P}_{44}$ on the fourth state of the local Hilbert space for different values of $\chi$. The commensurate-incommensurate transition at $J_2 \approx 2.0$ is clearly visible.}
    \label{fig:CorrP44}
\end{figure}

Finally, in Fig.~\ref{fig:CorrP44}, we represent the connected correlations of the projector $\hat{P}_{44}$ on the local state $(0, 0, 0, 1)$. The transition towards an incommensurate state at $J_2 \approx 2.0$ is directly visible in the oscillations of the exponentially decaying correlation functions, even though the corresponding eigenvalues of the transfer matrix are subdominant.\\

\section{SU(2) case}
\label{app:su2}
\begin{figure}[!htb]
\includegraphics[width=\linewidth,clip]{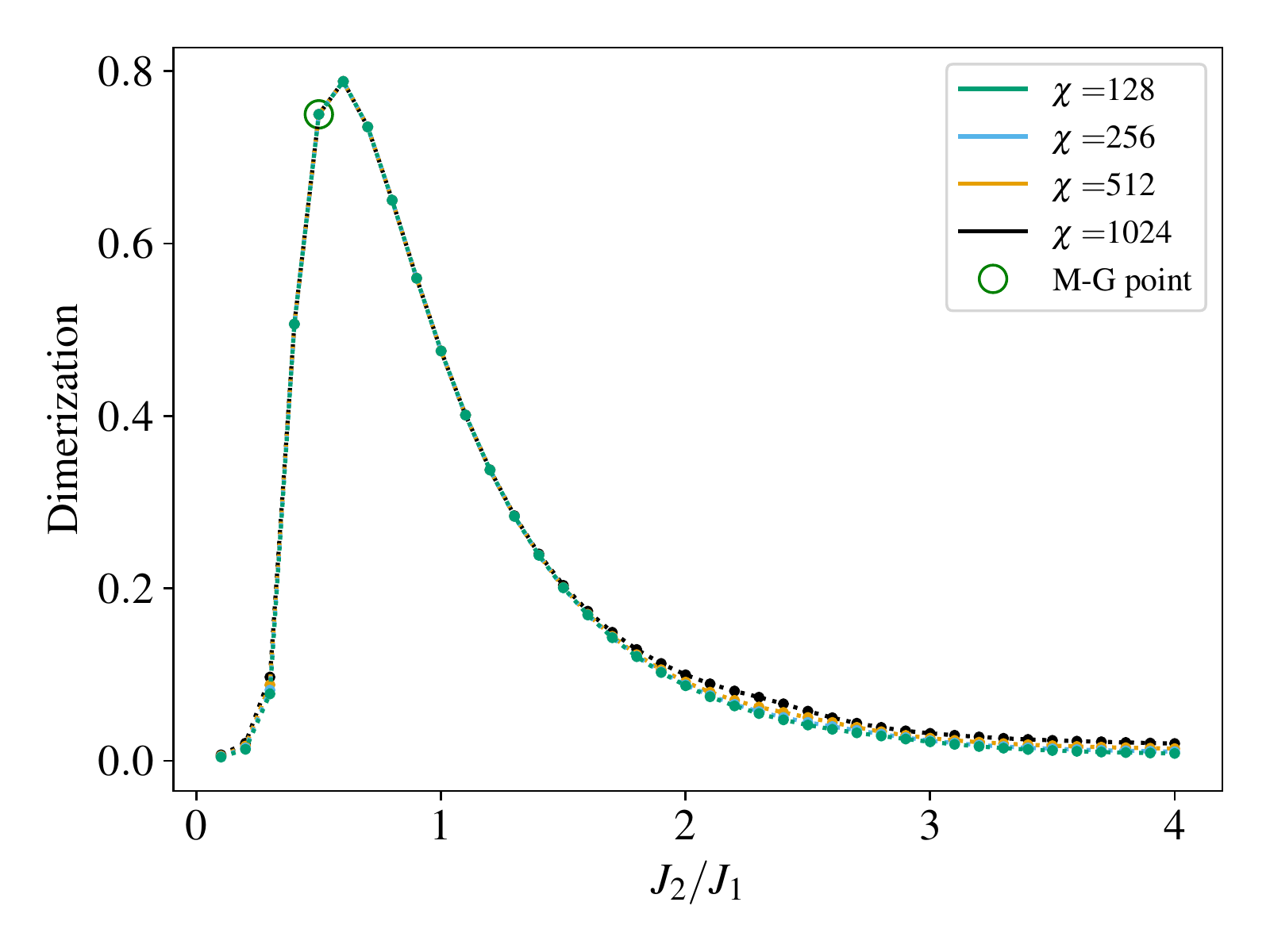}
    \caption{Dimerization in the SU(2) $J_1-J_2$ Heisenberg spin chain as a function of $J_2/J_1$. At $J_2\simeq 0.24$, a gapped dimerized phase opens. The exact product state described by Majumdar and Ghosh~\cite{MajumdarGhosh69} is reached at $J_2 = 0.5 J_1$. Small bond dimensions are enough to capture the physics of the chain.}
    \label{fig:dmrg_su2}
\end{figure}

\begin{figure}[!htb]
\includegraphics[width=\linewidth,clip]{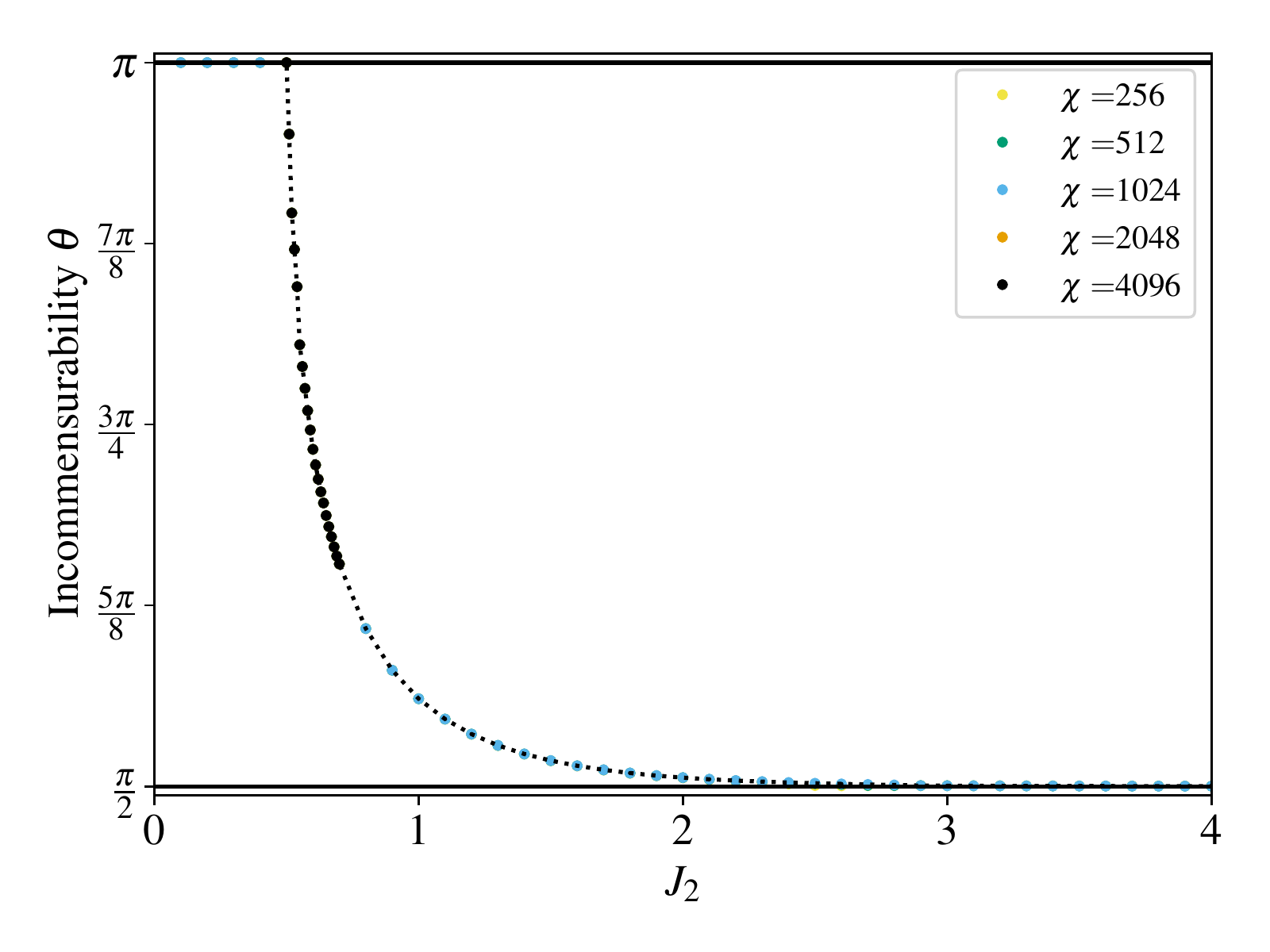}
    \caption{Incommensurate angle for the SU(2) $J_1$-$J_2$ Heisenberg spin chain, measured as in Eq.~\eqref{eq:defInc}. The incommensurability opens at the Majumdar-Ghosh point. Note that due to our unit cell, the angle is here defined modulo $\pi/2$. As we know that in the $J_2 = 0$ limit, the system is fully antiferromagnetic, we chose $\theta = \pi$ in the critical phase.}
    \label{fig:incommensuration_SU2}
\end{figure}

For completeness, we review the well-studied case of SU(2) case, namely a spin-1/2 zigzag $J_1-J_2$ chain. For $J_2=0$ this is the simple critical Heisenberg chain (with central charge $c=1$). Using field-theory and exact diagonalization, the second-neighbour spin exchange opens a gap and a spontaneous dimerization appears when $J_2/J_1 \simeq 0.241167$, which can be found with high accuracy~\cite{Eggert-96}. Then the dimerized phase persists for all larger $J_2$ couplings, including, in particular, the exact solution found by Majumdar and Ghosh~\cite{MajumdarGhosh69} for $J_2/J_1=1/2$. For reference, Fig.~\ref{fig:dmrg_su2} shows the measured dimerization obtained using iDMRG with relatively small bond dimensions. Quite interestingly, above this point, correlations become incommensurate~\cite{White-A-96} with a pitch angle reminiscent of the classical model, as shown in Fig.~\ref{fig:incommensuration_SU2}.

%%%%%%%% REFERENCES %%%%%%%%%%%%%%%%%%%%%%%%%%%%%%%%%%%
\bibliographystyle{apsrev4-1}
\bibliography{./references/SU3_SU3bar,./references/misc,./references/ref,./references/SUN,./references/topological_order,./references/field_theory}

\end{document}